%% file: koprowski.tex
\documentclass[useAMS]{mn2e}

\usepackage{graphicx,lscape,amssymb,amsmath,url}

\begin{document}
\hsize=6truein

%%%%%%%%%%%%%%%%%%%%%%%%%%%%%%%%%%%%%%%%%%%%%%%%

\title[A reassessment of properties of luminous SMGs]{A reassessment of the redshift distribution and physical properties of luminous (sub-)millimetre galaxies}
\author[M.\,P. Koprowski et al.]{M.\,P. Koprowski$^{1}$\thanks{E-mail: mpk@roe.ac.uk},
J.\,S. Dunlop$^{1}$, M.\,J. Micha{\l}owski$^{1,2}$\addtocounter {footnote} {1}\thanks{FWO Pegasus Marie Curie Fellow},
M. Cirasuolo$^{1,3}$,\and
R.\,A.\,A. Bowler$^{1}$\\
\setcounter{footnote}{1}$^{1}$SUPA\thanks{Scottish Universities Physics Alliance}, Institute for Astronomy, University of Edinburgh, Royal Observatory, Edinburgh, EH9 3HJ\\
$^{2}$Sterrenkundig Observatorium, Universiteit Gent, Krijgslaan 281 S9, 9000 Gent, Belgium\\
$^3$UK Astronomy Technology Centre, Royal Observatory, Edinburgh, EH9 3HJ\\
}

%\date{Accepted ???. Received ???; in original form ???}

\pagerange{\pageref{firstpage}--\pageref{lastpage}} \pubyear{2012}

\maketitle

\label{firstpage}

\begin{abstract}

Motivated by the current controversy over the redshift distribution and 
physical properties of luminous (sub-)mm sources, we have undertaken a new study 
of the brightest sample of unlensed (sub-)mm sources with pre-ALMA 
interferometric follow-up in the COSMOS field. Exploiting the very latest 
multi-frequency supporting data, we find that this sample displays 
a redshift distribution indistinguishable from that of the lensed 
sources uncovered with the South Pole Telescope (SPT), 
with $z_{median} \simeq 3.5$. We also find that, over the redshift 
range $z \simeq 2-6$ the median  
stellar mass of the most luminous (sub-)mm sources is $M_{\star} \simeq 3 
\times 10^{11}\,{\rm M_{\odot}}$, yielding a  
typical specific star-formation rate $sSFR \simeq 3\, {\rm Gyr^{-1}}$. 
Consistent with recent ALMA and SMA 
studies, we confirm that source blending is {\it not} a serious issue
in the study of luminous (sub-)mm sources uncovered by ground-based, 
single-dish surveys; only $\simeq 10-15$\% of bright ($S_{850} 
\simeq 5 - 10$\,mJy) (sub-)mm sources arise from significant (i.e. $>20$\%) 
blends, and so our conclusions are largely unaffected by whether we 
adopt the original single-dish
mm/sub-mm flux-densities/positions, or the interferometric data.
Our results suggest that apparent disagreements over the 
redshift distribution of 
(sub-)mm sources are a result of ``down-sizing'' in
dust-enshrouded star-formation, consistent with 
existing knowledge of the star-formation histories of massive galaxies.
They also indicate that extreme star-forming galaxies at high redshift are,
on average, subject to the same star-formation rate-limiting processes 
as less luminous objects, and lie on the ``main sequence'' of 
star-forming galaxies at $z>3$.

\end{abstract}


\section{Introduction}

Since their discovery 15 years ago in the first blank-field SCUBA surveys at the James Clerk Maxwell Telescope (JCMT), 
it has been known that sub-mm sources selected at high galactic latitudes are luminous dust-enshrouded star-forming 
galaxies, primarily located at high redshifts ($1 < z < 5$; Hughes et al. 1998; Barger et al. 1998). 
Indeed, in a hint of things to come, it was quickly realised that the brightest sub-mm source uncovered in the first 
850\,$\mu$m image of the Hubble Deep Field North, HDF850.1, was not visible in the ultra-deep 
{\it Hubble Space Telescope (HST)} optical imaging then 
available, and a number of follow-up studies suggested that it most likely lay at $z > 4$ (Downes et al. 1999; 
Dunlop et al. 2004; Cowie et al. 2009). Recently, in an impressive demonstration of the ever-improving capabilities 
of mm/sub-mm spectroscopy, HDF850.1 has been revealed to lie at $z = 5.2$ (Walter et al. 2012).

Despite the fairly extreme redshift of the first blank-field sub-mm source ever discovered, improved
and expanded sub-mm/mm surveys over the last decade undertaken with SCUBA, LABOCA, AzTEC 
and MAMBO have generally yielded a consistent picture, whereby sources selected at $S_{850} \simeq  
5$\,mJy display a redshift distribution which peaks at $z \simeq 2.5$, albeit with a 
significant lower-redshift tail down to $z \simeq 1$, and a high-redshift tail 
extending up to $z \simeq 4 -5$. In general this information has been gleaned from either 
optical spectroscopic redshifts (e.g. Chapman et al. 2003, 2005) or from (more complete, but less accurate) 
optical-infrared photometric redshifts (e.g. Clements et al. 208; Dye et al. 2008; Chapin et al. 2009;  
Dunlop et al. 2010; Wardlow et al. 2011; Micha{\l}owski et al. 2012a) derived for the galaxy counterparts identified via the improved 
spatial information provided by radio and/or {\it Spitzer} observations of the (sub-)mm sources
(e.g. Ivison et al. 2007; Biggs et al. 2011; Micha{\l}owski et al. 2012a; Yun et al. 2012).
The determination of redshifts from optical spectroscopy is well known to be difficult 
in the ``redshift desert'' at $1.5 < z < 2.0$ (due to the the lack of emission lines 
accessible to silicon-based detectors) and even at higher redshifts success is by no means 
guaranteed for sub-mm galaxies, given the ease with which Lyman-$\alpha$ emission can be extinguished 
by dust. Nevertheless, a sufficient number of spectroscopic redshifts have been measured to 
confirm the reliability of photometric redshift determination for sub-mm sources, and typically 
$\simeq 80$\% of sub-mm sources in blank-field surveys can now be successfully associated 
with a galaxy counterpart (e.g. Ivison et al. 2007; Lindner et al. 2011; Micha{\l}owski et al. 2012a). Thus, despite the 
fact that radio and mid-infrared galaxy counterpart detection becomes increasingly difficult with 
increasing redshift (unlike sub-mm/mm detection), there appears to be limited room for a 
substantial extreme-redshift population in the typical sub-mm/mm galaxy samples studied to date.
Indeed, the relatively modest disagreements between the redshift distributions of existing
sub-mm galaxy samples can be attributed to cosmic variance (Micha{\l}owski et al. 2012a).

Now, however, a new generation of facilities is being utilised. First, {\it Herschel} 
and the South Pole Telescope (SPT) have now delivered sufficiently large far-infrared/mm maps 
to uncover examples of rare, very bright, generally lensed objects, for which follow-up molecular 
spectroscopy has proved feasible with ALMA and the latest generation of wide bandwidth 
redshift receivers on single-dish (sub-)mm telescopes. 
For example, pre-selection of red sources from {\it Herschel} data 
has yielded a new redshift record of $z = 6.34$ for a sub-mm selected galaxy 
(Riechers et al. 2013), while ALMA follow-up of a bright sample of lensed sources uncovered 
with the SPT has yielded a redshift distribution which apparently peaks at $z > 3$
(Vieira et al. 2013; Weiss et al. 2013). In parallel with these sub-mm/mm spectroscopic 
studies of bright lensed sources, ALMA has also recently been used to undertake a systematic 
imaging study of unlensed sources in the Chandra Deep Field South
(Karim et al. 2013; Hodge et al. 2013), as originally uncovered in the LABOCA LESS survey (Weiss et al. 2009).

These new studies have produced results which some have regarded as casting doubt on our existing 
knowledge of the (sub-)mm source population. First, it has been claimed that the 
(apparently robustly established) redshift distribution of 
(sub-)mm sources has been biased low (Vieira et al. 2013), questioning the reliability 
of the aforementioned galaxy identification techniques based on the supporting radio-near/mid-infrared imaging.
Second, it has been suggested that a substantial fraction of bright (sub-)mm sources in single-dish
surveys arise from blends, raising additional concerns about the effectiveness of identification 
methods applied to large-beam sub-mm maps (Wang et al. 2011; Karim et al. 2013; Hodge et al. 2013).

The first of these claims might seem surprising, 
given the high completeness of galaxy identifications in previous
blank-field surveys, and the robustness of photometric redshifts
(consistently yielding $z_{median} \simeq 2.5$). Nevertheless, by the end of 2012, 
over ten sub-mm galaxies had already been spectroscopically confirmed at $z > 4$
(Coppin et al. 2009; Capak et al. 2008, 2011; Schinnerer et al. 2008; Daddi et al. 2009a,b; 
Knudsen et al. 2009; Riechers et al. 2010; Cox et al. 2011; Smolcic et al. 2011; 
Combes et al. 2012; Walter et al. 2012),  and it has been suggested by several 
authors that the most luminous sub-mm/mm galaxies appeared to lie at preferentially higher redshifts 
than their more moderate luminosity counterparts (e.g. Ivison et al. 2002; Wall, Pope
\& Scott 2008; Dunlop 2011; Micha{\l}owski et al. 2012a). 
The second claim, regarding prevalent source blending, seems equally surprising given that previous sub-mm/mm interferometry with the SMA 
and PdBI interferometers had suggested that serious multiplicity was not a big issue
(e.g. Iono et al. 2006; Younger et al. 2007, 2008, 2009; Hatsukade et al. 2010).

Motivated by this controversy and confusion, and by the ever-improving multi-frequency dataset in the 
Cosmological Evolution Survey (COSMOS) field (including UltraVISTA:  McCracken et al. 2012; Bowler et al. 
2012), we have undertaken a fresh investigation of the properties of bright (but unlensed) sub-mm/mm galaxies 
as selected from the largest flux-limited sub-mm sample  with pre-ALMA interferometric 
follow-up observations. Our sample consists of the 30 brightest sub-mm/mm sources 
in the COSMOS field which were originally uncovered with AzTEC and LABOCA, 
and which have subsequently been imaged with the Submillimeter Array (SMA) 
(Younger et al. 2007, 2009) and the Plateau de Bure Interferometer (PdBI) 
(Smolcic et al. 2012). Our aim was to combine the $\simeq 0.2$ positional 
accuracy delivered by the sub-mm/mm interferometry,
with the latest Subaru, UltraVISTA and {\it Spitzer} optical-infrared 
photometry to unambiguously establish the galaxy identifications, redshifts ($z$), 
stellar masses ($M_{\star}$) and specific star-formation rates ($sSFR$) for a 
well-defined sample of bright sub-mm sources. At the same time we have 
taken the opportunity to revisit the issue of source multiplicity, and the robustness 
of galaxy identifications established using the statistical associations with 
radio/infrared sources which would have been deduced 
based on the original single-dish sub-mm/mm positions.

The remainder of this paper is structured as follows. In Section 2 we describe 
the published (sub-)mm samples in the COSMOS field with interferometric follow-up, 
and summarize the latest multi-frequency data that 
we have used to uncover and study the galaxies which produce 
the detected sub-mm/mm emission. Next, in Section 3, we describe the process of galaxy identification, 
and the extraction of robust optical-infared multi-wavelength photometry. Then, in Section 4 we 
present and discuss the derived properties of the galaxies, with special emphasis on the derived 
redshift distribution of bright (sub-)mm sources, and the stellar masses of the associated galaxies. 
In Section 5 we consider further our findings in the context of the latest {\it Herschel}/SPT/ALMA studies 
detailed above, and include a reassessement of how reliably 
galaxy counterparts can actually be established purely on the basis of the 
original single-dish sub-mm/mm maps (and hence to what extent higher-resolution
sub-mm/mm imaging impacts on our understanding of the sub-mm galaxy population). 
Our conclusions are summarized in Section 6.

Throughout we use the AB magnitude system (Oke 1974), and assume a flat cosmology with 
$\Omega_m=0.3, \Omega_{\Lambda}=0.7$ and $H_0=70$ km s$^{-1}$Mpc$^{-1}$.

\section{Data}

The AzTEC/COSMOS survey covers 0.15\,deg$^2$ of the COSMOS field at 1.1\,mm with an rms noise of 1.3\,mJy\,beam$^{-1}$ (Scott et al. 2008).
The published AzTEC/COSMOS catalogue consists of 44 sources with $S/N \geq 3.5\sigma$. The brightest fifteen of these sources were then followed up with 
the SMA (Younger et al. 2007, 2009), effectively yielding a flux-limited sample of millimetre selected galaxies with refined positions. 
All fifteen of these sources were detected with the SMA, providing sub-millimetre positions accurate to $\simeq 0.2$\,arcsec (see Table 3). 
Two of the sources were split by the SMA into two distinct components; AzTEC11 was subdivided into north and south components and AzTEC14 into west and east. 
In the case of AzTEC11 however, as can be seen from figure 1 of Younger et al. (2009), the resolution of the SMA image is not high enough to clearly 
separate the components. For this reason we decided to continue to treat AzTEC11 as a single (albeit somewhat extended) galaxy for the 
purpose of this study.

The LABOCA/COSMOS survey covers the inner $\simeq 0.7$\,deg$^2$ of the COSMOS field, delivering a sub-millimetre 
map at $\lambda = 870\,\mu$m with an rms noise level of 1.5\,mJy\,beam$^{-1}$ 
(Navarrete et al. in preparation). The 28 brightest 870\,$\mu$m sources were chosen for IRAM PdBI follow-up observations with the requirement 
that the signal-to-noise S/N$_{\rm{LABOCA}}\gtrsim 3.8$ (Smolcic et al. 2012).
Most of these were detected with the IRAM interferometer. To create a well-defined and (near) flux-limited sample 
for the present study we selected the 16 objects with S/N$_{\rm{PdBI}}\gtrsim 4.0$. These are listed 
in Table 4. However, as described in the notes on individual sources in Appendix A, 
the PdBI position of COSLA-38 is so far from the original LABOCA position, and so close to the edge of the 
beam that it is hard to be confident it is the same source. For this reason we have excluded 
COSLA-38, and all further analysis is thus performed on a final sample of 30 (sub-)mm sources.

\begin{table}
\caption{A summary of the optical and near-infrared imaging data utilised in this study.
Column 1 gives the filter bandpass names, column
2 their effective wavelengths, column 3 the FWHM of the bandpasses, column
4 gives the 5$\sigma$ photometric depths (AB mag) within a 2-arcsec
diameter aperture and column 5 gives the seeing in arcsec. The $u,g,r,i$ imaging was delivered by the CFHT Legacy Survey, 
the $z'$ imaging was obtained with the refurbished Suprime-Cam on Subaru (Bowler et al. 2012; 
Furusawa et al., in preparation) while the  
$Y,J,H,K_s$ imaging was provided by UltraVISTA DR1 (McCracken et al. 2012).}
\label{tab:filters1}
\begin{center}
\begin{tabular}{ccccc}
\hline
filter  &  $\lambda_{eff}$/nm  &  $FWHM$/nm  &  5$\sigma$/AB mag &  seeing/$\prime \prime$ \\
\hline
$u$     &  381.1  &  65.2  &  26.9  &  0.80 \\
$g$     &  486.2  &  143.6 &  27.0  &  0.65 \\
$r$     &  625.8  &  121.7 &  26.6  &  0.65 \\
$i$     &  769.0  &  137.0 &  26.4  &  0.65 \\
$z'$    &  903.7  &  85.6  &  26.3  &  1.15 \\
$Y$     &  1020   &  100   &  24.7  &  0.82 \\
$J$     &  1250   &  180   &  24.5  &  0.79 \\
$H$     &  1650   &  300   &  24.0  &  0.76 \\
$K_s$   &  2150   &  300   &  23.8  &  0.75 \\
\hline 
\end{tabular}
\end{center}
\end{table}

\begin{table}
\caption{A summary of the wider-area Subaru optical imaging (Taniguchi et al. 2007) utilised in the 
study of AzTEC7 and AzTEC12. 
Column 1 gives the filter bandpass names, column 
2 their effective wavelengths, column 3 the FWHM of the bandpasses, column 
4 gives the 5$\sigma$ photometric depths (AB mag) within a 2-arcsec 
diameter aperture and column 5 gives the seeing in arcsec.}
\label{tab:filters2}
\begin{center}
\begin{tabular}{ccccc}
\hline
filter  &  $\lambda_{eff}$/nm  &  $FWHM$/nm  &  5$\sigma$/AB mag &  seeing/$\prime \prime$ \\
\hline
$B$    &  446.0  &  89.7  &  27.14  &  0.95 \\
$V$    &  548.4  &  94.6  &  26.75  &  1.33 \\
$g'$   &  478.0  &  126.5 &  27.26  &  1.58 \\
$i'$   &  764.1  &  149.7 &  26.08  &  0.95 \\
$r'$   &  629.5  &  138.2 &  26.76  &  1.05 \\
$z'$   &  903.7  &  85.6  &  26.00  &  1.15 \\
\hline 
\end{tabular}
\end{center}
\end{table}

We used the refined positions provided by the SMA and PdBI interferometry to identify galaxy counterparts in the available
multi-frequency imaging. The location of the AzTEC/SMA and LABOCA/PdBI sources within the key available 
multi-wavelength imaging in the COSMOS field is illustrated in Fig.\,1. This imaging consists of the public IRAC imaging obtained via the S-COSMOS survey (Sanders et al. 2007),
the new near-infrared imaging provided by UltraVISTA DR1 (McCracken et al. 2012), and optical imaging from the CFHT Legacy Survey (Gwyn et al. 2011), 
and Subaru (Taniguchi et al. 2007; Furusawa et al. in preparation). The details of this imaging are summarized in
Table \ref{tab:filters1} and Table \ref{tab:filters2}, with the latter table being relevant for AzTEC7 and AzTEC12 which lie just outside the 
deep CFHT MegaCam pointing (see Fig.\,1), and thus required use of the (somewhat shallower) Subaru imaging available over the whole COSMOS field.

\begin{figure}
\begin{center}
\includegraphics[scale=0.8]{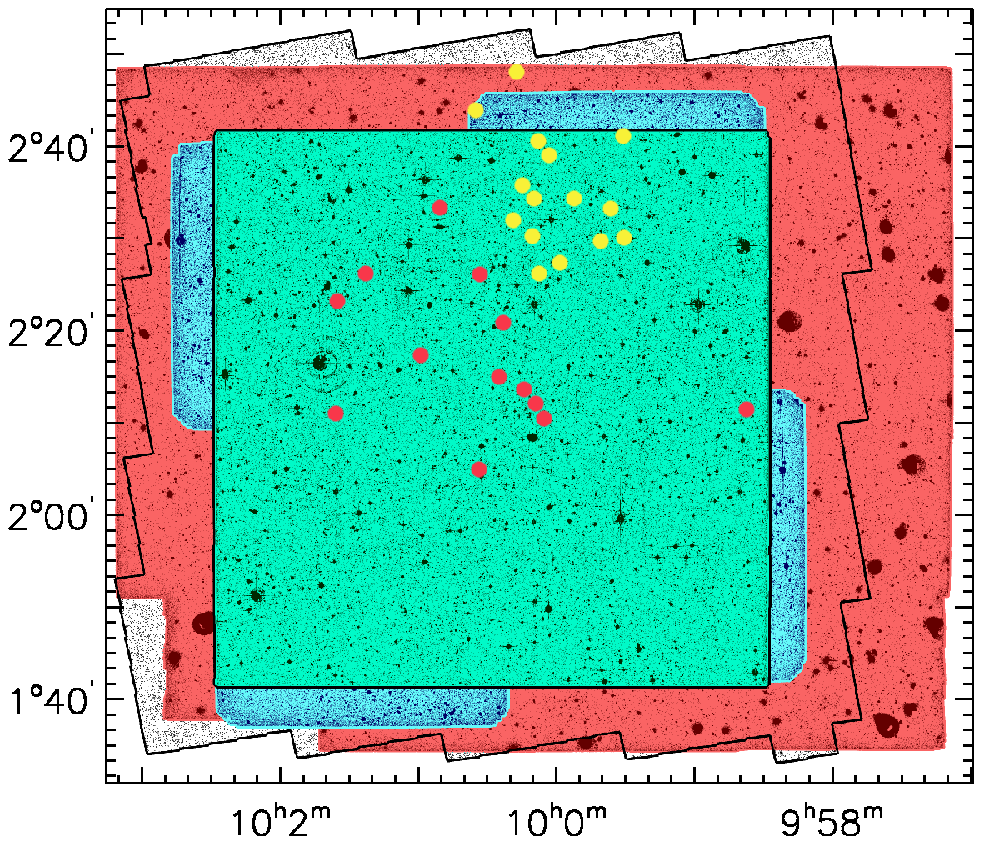}
\end{center}
\caption{The location of the 30 (sub-)mm sources studied here within the 
multi-band coverage of the COSMOS field. The $x$ and $y$ axes are RA and Dec respectively. From the outside, the red area is the 1.5\,deg$^2$ 
UltraVISTA field, the irregular black outline delineates the HST/ACS f$_{814}$-band imaging, the blue region is the Subaru $z'$-band Suprime-Cam mosiac, 
and the innermost green area marks the CFHTLS D2 optical data. Yellow and red dots indicate 
the positions of the AzTEC and LABOCA sources respectively (figure adapted from Bowler et al. 2012).}
\label{fig:COSMOS}
\end{figure}

\begin{table*}
 \begin{normalsize}
 \begin{center}
 \caption{The fifteen brightest COSMOS AzTEC mm sources chosen for SMA interferometric follow-up observations which were utilised in the 
present study. Column 1 gives the source name, column 2 the SMA position, column 3 the SMA $890\,\mu$m signal-to-noise ratio, 
column 4 the AzTEC 1.1\,mm signal-to-noise ratio (Younger et al. 2007, 2009), colum 5 the SMA flux density,
column 6 the de-boosted AzTEC $1.1\,$mm flux density (Scott et al. 2008). AzTEC14 was 
resolved by the SMA into the east and west components. AzTEC11, even though it was also just resolved by the SMA into two components, is treated here 
as a single, extended SMG with an $890\,\mu$m flux density which is the sum of the flux densities of both components 
(table 1 of Younger et al. 2007).}
 \label{tab:SMA} 
 \begin{tabular}{lccccc}
  \hline
SMA ID      & SMA coords (J2000)                 & S/N & S/N & F$_{890\,\mu {\rm m}}$ & F$_{1.1\,{\rm mm}}$ \\
            & \phantom{A}RA\phantom{AAAAAAAA}Dec & SMA  & AzTEC  &  /mJy   & /mJy\\
 \hline
AzTEC1      & $09:59:42.86 \hspace{3mm} +02:29:38.2$  & $14.2$ & $8.3$ & $15.6\pm1.1$          & $9.3^{+1.3}_{-1.3}$   \\
AzTEC2      & $10:00:08.05 \hspace{3mm} +02:26:12.2$  & $12.4$ & $7.4$ & $12.4\pm1.0$          & $8.3^{+1.3}_{-1.3}$   \\
AzTEC3      & $10:00:20.70 \hspace{3mm} +02:35:20.5$  & $\phantom{1}5.8$  & $5.9$ & $\phantom{1}8.7\pm1.5$           & $5.9^{+1.3}_{-1.3}$   \\
AzTEC4      & $09:59:31.72 \hspace{3mm} +02:30:44.0$  & $\phantom{1}7.5$  & $5.3$ & $14.4\pm1.9$          & $5.2^{+1.3}_{-1.4}$   \\
AzTEC5      & $10:00:19.75 \hspace{3mm} +02:32:04.4$  & $\phantom{1}7.1$  & $6.2$ & $\phantom{1}9.3\pm1.3$           & $6.5^{+1.2}_{-1.4}$   \\
AzTEC6      & $10:00:06.50 \hspace{3mm} +02:38:37.7$  & $\phantom{1}6.6$  & $6.3$ & $\phantom{1}8.6\pm1.3$           & $6.3^{+1.3}_{-1.2}$   \\
AzTEC7      & $10:00:18.06 \hspace{3mm} +02:48:30.5$  & $\phantom{1}8.0$  & $6.4$ & $12.0\pm1.5$          & $7.1^{+1.4}_{-1.4}$   \\
AzTEC8      & $09:59:59.34 \hspace{3mm} +02:34:41.0$  & $10.9$ & $5.7$ & $19.7\pm1.8$          & $5.5^{+1.3}_{-1.3}$   \\
AzTEC9      & $09:59:57.25 \hspace{3mm} +02:27:30.6$  & $\phantom{1}4.1$  & $5.6$ & $\phantom{1}9.0\pm2.2$           & $5.8^{+1.3}_{-1.5}$   \\
AzTEC10     & $09:59:30.76 \hspace{3mm} +02:40:33.9$  & $\phantom{1}5.3$  & $5.1$ & $\phantom{1}5.3\pm1.0$           & $4.7^{+1.3}_{-1.3}$   \\
AzTEC11     & $10:00:08.91 \hspace{3mm} +02:40:10.2$  & $\phantom{1}8.2$  & $5.1$ & $14.4\pm1.9$          & $4.7^{+1.3}_{-1.3}$   \\
AzTEC12     & $10:00:35.29 \hspace{3mm} +02:43:53.4$  & $\phantom{1}7.5$  & $4.8$ & $13.5\pm1.8$          & $4.5^{+1.3}_{-1.5}$   \\
AzTEC13     & $09:59:37.05 \hspace{3mm} +02:33:20.0$  & $\phantom{1}4.5$  & $4.8$ & $\phantom{1}8.2\pm1.8$           & $4.4^{+1.3}_{-1.4}$   \\ 
AzTEC14     & ...          \hspace{14mm} ...          & ...    & $4.7$ & ...                   & $4.3^{-1.4}_{-1.4}$   \\
AzTEC14.E   & $10:00:10.03 \hspace{3mm} +02:30:14.7$  & $\phantom{1}5.0$  & ...   & $\phantom{1}5.0\pm1.0$           & ...                   \\
AzTEC14.W   & $10:00:09.63 \hspace{3mm} +02:30:18.0$  & $\phantom{1}3.9$  & ...   & $\phantom{1}3.9\pm1.0$           & ...                   \\
AzTEC15     & $10:00:12.89 \hspace{3mm} +02:34:35.7$  & $\phantom{1}4.4$  & $4.6$ & $\phantom{1}4.4\pm1.0$           & $4.2^{+1.3}_{-1.4}$   \\
\hline
\end{tabular}
\end{center}
\end{normalsize}
\end{table*}

\begin{table*}
 \begin{normalsize}
 \begin{center}
 \caption{The sixteen brightest COSMOS LABOCA sub-mm sources which 
were followed up with the IRAM PdBI and are utilised here. Column 1 gives the source name, column 2 the PdBI position, 
columns 3 and 4 give the PdBI and LABOCA signal-to-noise ratios, while columns 5 and 6 give the PdBI and LABOCA flux densities.
(Smolcic et al. 2012). Note that COSLA-38 was excluded from the analysis presented here due to the very large offset
between the PdBI and LABOCA positions - see Notes on Individual Objects in Appendix A}
 \label{tab:PdBI} 
 \begin{tabular}{lccccc}
  \hline
PdBI ID      & PdBI coords (J2000)                 & S/N & S/N & F$_{\rm 1.3 mm}$ & F$_{\rm 870\mu m}$ \\
            & \phantom{A}RA\phantom{AAAAAAAA}Dec & PdBI  & LABOCA  &  /mJy   & /mJy\\
\hline
COSLA-5      & $10:00:59.521 \hspace{3mm} +02:17:02.57$  & $4.1$  & $5.0$ & $2.04\pm0.49$            & $12.5\pm2.6$   \\
COSLA-6N     & $10:01:23.640 \hspace{3mm} +02:26:08.42$  & $5.4$  & $4.7$ & $2.66\pm0.49$            & $16.0\pm3.3$   \\
COSLA-6S     & $10:01:23.570 \hspace{3mm} +02:26:03.62$  & $4.8$  & $4.7$ & $3.08\pm0.65$            & $16.0\pm3.3$   \\
COSLA-8      & $10:00:25.550 \hspace{3mm} +02:15:08.44$  & $4.2$  & $4.6$ & $2.65\pm0.62$            & $\phantom{1}6.9\pm1.6$    \\
COSLA-16N    & $10:00:51.585 \hspace{3mm} +02:33:33.56$  & $4.3$  & $4.2$ & $1.39\pm0.32$            & $14.0\pm3.6$   \\
COSLA-17N    & $10:01:36.811 \hspace{3mm} +02:11:09.66$  & $4.6$  & $4.2$ & $3.55\pm0.77$            & $12.5\pm3.2$   \\
COSLA-17S    & $10:01:36.772 \hspace{3mm} +02:11:04.87$  & $5.3$  & $4.2$ & $3.02\pm0.57$            & $12.5\pm3.2$   \\
COSLA-18     & $10:00:43.190 \hspace{3mm} +02:05:19.17$  & $4.5$  & $4.2$ & $2.15\pm0.48$            & $10.0\pm2.6$   \\
COSLA-19     & $10:00:08.226 \hspace{3mm} +02:11:50.68$  & $4.1$  & $4.1$ & $3.17\pm0.76$            & $\phantom{1}6.7\pm1.8$    \\
COSLA-23N    & $10:00:10.161 \hspace{3mm} +02:13:34.95$  & $7.3$  & $3.9$ & $3.42\pm0.47$            & $\phantom{1}6.4\pm1.6$    \\
COSLA-23S    & $10:00:10.070 \hspace{3mm} +02:13:26.87$  & $6.2$  & $3.9$ & $3.70\pm0.60$            & $\phantom{1}6.4\pm1.6$    \\
COSLA-35     & $10:00:23.651 \hspace{3mm} +02:21:55.22$  & $4.2$  & $3.8$ & $2.15\pm0.51$            & $\phantom{1}8.2\pm2.2$    \\
COSLA-38     & $10:00:12.590 \hspace{3mm} +02:14:44.31$  & $4.4$  & $3.7$ & $8.19\pm1.85$            & $\phantom{1}5.8\pm1.6$    \\ 
COSLA-47     & $10:00:33.350 \hspace{3mm} +02:26:01.66$  & $5.3$  & $3.6$ & $3.11\pm0.59$            & $\phantom{1}9.0\pm2.8$    \\
COSLA-54     & $09:58:37.989 \hspace{3mm} +02:14:08.52$  & $5.0$  & $3.6$ & $3.26\pm0.65$            & $11.6\pm4.1$   \\
COSLA-128    & $10:01:37.990 \hspace{3mm} +02:23:26.50$  & $4.8$  & $3.1$ & $4.50\pm0.94$            & $11.0\pm3.5$   \\
\hline
\end{tabular}
\end{center}
\end{normalsize}
\end{table*}

\section{Galaxy counterparts and multi-wavelength photometry}

Initially we searched for galaxy counterparts in the UltraVISTA DR1 $K_s$-band imaging, using a (deliberately generous) search radius of 3\,arcsec 
around the interferometric (sub-)mm positions. Near-infrared counterparts were found for all of the (sub-)mm sources except for AzTEC14.W, COSLA-6N, 
COSLA-17S and COSLA-128. However, as can be seen in Fig.\,2, for AzTEC2 (A2.S), 13, 14.E,
COSLA-8, 19 and 23S the (sub-)mm to $K_s$ positional offset is too large for the association to be trusted.
Also, for the reasons detailed in the `Notes on individual objects' in the appendix, the optical/infrared counterparts labelled  
A2.N , A6 and C5 were also not deemed reliable. This leaves a total of 18/30 (sub-)mm sources with robust near-infrared galaxy counterparts
(note that in Section 5.2 we discuss the extent to which the same galaxy counterparts would have been identified without the 
availability of (sub-)mm interferometric observations).

\begin{figure}
\begin{center}
\includegraphics[scale=0.45,angle=270]{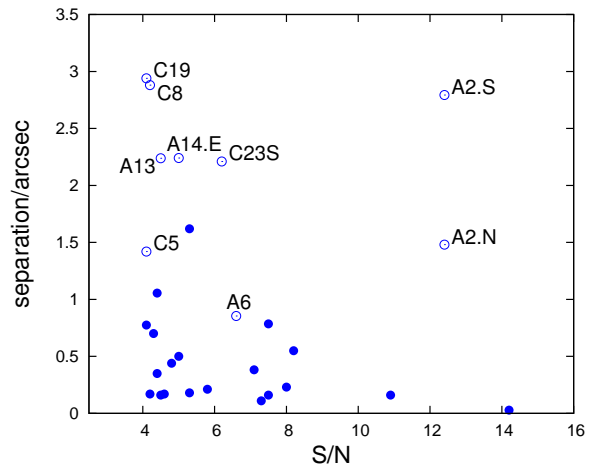}
\end{center}
\caption{The interferometric S/N of each (sub-)mm detection is plotted here as a function of angular separation between 
the (sub-)mm interferometric position and the nearest potential near-infrared/optical counterpart in the available imaging. The empty circles 
represent objects for which we regard the multi-frequency match as incorrect given the positional 
accuracy delivered by the interferometry (i.e. all objects with a separation $> 2$\,arcsec). AzTEC2 was initially matched to a bright foreground galaxy (A2.S) 
in the wings of which a fainter, possibly lensed object was discovered (A2.N) after careful image analysis. However, 
because the radio counterpart of AzTEC2 is exactly at the position of the SMA ID, both these possible near-infrared counterparts can be excluded.
COSLA-5 was matched to an optical object (C5), as was AzTEC6 (A6), for which Smolcic et al. (2012) derived 
photometric redshifts of $z_{est} \simeq 0.85$ and $z_{est} \simeq 0.82$ respectively.
However, these relatively low-redshift possible identifications can be excluded due to the lack of any radio detections in the available VLA 1.4\,GHz imaging, which securely places
the (sub-)mm sources at higher redshifts (at least $z > 1.5$; see Fig.\,4, and Notes on Individual Objects in Appendix A). All the unlabelled objects are summarised in 
Tables \ref{tab:photo1} and \ref{tab:photo2}. The blue filled dot with a separation of 1.62\,arcsec is our optical counterpart for AzTEC10, which we selected on the 
basis of $8\,\mu$m flux density and $i-K$ colour. The filled blue dot with a separation of 1.05\,arcsec indicates our chosen identification for AzTEC15.}
\label{fig:ston}
\end{figure}

After ensuring that all the optical--infrared imaging was accurately astrometrically aligned to the $K_s$-band imaging
(see Bowler et al. 2012), multi-band aperture photometry was performed at all available wavelengths through 2-arcsec 
diameter apertures, with multiple 2-arcsec diameter apertures placed on blank-sky regions 
within $\simeq 30$\,arcsec of the source in order to reliably estimate the local photometric uncertainty in each band. 
With the obvious exception of the IRAC imaging, the imaging data are fairly well matched in terms of seeing 
quality, but all aperture magnitudes were subsequently corrected to total utilising the measured point spread function
in each band. Photometry in the IRAC bands was taken from the S-COSMOS imaging, again corrected to total 
assuming the sources were not significantly resolved at IRAC wavelengths.
The final multi-band photometry measured for the 18 sources with reliable optical--infrared galaxy counterparts 
is detailed in Tables B2 and B3.

\section{Source properties}

\subsection{Photometric redshifts}

The multi-band photometry described above was used to 
derive photometric redshifts using a $\chi^2$ minimization method (Cirasuolo et al. 2007, 2010) with a code based on the H{\small YPER}Z package 
(Bolzonella et al. 2000). To create templates of galaxies, the stellar population synthesis models of Bruzual \& Charlot (2003) were applied, using the 
Chabrier (2003) stellar initial mass function (IMF) with a lower and upper mass cut-off of $0.1$ and $100\,{\rm M_{\odot}}$ respectively. 
A double-burst star-formation history with a fixed solar metallicity was used. Dust reddening was taken into account using the Calzetti (2000) 
law within the range $0 \leq A_V\leq 6$. The HI absorption along the line of sight was applied according to Madau (1995).

For the (sub-)mm sources for which no optical near-infrared counterpart was found in the available imaging, long-wavelength photometric redshift estimates 
were derived from their 24\,$\mu$m to 20\,cm SEDs (including the 
radio flux densities given by Smolcic et al. 2012)
using the average sub-mm galaxy spectral template derived by Micha{\l}owski et al. (2010). Given the potential complications
of dust temperature varying with redshift (e.g. Aretxaga et al. 2007; Amblard et al. 2010; Hwang et al. 2010), we experimented 
with various template libraries, but found that the strongest 
correlation between redshifts derived from the long-wavelength data and the known optical--near-infrared redshifts (either 
spectroscopic or photometrically estimated) was achieved by fitting the long-wavelength data with this average template (see Fig.\,3). 
Thus, treating the shorter-wavelength redshift information as a training set, we adopted values for $z_{LW}$ based on fitting the 
far-infrared$-$radio data with the Micha{\l}owski et al. (2010) template, and these are the values listed in column 4 of Table 5.

\begin{figure}
\begin{center}
\includegraphics[scale=0.7,angle=270]{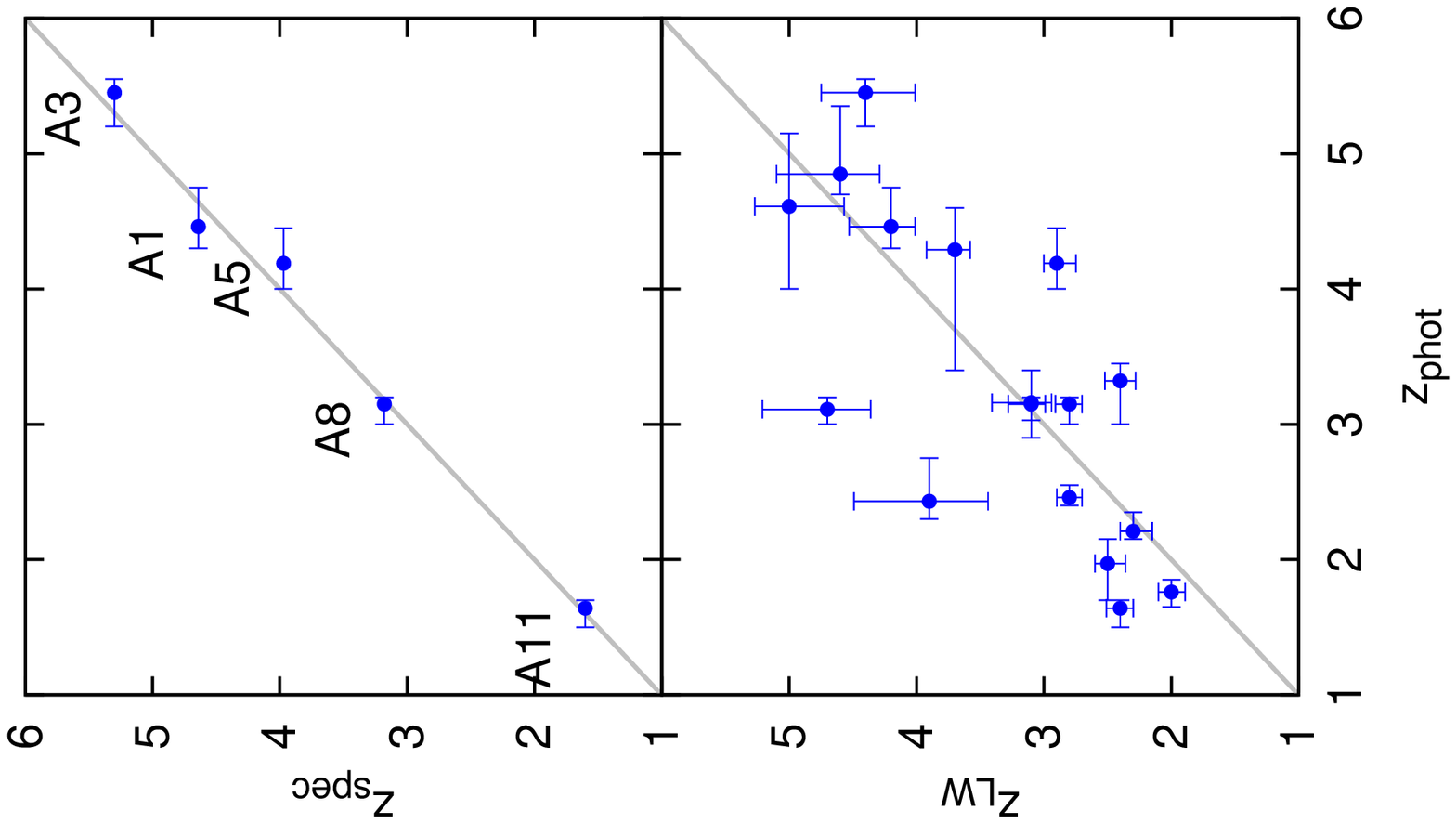}
\end{center}
\caption{Upper panel: our optical/near-infrared photometric redshifts plotted versus the spectroscopic redshifts for the five sources
with reliable spectroscopy (Smolcic et al. 2012), demonstrating the accuracy of $z_{phot}$. Lower panel: the optical/near-infrared
photometric redshifts ($z_{phot}$) are compared with our long-wavelength mm/radio estimates ($z_{LW}$) for those objects for which both
measurements are possible (see Table \ref{tab:redshifts}) in order to check for accuracy and potential bias; the significantly
greater uncertainty in $z_{LW}$ is apparent, but the mean value of $z_{phot}/z_{LW}$ is $1.2\pm 0.36$,  consistent with unity, and thus
indicating no major systematic bias.}
\label{fig:z}
\end{figure}

\begin{table*}
\begin{normalsize}
\caption{Spectrocopic redshifts ($z_{spec}$), optical/near-infrared 
photometric redshifts ($z_{phot}$), `long-wavelength' 
(sub-)mm/radio redshift estimates ($z_{LW}$), Smolcic et al. (2012) redshifts ($z_S$) and 
our stellar masses calculations ($M_*$) for the (sub-)mm galaxies in our final 30-source 
COSMOS sample. Note that stellar masses can only be 
estimated for the 18 sources for which an optical/near-infrared counterpart 
was secured in the available imaging data. Errors on the photometric redshifts 
were derived from the redshift values corresponding to $\chi^2$ values 
higher by $\Delta \chi^2 = 1$ from the minimum-$\chi^2$ solution (see 
Appendix D) and these photometric redshift errors are propagated through 
to the derived random errors on the stellar masses 
(which they dominate). In the case of the Smolcic et al. (2012) redshifts,
the values without errors are the optical spectroscopic redshifts 
for their chosen galaxy identifications (albeit we reject several of these as 
implausible for the (sub-)mm sources; see Fig.\,4) and the two lower 
limits are mm-to-radio estimates (which are clearly consistent with our own estimates of 
$z_{LW}$).}
\setlength{\tabcolsep}{3.5 mm}
\label{tab:redshifts}
\begin{center}
\begin{tabular}{lccclc}
\hline
Source    & $z_{spec}$ &   $z_{phot}$     &   $z_{LW}$              &  $z_S$         &  $\log_{10}(M_*/{\rm M_{\odot}})$\\
\hline
AzTEC1    & 4.64 & $4.46^{+0.29}_{-0.16}$ & $4.20^{+0.33}_{-0.19}$  & $4.26^{+0.17}_{-0.20}$  & $11.30^{+0.04}_{-0.03}$ \\
AzTEC2    & -    & -                      & $3.60^{+0.13}_{-0.18}$  & 1.125                   & -                       \\
AzTEC3    & 5.30 & $5.45^{+0.10}_{-0.25}$ & $4.40^{+0.35}_{-0.39}$  & 5.299                   & $10.93^{+0.01}_{-0.03}$ \\
AzTEC4    & -    & $4.61^{+0.54}_{-0.61}$ & $5.00^{+0.27}_{-0.43}$  & $4.10^{+0.43}_{-1.11}$  & $11.53^{+0.08}_{-0.10}$ \\
AzTEC5    & 3.97 & $4.19^{+0.26}_{-0.19}$ & $2.90^{+0.10}_{-0.15}$  & 3.971                   & $11.49^{+0.04}_{-0.03}$ \\
AzTEC6    & -    & -                      & $3.86^{+4.91}_{-0.92}$  & 0.802                   & -                       \\
AzTEC7    & -    & $1.76^{+0.09}_{-0.11}$ & $2.00^{+0.10}_{-0.11}$  & $2.30^{+0.10}_{-0.10}$  & $11.56^{+0.03}_{-0.04}$ \\
AzTEC8    & 3.18 & $3.15^{+0.05}_{-0.15}$ & $2.80^{+0.11}_{-0.10}$  & 3.179                   & $11.23^{+0.01}_{-0.03}$ \\
AzTEC9    & -    & $4.85^{+0.50}_{-0.15}$ & $4.60^{+0.50}_{-0.31}$  & 1.357                   & $11.02^{+0.07}_{-0.02}$ \\
AzTEC10   & -    & $5.00^{+2.00}_{-0.50}$ & $4.90^{+0.60}_{-0.41}$  & $2.79^{+1.86}_{-1.29}$  & $11.76^{+0.25}_{-0.08}$ \\
AzTEC11   & 1.60 & $1.64^{+0.06}_{-0.14}$ & $2.40^{+0.11}_{-0.10}$  & 1.599                   & $10.95^{+0.02}_{-0.05}$ \\
AzTEC12   & -    & $2.46^{+0.09}_{-0.06}$ & $2.80^{+0.10}_{-0.10}$  & $2.54^{+0.13}_{-0.33}$  & $11.35^{+0.02}_{-0.02}$ \\
AzTEC13   & -    & -                      & $4.70^{+1.25}_{-1.04}$  & $>3.59$                 & -                       \\
AzTEC14   & -    & -                      & $3.38^{+1.00}_{-0.54}$  & $>3.03$                 & -                       \\
AzTEC15   & -    & $2.43^{+0.32}_{-0.13}$ & $3.90^{+0.59}_{-0.46}$  & $3.01^{+0.12}_{-0.36}$  & $11.19^{+0.08}_{-0.03}$ \\
COSLA-5   & -    & -                      & $2.50^{+0.26}_{-0.17}$  & $0.85^{+0.07}_{-0.06}$  & -                       \\
COSLA-6N  & -    & -                      & $3.72^{+1.42}_{-0.63}$  & $4.01^{+1.51}_{-0.83}$  & -                       \\
COSLA-6S  & -    & -                      & $4.05^{+1.70}_{-0.71}$  & $0.48^{+0.19}_{-0.22}$  & -                       \\
COSLA-8   & -    & -                      & $1.90^{+0.11}_{-0.22}$  & $1.83^{+0.41}_{-1.31}$  & -                       \\
COSLA-16N & -    & $2.21^{+0.14}_{-0.06}$ & $2.30^{+0.10}_{-0.15}$  & $2.16^{+0.12}_{-0.25}$  & $11.38^{+0.04}_{-0.02}$ \\
COSLA-17N & -    & $3.11^{+0.09}_{-0.11}$ & $4.70^{+0.51}_{-0.34}$  & $3.37^{+0.14}_{-0.22}$  & $11.09^{+0.02}_{-0.02}$ \\
COSLA-17S & -    & -                      & $3.94^{+1.64}_{-0.70}$  & $0.70^{+0.21}_{-0.22}$  & -                       \\
COSLA-18  & -    & $1.97^{+0.18}_{-0.27}$ & $2.50^{+0.10}_{-0.14}$  & $2.90^{+0.31}_{-0.43}$  & $11.37^{+0.05}_{-0.08}$ \\
COSLA-19  & -    & -                      & $3.50^{+0.34}_{-0.34}$  & $3.98^{+1.62}_{-0.90}$  & -                       \\
COSLA-23N & -    & $4.29^{+0.31}_{-0.89}$ & $3.70^{+0.22}_{-0.12}$  & $4.00^{+0.67}_{-0.90}$  & $11.53^{+0.05}_{-0.16}$ \\
COSLA-23S & -    & -                      & $4.80^{+2.25}_{-0.86}$  & $2.58^{+1.52}_{-2.48}$  & -                       \\
COSLA-35  & -    & $3.16^{+0.24}_{-0.26}$ & $3.10^{+0.31}_{-0.16}$  & $1.91^{+1.75}_{-0.64}$  & $11.46^{+0.05}_{-0.06}$ \\
COSLA-47  & -    & $3.32^{+0.13}_{-0.32}$ & $2.40^{+0.12}_{-0.12}$  & $2.36^{+0.24}_{-0.24}$  & $11.54^{+0.03}_{-0.07}$ \\
COSLA-54  & -    & $3.15^{+0.05}_{-0.15}$ & $3.10^{+0.18}_{-0.11}$  & $2.64^{+0.38}_{-0.26}$  & $11.62^{+0.01}_{-0.03}$ \\
COSLA-128 & -    & -                      & $4.90^{+2.27}_{-0.90}$  & $0.10^{+0.19}_{-0.00}$  & -                       \\
\hline 
\end{tabular}
\end{center}
\end{normalsize}
\end{table*}

The resulting redshift measurements and estimates are summarised in Table \ref{tab:redshifts}.
As a basic test of the reliability of our 
redshift estimates we compare (in Fig.\,\ref{fig:z}) our photometric redshifts with the spectroscopic measurements for the five sources 
in our sample for which reliable optical spectroscopy of the current galaxy counterparts has been obtained (Smolcic et al. 2012); the mean offset 
is $\Delta z/(1+z_{spec})=0.009\pm 0.026$, consistent with zero. In the lower panel of this figure we compare our optical/near-infrared
photometric redshift estimates with our long-wavelength photometric redshifts for those sources for which both estimates are available.
This shows that the $z_{LW}$ redshift estimates are certainly consistent with the optical/near-infrared photometric redshifts, 
albeit with more scatter and with a trend for some high-redshift sources to have redshift underestimated by $z_{LW}$. 
This suggests that at least some of the most distant (sub-)mm galaxies in our sample may have higher dust temperatures 
compared to the average $z \simeq 2-3$ (sub-)mm galaxies SED template utilised here to derive $z_{LW}$.

In Fig.\,4 we plot our objects on the redshift$-$millimetre/radio flux-density ratio plane, both using 
our own final redshifts (from Table 5) and using the redshifts given for these same objects 
by Smolcic et al. (2012) (given in column 4 of our Table 5). We plot the redshift information in this way both to clarify the extent 
to which our redshift estimates differ from those adopted by Smolcic et al. (2012) on a source-by-source 
basis, and to demonstrate that all our adopted redshifts ($z_{spec}$, or failing that $z_{phot}$ or failing that $z_{LW}$) 
are consistent with the anticipated redshift dependence of the millimetre/radio 
flux-density ratio displayed by a reasonable range of template long-wavelength SEDs (as detailed in the plot 
legend). This plot serves to emphasize that the redshifts given for at least 6 (and more likely 8) of 
these (sub-)mm sources by Smolcic et al. (2012) are clearly incorrect, as the resulting flux-density
ratios are inconsistent with (i.e. much larger than) even extreme choices of cool SEDs at the relevant redshifts. 
The interested reader can find the details 
for these differences in the Notes on Individual Objects given in Appendix A, which can be usefully read 
in conjunction with Fig.\'4.

\begin{figure}
\begin{center}
\includegraphics[scale=0.47,angle=0]{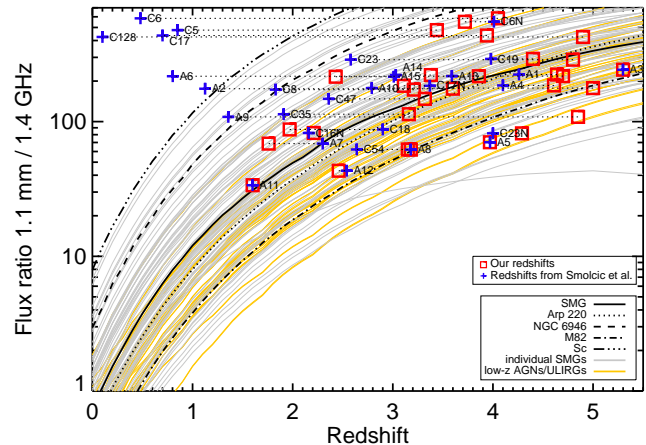}
\end{center}
\caption{The millimetre/radio flux-density ratio of the 30 COSMOS (sub-)mm sources plotted against 
their redshifts as derived in the present study (red squares) and in the previous study by Smolcic
et al. (2012) (blue crosses). These data points showing the positions of the 
individual sources on this diagram are overlaid on a range of curves indicating the 
expected redshift dependence of the 
observed value of the 1.1mm/1.4GHz flux-density ratio as derived from a wide range of observed galaxy SEDs 
(figure adapted from Micha{\l}owski et al. 2012a).
This plot serves to illustrate three key points. First, it shows that the redshifts derived here
(whether spectroscopic redshifts, optical--near-infrared photometric estimates, 
or long-wavelength SED fits) all result in reasonable values for the mm/radio flux-density ratios.
Second, it is clear that the redshifts adopted by Smolcic et al. (2012) for at least 
six of the sources are implausible, in the sense that they are inconsistent with 
the form of any plausible long-wavelength SED. Third, by connecting the alternative redshift 
estimates of each source with dotted lines, it is made clear which sources have had their 
redshifts most dramatically revised in the current work (see also the notes on individual 
sources in Appendix A).}
\label{fig:compare}
\end{figure}

\subsection{Redshift distribution}

\begin{figure*}
\begin{center}
\begin{tabular}{cc}
\includegraphics[scale=0.45,angle=270]{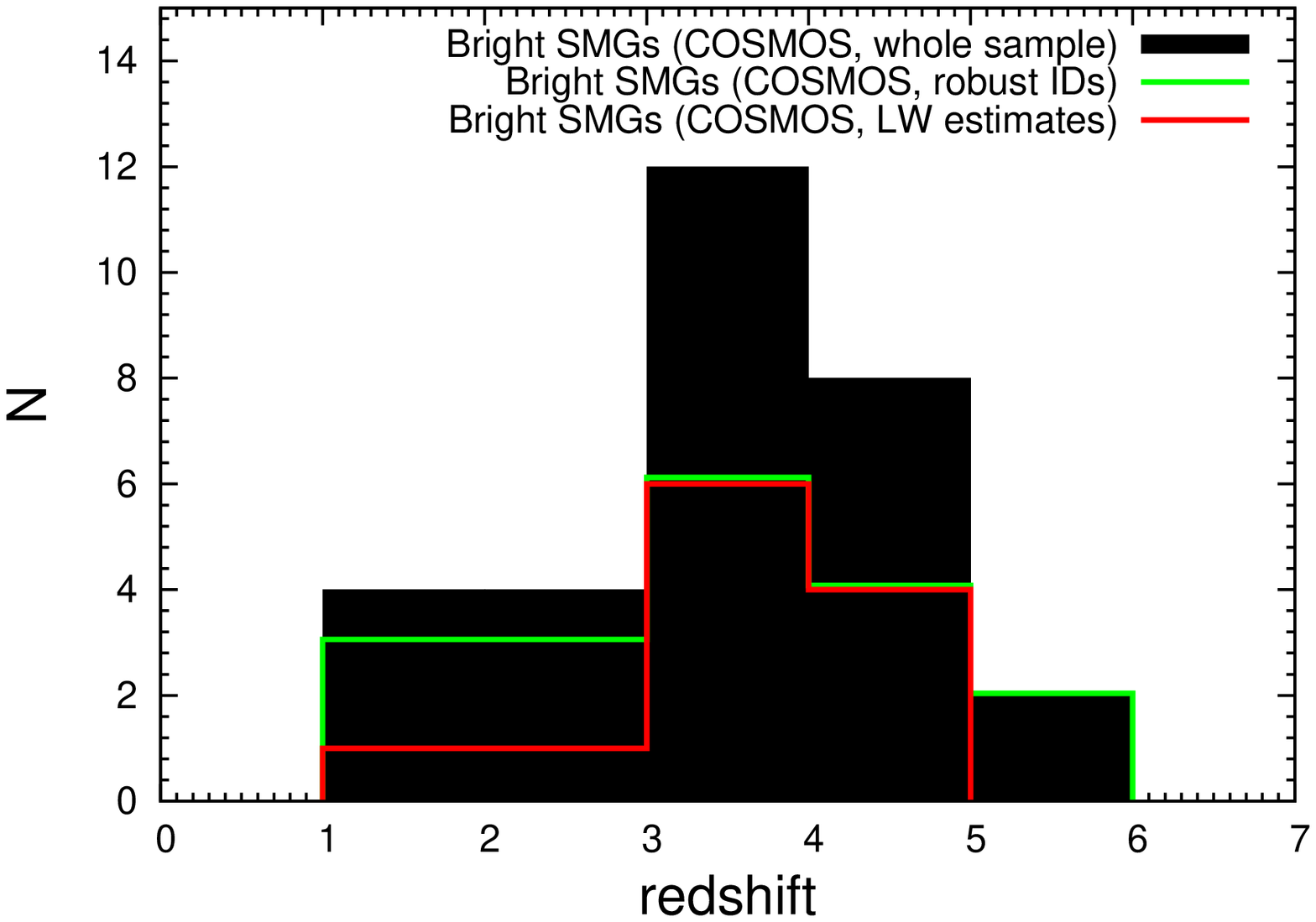}
\includegraphics[scale=0.45,angle=270]{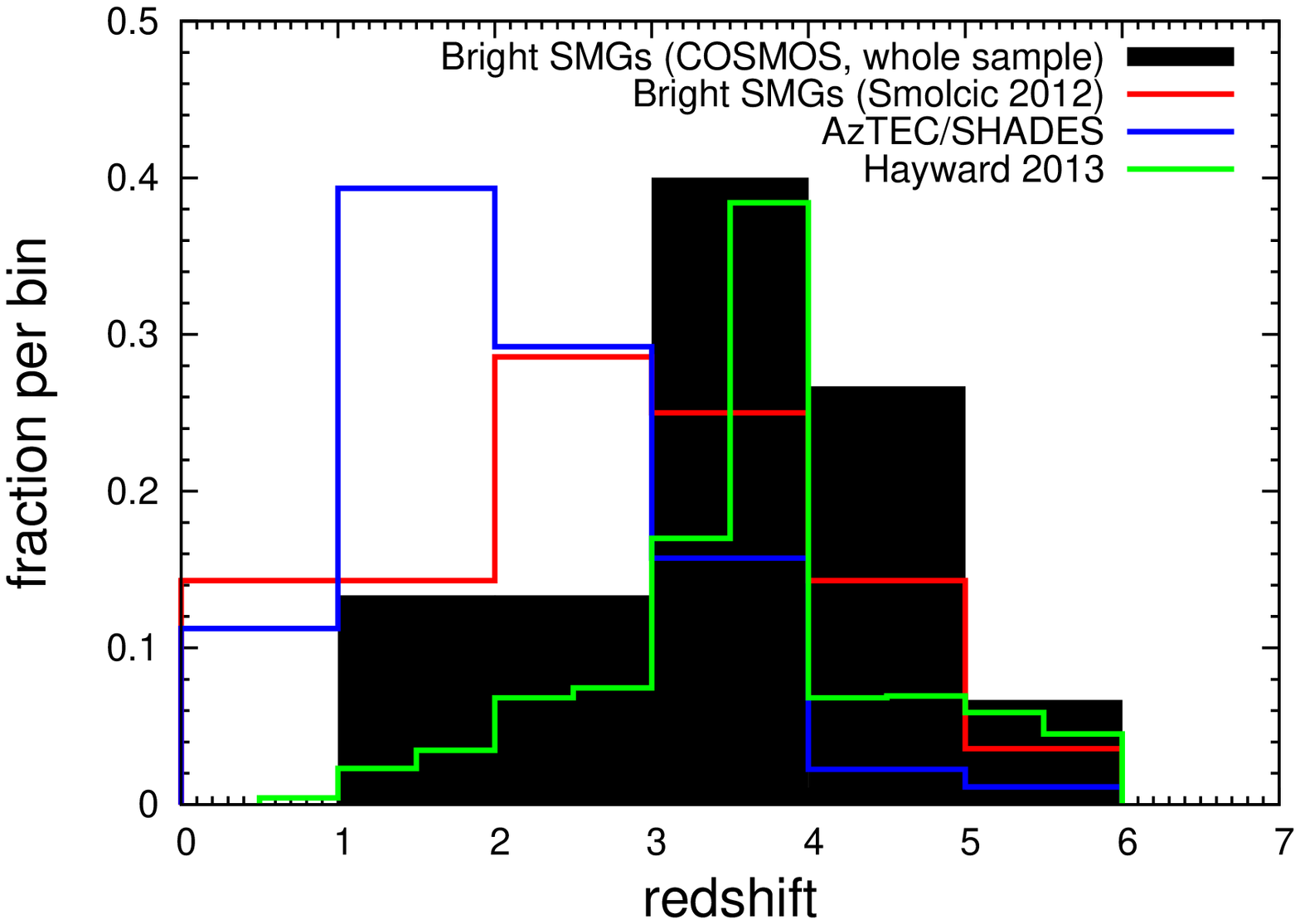}
\end{tabular}
\end{center}
\caption{{\bf Left panel}: The redshift distribution of our full 30-source sample of
luminous (sub-)mm sources in the COSMOS field (Table \ref{tab:redshifts}).
The mean redshift is $\bar{z}=3.53\pm0.19$. Where available, optical
spectroscopic redshifts ($z_{spec}$) have been used (5 sources), with
optical/near-infrared photometric estimates ($z_{phot}$) then used
where judged robust (13 sources),
and long-wavelength redshift estimates ($z_{LW}$) adopted for the remaining
objects (12 sources). {\bf Right panel}
Redshift distribution for the whole COSMOS sample with overlaid distributions derived for the
COSMOS sources by Smolcic et al. (2012) ($\bar{z}=2.8\pm 0.3$), and
for the robust galaxy identifications
in the AzTEC/SHADES survey presented by Micha{\l}owski et al. (2012a)
($\bar{z}=2.0\pm 0.1$).
In addition we plot the Hayward et al. (2013) simulated redshift
distribution for mm-selected sources with F$_{1.1\,{\rm mm}}>4$\,mJy,
which is consistent with the observed redshift distribution presented here
for comparably luminous sources.}
\label{fig:redshifts}
\end{figure*}

The differential redshift distribution derived for our complete 30-source sample is presented in Fig.\,\ref{fig:redshifts}, 
where it is compared with several recently-published redshift distributions for (sub-)mm source samples.
The median redshift derived for our COSMOS sample is $z_{med}=3.44\pm 0.16$, 
whereas for the AzTEC/SHADES sample it is $z_{med}=1.89\pm 0.06$ (Micha{\l}owski et al. 2012a), 
and for the sample of Chapman et al. (2005), $z_{med}=2.14\pm 0.06$. Clearly, the redshift distribution 
of our (sub-)mm sample lies at somewhat 
higher redshift than the majority of recently-published redshift distributions for (sub-)mm selected samples.
In part this could be due to the fact that there are no obvious biases in the identification techniques used here, whereas
several previously-published redshift distributions contain only sources with robust radio identifications.
However, as we explore further below, it may also be due to the fact that the sample considered here is confined 
to significantly more luminous (sub-)mm sources than, for example, the source samples considered by Micha{\l}owski et al. (2012a), 
or Yun et al. (2012), or Simpson et al. (2014). We re-emphasize that, despite the fact that most of 
the (sub-)mm sources are in common, our redshift distribution lies at significantly higher redshift 
than that published by Smolcic et al. (2012); as discussed above (and detailed in Fig.\,4) 
in part this is undoubtedly due to 
our rejection of several of the lower-redshift candidate identifications proposed by Smolcic et al. (2012), but it is also 
in part a result of our deliberate exclusion of some of the less luminous LABOCA/PdBI sources in an effort to achieve 
a homogenous bright source sample.

\begin{figure}
\begin{center}
\includegraphics[scale=0.45,angle=270]{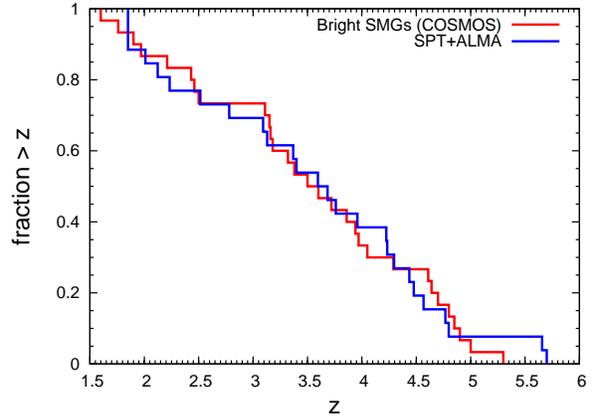}
\end{center}
\caption{A comparison of the our estimated cumulative redshift distribution 
for the bright 30-source COSMOS sample considered here, and that published by 
Vieira et al. (2013) from ALMA follow-up CO spectroscopy of the lensed mm sources uncovered by 
the SPT. It is visually obvious that the redshift distributions are indistinguishable, and indeed 
application of the K-S test yields a significance value $p = 0.991$.}
\label{fig:vieira}
\end{figure}

Interestingly, as shown in Fig.\,\ref{fig:vieira}, the redshift distribution derived here is basically identical to that 
produced by Vieira et al. (2013) from their ALMA follow-up CO spectroscopy of the lensed mm-selected galaxy sample
from the SPT (the K-S test yields $p = 0.991$). This is potentially important because, until now, it has been claimed that 
the SPT redshift distribution is inconsistent with any (sub-)mm source redshift distribution derived without the
benefit of ALMA CO spectroscopy (see Vieira et al. 2013). 

It is reassuring that these two redshift distributions are so clearly consistent, as it 
is hard to imagine that our rather robust and well-validated photometric redshift 
estimation techniques should yield a significantly biased redshift distribution. However, it needs to be explained
why the sample studied here yields a redshift distribution consistent with the SPT results, while most 
other studies of (sub-)mm galaxies clearly do not. As justified further below, we believe there is good evidence
that this is primarily a result of `downsizing' in the star-forming population, and that both our COSMOS sample and 
the SPT sample are biased to significantly higher-luminosity sources than most other studies
(e.g. Michalowski et al. 2012a; Simpson et al. 2014; Swinbank et al. 2014). Of course, part 
of the reason the SPT sources are so {\it apparently} bright is that they are lensed, but it transpires 
that in general the lensing factors are not sufficiently extreme to remove the overall bias of the 
bright/large SPT survey towards the most intrinisically luminous mm sources (for example, the de-lensed 
860\,$\mu$m flux densities of four SPT sources with completed 
lens modelling reported by Hezaveh et al. (2013) are 
5, 6, 16, and 23\,mJy).

The above comparison and discussion suggests that there is a correlation between 
(sub-)mm luminosity and mean redshift, in the sense that more luminous sources lie, on average, 
at systematically higher redshifts. Such a correlation has been suggested before (e.g. Dunlop et al. 1994; Ivison et al. 1998; 
Dunlop 2011; Micha{\l}owski et al. 2012a; Smolcic et al. 2012) 
and, as discussed above, provides arguably the most natural explanation for the consistency
of the redshift distribution presented here with that derived from the bright SPT surveys.

\begin{figure}
\begin{center}
\includegraphics[scale=0.57,angle=270]{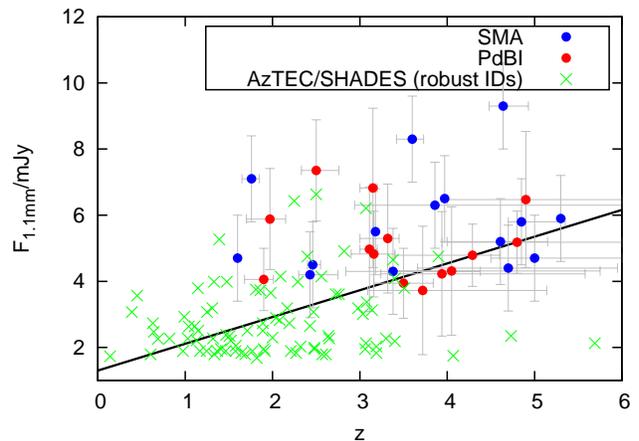}
\end{center}
\caption{1.1\,mm flux density versus redshift. Red and blue dots represent our LABOCA and AzTEC samples respectively. Black crosses are
AzTEC/SHADES sources with robust galaxy counterparts (Micha{\l}owski et al. 2012b). The fluxes are those measured by the single dish facilities, with
LABOCA 870\,$\mu$m flux densities converted to 1.1\,mm estimated measurements assuming the mean sub-mm galaxy SED template of
Micha{\l}owski et al. (2010). The blue line is the best-fitting
straight line; $F_{1.1\,{\rm mm}}=(0.73\pm 0.12)z+(1.73\pm0.33)$. The Spearman correlation coefficient is 0.4557;
the resulting significance level ($p$) is less than $10^{-6}$, indicating
a highly significant correlation between redshift and mm flux density (and hence luminosity).}
\label{fig:correlation}
\end{figure}

In an attempt to better establish the statistical evidence for this, we plot in 
Fig.\,\ref{fig:correlation} the 1.1\,mm flux density for the sources studied here and in the 
SHADES AzTEC survey (Micha{\l}owski et al. 2012a) versus their redshifts.
A correlation is apparent, and calculation of the Spearman rank coefficient for the 
flux-redshift correlation is yields 0.4557, rejecting the null hypothesis of no correlation
with a significance value $p < 10^{-6}$. 
However, this result is potentially biased by the 
fact that it includes only the identified sources in the AzTEC/SHADES sample. When the AzTEC/SHADES sources 
with no secure identifications/redshifts are included (with redshifts scattered randomly between the lower limit 
implied by the mm/radio flux ratio and $z=6$), the Spearman rank coefficient drops to 0.116, yielding $p = 0.025$.
We thus conclude that the data do indeed support the existence of a correlation between 
(sub-)mm luminosity and typical redshift, but that more dynamic range and improved redshift completeness for the 
fainter samples is required to establish the significance and form
of this relation beyond doubt.

\subsection{Stellar masses and specific star formation rates}

For the 18 galaxies for which we secured a robust optical-infrared identification, 
we were able to use the results of the two-component SED fitting which was used to obtain 
photometric redshifts (see Section 4.2) to obtain an estimate of the stellar mass of each (sub-)mm selected
galaxy. As described in Micha{\l}owski et al. (2012b), we assumed a Chabrier (2003) stellar IMF, 
and the stellar masses are based on the models of Bruzual \& Charlot (2003) adopting
a two-component star-formation history. Where a robust spectroscopic  
redshift was available we adopted it, but otherwise derived the mass based on the 
photometric redshift. The results are tabulated in the final column of Table 3.
The median stellar mass is $M_{\star} \simeq 2.2 \times 10^{11}\,{\rm M_{\odot}}$, in excellent agreement with the average stellar 
mass of $z \simeq 2$ sub-mm galaxies by Micha{\l}owski et al. (2012b).

We also used the redshifts and (sub-)mm flux densities of the identified sources
to estimate their star-formation rates ($SFR$).
The $SFR$s were calculated from the (sub-)mm flux densities assuming the average
(sub-)mm  SED template of Micha{\l}owski et al. (2010). Due to the negative K-correction,
a flux density of  $1$\,mJy at $\lambda \simeq 1$\,mm corresponds 
approximately to a total (bolometric) infrared
luminosity of $\simeq10^{12}\,{\rm L_{\odot}}$ at $z>1$, which
converts to a $SFR \simeq 100\,{\rm M_{\odot} yr}^{-1}$
after converting to a Chabrier (2003) IMF (Kennicutt 1998).

Armed with stellar masses and estimates of $SFR$, we have then proceeded to derive 
the specific star-formation rate of each source ($sSFR$).
The results are plotted in Fig. 7, where we show both the values derived from the original single-dish
measurements, and those derived assuming the 
interferometric flux densities. While individual values vary (see figure 
caption for details), it can be seen that 
in both cases the median value is $sSFR \simeq 2.5\,{\rm Gyr^{-1}}$. This is 
essentially identical to the average $sSFR$ displayed by `normal' star-forming 
galaxies on the `main sequence' of star formation at $z > 2$ (e.g. Gonzalez et al.
2010; but see also Stark et al. 2013) and is again consistent with the findings 
of Micha{\l}owski et al. (2012b); while some subset of (sub-)mm selected 
galaxies might display values $sSFR$ which place them above the main sequence, 
in general they display star-formation rates which are perfectly consistent 
with the main-sequence expectation based on their high stellar masses (see also Roseboom et al. 2013).

\begin{figure*}
\begin{center}
\begin{tabular}{cc}
\includegraphics[scale=0.47,angle=270]{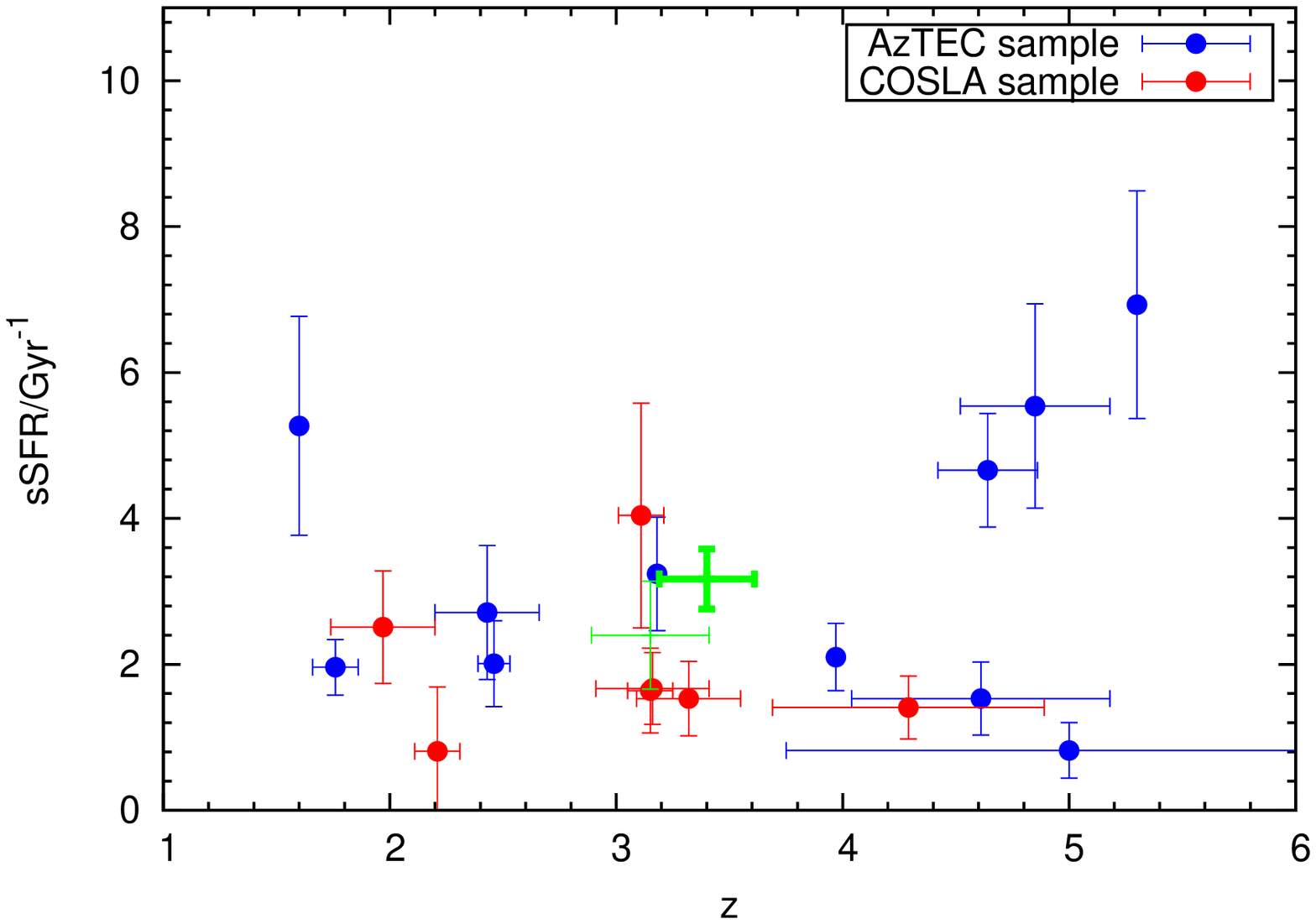}
\includegraphics[scale=0.47,angle=270]{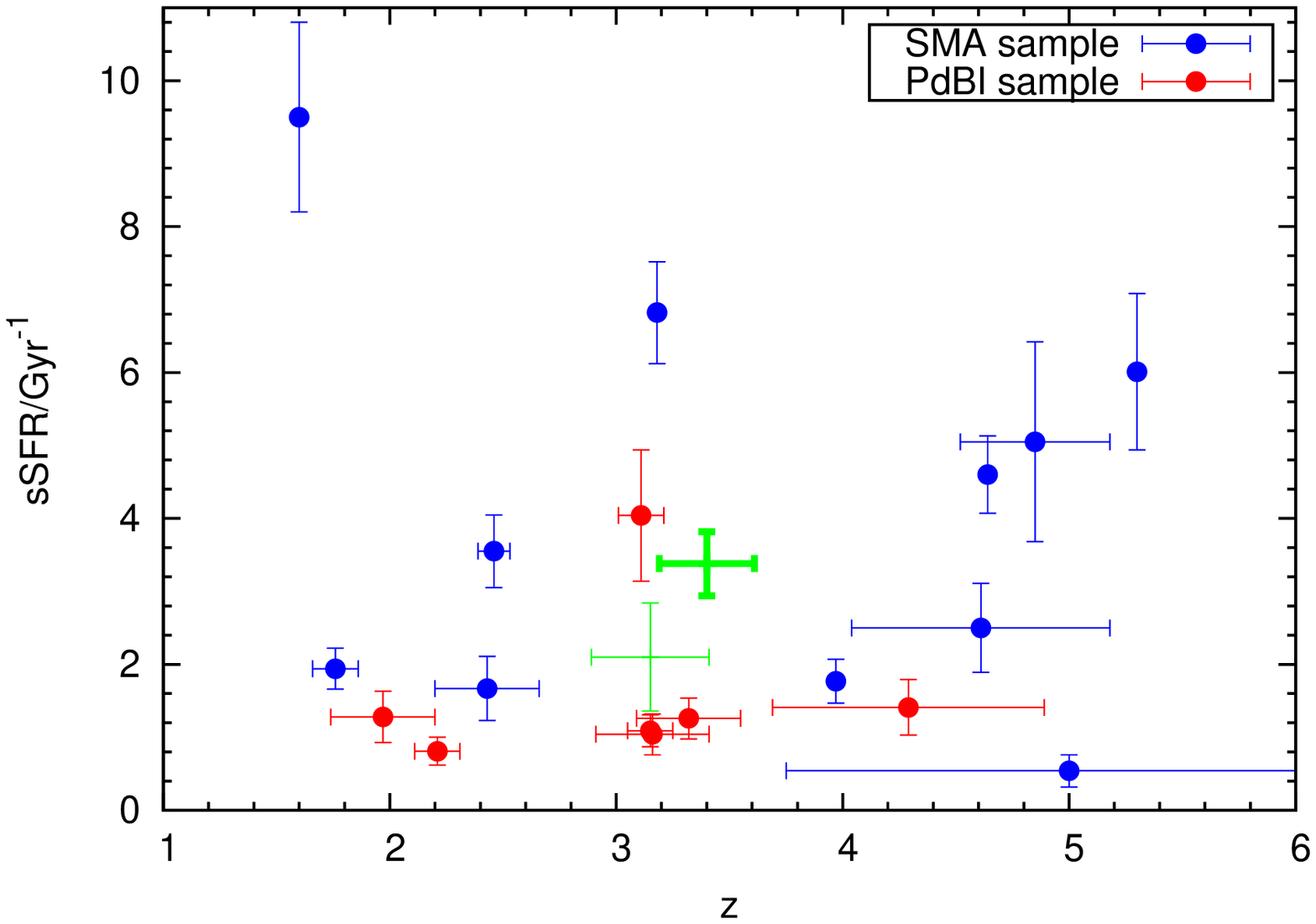}
\end{tabular}
\end{center}
\caption{Specific star-formation rate ($sSFR$) versus redshift. 
The left-hand panel shows $sSFR$ values based on AzTEC (blue dots) 
and LABOCA (red dots) flux densities, while in the right-hand panel we plot $sSFR$ values based on SMA (blue dots) and PdBI 
(red dots) interferometric flux densities. The green points with error bars show the median (thinner error bars)
and mean (thicker error bars) values of $sSFR$ and $z$ in each panel; in the left-hand panel the median $sSFR = 2.40 \pm 0.74$\,Gyr$^{-1}$ 
(mean $sSFR = 3.17 \pm 0.41$\,Gyr$^{-1}$) 
while in the right-hand panel median $sSFR = 2.1 \pm 0.74$\,Gyr$^{-1}$ (mean $sSFR = 3.38 \pm 0.44$\,Gyr$^{-1}$). We conclude that the typical 
value of $sSFR \simeq 2.5$\,Gyr$^{-1}$, consistent with the `main sequence' of star-forming galaxies at $z > 2$, and that 
this conclusion is basically unaffected by whether we adopt the single-dish or interferometric measurements 
of (sub-)mm flux density. Errors on $sSFR$ are dominated by the combined effects of the uncertainties 
in stellar mass (see Table 5) and the uncertainties in the long-wavelength flux-density measurements. 
Errors in redshifts are as given in Table 5, with no horizontal error bar visible for those sources 
with spectroscopic redshift measurements.}
\label{fig:sSFR}
\end{figure*}

\section{Single dish versus interferometric measurements}

\subsection{Multiplicity and number counts}

Recently, ALMA observations of $122$ 870\,$\mu$m sources in the Extended Chandra Deep Field South 
(ECDFS) from the Laboca LESS survey (Weiss et al. 2009)
have been presented, first by Karim et al. (2013), and then in more detail by Hodge et al. (2013). 
This sample includes twelve bright objects with 
original single-dish flux-density measurements of $S_{870} > 9$\,mJy. From this `ALESS' study, Karim et al. (2013) reported that 
source multiplicity is common, and that most bright (sub-)mm sources uncovered in single-dish surveys to date are in fact 
artificial, resulting from blends of fainter (albeit sometimes physically associated) sources within the original single-dish beam. Indeed, 
Karim et al. (2013) went so far as to claim that $S_{870} > 9$\,mJy may represent a physical limit to the luminosity 
of a star-forming galaxy.

However, it is clear that this conclusion is at odds with the sample under study here, in which nine objects retain flux-densities 
$S_{870} > 9$\,mJy within a single component in the high-resolution interferometric follow-up. It also runs contrary to 
the results of various other SMA follow-up studies of SCUBA sources, which have generally suggested that (sub-)mm source multiplicity is rare
(e.g. Downes et al. 1999; Iono et al. 2006; Younger et al. 2007, 2008, 2009; Cowie et al. 2009; Hatsukade et al. 2010)

A more detailed account of the ALESS results has now been published by Hodge et al. (2013), facilitating an assessment of the 
prevalence of multiplicity. In fact, contrary to the claims advanced in Karim et al. (2013) 
(and repeated in the abstract of Hodge et al. 2013), the ALMA results show that 
significant multiplicity is not common at all, consistent with previous 
studies (including the sample under study here). Specifically, for the 20 brightest LESS sources for which 
Hodge et al. (2013) report ALMA results, only 5 reveal multiple ALMA subcomponents, 
and in only 2 of these 5 does the secondary component contribute $> 20$\% of the flux density, thereby 
potentially significantly distorting the flux density and/or position of the 
original single-beam LABOCA source. Moreover, table 3 from Hodge et al. (2013) confirms 
that for the brightest 20 LESS sources, the radio identification technique 
in fact already yielded the correct galaxy counterpart in 17/20 cases (Biggs et al. 2011).

Thus the ALMA results in fact confirm that multiplicity is {\it not} common, with 
only $\simeq 10$\% of bright sources showing a significant (e.g. $> 20$\%) 
flux contribution from a secondary component. This result 
is confirmed by recent reports of SMA follow-up of SCUBA2 sources, which conclude 
that only $\simeq 12$\% of the 850\,$\mu$m sources in SCUBA2 samples arise from blends of multiple 
fainter sources (Chen et al. 2013).

In the present study we have also investigated whether there is any evidence that, on average, significantly
less (sub-)mm flux-density is returned by the interferometric observations as compared to the original 
single-dish measurements. Here this is complicated by the fact that the AzTEC sources were followed up
with (SMA) interferometry at shorter wavelengths, while the COSLA sources were followed up with (PdBI) 
interferometry at longer wavelengths. However, at least this brings some symmetry to the problem, potentially 
ameliorating somewhat any biases introduced by an incorrect choice of long-wavelength SED when performing the 
necessary k-corrections. In addition, we have performed this test with two different 
long-wavelength SED templates. Using the average SMG template described in Section 4.1 (applied 
at the relevant redshifts), we find that 
the mean interferometric/single-dish flux-density ratio for the 30 sources is $F_{int}/F_{single} = 0.96 \pm 0.09$ (median $F_{int}/F_{single} = 0.89$). 
Using an Arp220 template, we find 
that mean $F_{int}/F_{single} = 0.98 \pm 0.08$ (median $F_{int}/F_{single} = 0.90$). Thus, while we acknowledge that the current sample is not ideal for this test, we find no
significant evidence that either multiplicity or very extended emission is (on average) present at a level 
than can distort the true flux density of the sources in the large-beam single-dish 
measurements (at least with the beam sizes utilised here) by more than $\simeq 10$\%.

In summary, it now appears extremely unlikely that the number counts of (sub-)mm 
sources derived from single-dish surveys (e.g. Coppin et al. 2006; Austermann et al. 2010; Scott et al. 2012) 
have been significantly distorted by source blending,
and the new interferometry results reinforce the success of previous galaxy counterpart identification 
programs which have concluded that $\simeq 80$\% of (sub-)mm sources 
can have their galaxy counterparts correctly identified via sufficiently deep ancillary radio 
and/or Spitzer data. For completeness, we now explore this issue further, focussing on what 
conclusions would be drawn from the 30-source sample considered here, both with and without 
the extra information provided by interferometric follow-up.

\subsection{The reliability of (sub-)mm galaxy identifications}

Given the afore-mentioned success of the pre-ALMA LESS identification program (Biggs et al. 2011), it is of interest
to consider the extent to which the galaxy counterparts in the present COSMOS (sub-)mm sample would have been 
successfully identified without the assistance of the SMA and PdBI interferometric follow-up. 
 
In the fifteen years since the discovery of (sub-)mm sources, several methods have been proposed to identify 
their galaxy counterparts in the face of the relatively poor positional accuracy provided by single-dish (sub-)mm imaging. 
As already discussed, deep radio (generally 1.4\,GHz VLA) imaging and deep mid-infrared (generally 24\,$\mu$m {\it Spitzer} MIPS) imaging
have proved particularly powerful in identifying galaxy counterparts, due to the fact these wavelengths also 
trace star-formation activity (e.g. Ivison et al. 2010), provide improved positional accuracy (especially at radio wavelengths) 
and yield source densities on the sky which are generally low enough to yield statistically-significant associations
(e.g. Ivison et al. 2002, 2007; Dunlop et al. 2010; Biggs et al. 2011; Wardlow et al. 2011; Yun et al. 2012; Micha{\l}owski et al. 2012a). 
It has also been found that (sub-)mm sources generally display very red optical-infrared ($i-K$) 
colours (e.g. Smail et al. 2004; Ashby et al. 2006; Micha{\l}owski et al. 2012a; Yun et al. 2012),  
apparently caused by a combination of dust obscuration and the presence of underlying massive evolved stellar populations (Micha{\l}owski et al. 2012b). 
Finally, it is now also well-established that (sub-)mm galaxies are among the brightest galaxies at rest-frame near-infrared wavelengths, 
again due to their large stellar masses. At high redshifts this manifests itself as (sub-)mm galaxies appearing
to be among the apparently brightest objects in {\it Spitzer} 8\,$\mu$m IRAC imaging 
(Pope et al. 2006, 2008; Dye et al. 2008; Hainline et al. 2009; Wardlow et al. 2011; Micha{\l}owski et al. 2012b; Targett et al. 2013). 

In order to test these methods we selected VLA 1.4\,GHz, {\it Spitzer} MIPS 24\,$\mu$m, IRAC 8\,$\mu$m, and red ($i-K>2$)
counterparts to the (sub-)mm galaxies in the COSMOS sample in a similar way to that presented in Micha{\l}owski et al. (2012a). 
Following the method outlined in Dunlop et al. (1989) and Ivison et al. (2007), we assessed the reliability of each potential
galaxy identification by calculating the corrected Poissonian probability, $p$, that each association 
could have been occurred by chance given our search parameters. Specifically, we applied this technique 
to the original pre-interferometric (sub-)mm source detections, using a search radius of 
$r_s=2.5\times 0.6\times \rm{FWHM}/(\rm{S/N})$, where FWHM is the full-width-half-maximum of the single-dish beam, 
and S/N is the signal:noise ratio of the original (deboosted) AzTEC or LABOCA detection.

Armed with interferometrically-refined coordinates from the subsequent SMA and PdBI observations, we can  
here test the success/reliability of such multi-frequency association methods directly. 

The results of this test of the identification process are summarised in Table B1. Additional 
details can be found in the caption to this table (see also the notes on individual objects 
in Appendix A), but the key result is that 16 of the 30 sources would have been successfully
identified on the basis of the single-dish (sub-)mm positions and the available multi-frequency
follow-up imaging. These 16 objects (highlighted in bold in Table B1) are 15 of the 18 sources 
for which stellar masses are given in Table 3 (and for which 
the multi-frequency photometry is provided in Tables B2 and B3), 
plus AzTEC2, which is a purely radio identification
confirmed by the interferometric positions. This means that 16/19 = 84\% of the galaxy 
identifications achievable with the aid of the improved interferometric positional accuracy 
would be correctly identified on the basis of the original single-dish data. The three additional 
galaxy identifications secured with the aid of the SMA and PbBI data comprise new galaxy counterparts
for COSLA-54 and COSLA-17N, and a revised identification for AzTEC15 where a surprisingly large
positional shift is reported between the original AzTEC position and the SMA peak. 

Interestingly, three further identifications suggested by the single-dish positions are 
formally excluded by the interferometric data, but without the new positions yielding a new  
alternative identification. In two of these cases (COSLA-5 and COSLA-8) the proposed 
single-dish identification was statistically compelling but now appears unacceptable 
given the reduced error on the mm position delivered by PdBI. One possible explanation of such 
apparently conflicting conclusions is that both these objects could be lensed, 
and that the optical-infrared counterpart yielding the statistically
significant association is the lensing object. In our analysis we have, in effect, guarded against 
this possibility by adopting the long-wavelength redshift estimate for these objects. Finally,
the apparently significant identification of COSLA-128 listed in the last row of Table B1 is 
formally excluded by the PdBI follow-up, but this is primarily because the PdBI position is
$\simeq 11$\,arcsec from the LABOCA position (for reasons that are hard to explain).

In summary, while the interferometric observations clearly add important extra information 
on the AzTEC and LABOCA sources, for this luminous sample we find that $\simeq 80-85$\% 
of the galaxy identifications which are {\it achievable given the depth of the supporting multi-frequency
data} would have been successfully secured without the aid of the interferometric 
follow-up. In other 
words the main cause of failed identification is not blending or inadequate 
positional accuracy in the single-dish (sub-)mm positions, but supporting multi-wavelength 
data of inadequate depth to reveal the galaxy counterparts of the more high-redshift 
sources in the current sample. Of course, as the supporting data become deeper then 
the improved positional accuracy provided by interferometry (or, for example, SCUBA-2 
450\,$\mu$m imaging) will become increasingly valuable as the source densities in the 
supporting data rise.

For completeness, we show in Appendix C, Figs C1 and C2, how the locations of the sources
on the flux-density--redshift plane vary depending on whether one adopts 
the identifications based on single-dish or interferometric positions, and also whether one adopts the 
single-dish (Fig.\,C1) or interferometric (Fig.\,C2) flux densities.
The average (sub-)mm flux density inferred from the interferometry is only $\simeq 10$\% lower 
than the single-dish average, and in all four panels the average redshift of the 
identified sources lies just below $z = 3.5$ while the average redshift of the sources which 
currently lack optical-infrared is (as anticipated) slightly higher (but still at $z < 4$).
It is thus unsurprising that our main science results are little changed by whether we adopt
the single-dish or interferometric postions and flux densities in our analysis.

\section{Conclusions}

We have presented a new analysis of the brightest sample of unlensed (sub-)mm
sources with existing (pre-ALMA) interferometric (SMA or PdBI) 
follow-up observations. Because these sources lie within the 
COSMOS field, we have been able to exploit the latest Subaru, 
UltraVISTA and {\it Spitzer} optical-infrared photometry to 
better establish their redshifts ($z$), stellar masses ($M_{\star}$) 
and specific star-formation rates ($sSFR$). We have also explored the 
extent to which the supporting data in the field could have been 
used to reliably identify the galaxy counterparts {\it without} 
the improved positional accuracy provided by sub-mm/mm interferometry. 
We find that the bright (sub-)mm sources in the COSMOS field 
display a redshift distribution indistinguishable from that of the 
lensed SPT sources (Vieira et al 2013), 
peaking at $z_{median} \simeq 3.5$. 
We also find that the typical stellar mass of the most luminous 
(sub-)mm sources is independent of redshift for $z \simeq 2 - 5$, 
with median $M_{\star} \simeq 2 \times 10^{11}\,{\rm M_{\odot}}$ assuming a Chabrier (2003)
IMF. Consequently, 
their typical specific star-formation rates also remain approximately 
constant out to the highest redshifts probed, at $sSFR \simeq 2.5\, {\rm Gyr^{-1}}$. 
We note that, consistent with recent ALMA interferometric follow-up
of the LESS sub-mm sources (Hodge et al. 2013), and SMA follow-up of SCUBA2 sources
(Chen et al. 2013), source blending is {\it not} a serious issue
in the study of luminous (sub-)mm sources uncovered by ground-based, 
single-dish ($FWHM < 18$\,arcsec) surveys; only $\simeq 10-15$\% of bright 
($S_{850} \simeq 5 - 10$\,mJy) (sub-)mm sources arise from
significant blends, and so the conclusions 
of our study are largely unaffected by whether we adopt the original single-dish 
mm/sub-mm flux densities/positions, or the interferometric flux densities/positions. 
Our results suggest that apparent disagreements over the redshift distribution 
of (sub-)mm sources are simply a result of ``down-sizing'' in dust-enshrouded star-formation, 
consistent with existing knowledge of the star-formation histories of massive galaxies. 
They also indicate that bright (sub-)mm-selected galaxies at high redshift are, 
on average, subject to the same star-formation rate-limiting processes 
as less luminous objects, and lie on the ``main sequence'' of star-forming galaxies.

\section*{Acknowledgments}

MPK acknowledges the support of the UK Science and Technology Facilities
Council. JSD acknowledges the support of the Royal Society via a 
Wolfson Research Merit award, and the support of the European Research Council 
via the award of an Advanced Grant, and the 
contribution of the EC FP7 SPACE project
ASTRODEEP (Ref.No: 312725).
MJM acknowledges the support of the UK Science and Technology Facilities
Council, and the FWO Pegasus Marie Curie Fellowship. RAAB acknowledges 
the support of the European Research Council. MC acknowledges the support of 
the UK Science and Technology Facilities Council via an Advanced Fellowship.

This work is based in part on data products from observations made with ESO Telescopes at the 
La Silla Paranal Observatories under ESO programme ID 179.A-2005 and on data 
products produced by TERAPIX and the Cambridge Astronomy survey Unit 
on behalf of the UltraVISTA consortium. This study was based in part on 
observations obtained with MegaPrime/MegaCam, a joint project of CFHT and CEA/DAPNIA, 
at the Canada-France-Hawaii Telescope (CFHT) which is operated by the National 
Research Council (NRC) of Canada, the Institut National des Science de l'Univers of the 
Centre National de la Recherche Scientifique (CNRS) of France, and the University of Hawaii. 
This work is based in part on data products produced at TERAPIX and the Canadian Astronomy 
Data Centre as part of the Canada-France-Hawaii Telescope Legacy Survey, a collaborative 
project of NRC and CNRS. This work is based in part on observations made with the 
NASA/ESA {\it Hubble Space Telescope}, which is operated by the Association 
of Universities for Research in Astronomy, Inc, under NASA contract NAS5-26555.
This work is based in part on observations made with the {\it Spitzer Space Telescope}, 
which is operated by the Jet Propulsion Laboratory, California Institute of Technology 
under NASA contract 1407. We thank the staff of the Subaru telescope for their 
assistance with the $z'$-band imaging utilised here.
This research has made use of the NASA/IPAC Infrared Science Archive, 
which is operated by the Jet Propulsion Laboratory, California Institute of 
Technology, under contract with the National Aeronautics and Space 
Administration.

{}

\newpage
\mbox{}
\newpage

\appendix

\section{Notes on individual objects}

\textbf{AzTEC1.} A robust single identification only 0.03\,arcsec from the SMA position, which would also be selected by the $8\, \mu$m method based on the original AzTEC position.
Both $z_{phot}$ and $z_{LW}$ are in excellent agreement with the spectroscopic redshift of $z = 4.64$.

\noindent
\textbf{AzTEC2.} A secure radio and 24\,$\mu$m identification without a visible optical or $K$-band counterpart (and hence no stellar mass estimate
in Table 5).
An alternative object 1.4\,arcsec away from the SMA position was selected by Smolcic et al. (2012) and found to have a 
spectroscopic redshift $z = 1.125$. However, since the radio position is only 0.39\,arcsec from the SMA position and the mm/radio flux ratio yields  
a long-wavelength redshift estimate of $z_{LW} = 3.60$, this low-redshift object cannot be the correct identification 
(its mm/radio flux-density ratio is $\simeq 150$, inconsistent with such a low redshift; see Fig.\,4). The correct radio 
identification would still have been secured without the improved positional accuracy provided by the SMA interferometry.

\noindent
\textbf{AzTEC3.} Similar to AzTEC1, a robust single identification 0.21\,arcsec from the SMA position, which would also be selected by the $8\, \mu$m method based on the original AzTEC position.
Both $z_{phot}$ and $z_{LW}$ are in good agreement with the spectroscopic redshift of $z = 5.30$.

\noindent
\textbf{AzTEC4.} A robust single identification 0.78\,arcsec from the SMA position. 
$z_{phot}$ and $z_{LW}$ are in good agreement that the source has a redshift in the range $z = 4.5 - 5$.
This source would have been successfully identified on the basis of the original AzTEC position by both the $i-K$ and
$8\, \mu$m methods.

\noindent
\textbf{AzTEC5.} A robust single identification 0.38\,arcsec from the SMA position. $z_{phot}$ is in excellent agreement with the spectroscopic
redshift $z = 3.97$, while $z_{LW}$ is somewhat under-estimated. This source would have been securely identified using 
all four types of statistical association on the basis of the original AzTEC position.

\noindent
\textbf{AzTEC6.} Not identified with any method either using the AzTEC position or the refined 
SMA position. There is an optical object $\simeq 1$\,arcsec from the SMA position for which we find $z_{phot}=1.12$ (this is also the 
ID adopted by Smolcic et al. 2012, with $z_{spec}=0.82$), but as with AzTEC2 this optical counterpart can be excluded 
as the correct identification not just because of its relatively 
large positional offset, but also because its mm/radio flux-density ratio of $\simeq 150$ is inconsistent with $z < 1.5$ 
($z_{LW} \simeq 3.9$; see Fig.\,4). The lack of any optical-infrared counterpart means that no stellar mass estimate for this object can 
be included in Table 5.

\noindent
\textbf{AzTEC7.} A robust single identification 0.23\,arcsec from the SMA position. $z_{phot}$ and $z_{LW}$ are in good agreement that the source has a redshift 
$z \simeq 2$. Like AzTEC5, this source would have been securely identified using 
all four types of statistical association on the basis of the original AzTEC position.

\noindent
\textbf{AzTEC8.} A robust single identification 0.16\,arcsec from the SMA position. Both $z_{phot}$ and $z_{LW}$ are in good agreement with the spectroscopic redshift of $z = 3.18$.
This source would have been successfully identified on the basis of the original AzTEC position by both the $i-K$ and
$8\, \mu$m methods.

\noindent
\textbf{AzTEC9.} A robust single identification 0.77\,arcsec from the SMA position. 
Like AzTEC4, $z_{phot}$ and $z_{LW}$ are in good agreement that the source has a redshift in the range $z = 4.5 - 5$. The radio 
identification would have been correctly selected on the basis of the original AzTEC position.
Smolcic et al. (2012) selected a different object $\simeq 2.8$\,arcsec from the SMA position with a photometric redshift of $z_{phot} \simeq 1.07$ 
and a spectroscopic redshift $z = 1.357$. However, not only is such a large positional offset very unlikely, but
AzTEC9 has a large mm/radio flux ratio of $\simeq 100$, completely inconsistent with such a low redshift (see Fig.\,4). 
We therefore conclude that the counterpart selected by Smolcic et al. (2012) cannot be correct, and that the true identification 
is the higher redshift galaxy listed in Table 5.

\noindent
\textbf{AzTEC10.} There are three potential counterparts within 2\,arcsec of the SMA position. Using the SMA coordinates alone we would choose the closest and the brightest
one, but because of  the $8\,\mu$m flux and the very red $i-K$ colour of the more distant object ($\simeq 1.5$\,arcsec from the SMA
position), we chose it as the most likely identification. The photometric redshift determination yielded a very flat $\chi^2$ curve with a formal minimum at 
$z>7$. Even though such an extreme redshift is very unlikely, stacking the optical data shows that it is undetected in the 
optical wavebands suggesting $z>5$. Also our mm/radio estimate gives a redshift of $z_{LW} = 3.12$ 
(arguably biased low due to using a cold SED template appropriate for lower-redshift objects). Considering this, and the probability distribution 
for the optical-infrared $z_{phot}$, for this object we adopt a redshift $z \simeq 5$. This object would have been correctly identified 
using all but the radio identification technique on the basis of the original AzTEC position.

\noindent
\textbf{AzTEC11.} This source is split into two components by the SMA imaging, 
but it may be an extended object and therefore we continue to treat it as a single source. 
$z_{phot}$ is in excellent agreement with the spectroscopic
redshift $z = 1.60$, while this time $z_{LW}$ is somewhat over-estimated.
This source would have been securely identified using 
all four types of statistical association on the basis of the original AzTEC position.

\noindent
\textbf{AzTEC12.} A robust single identification 0.16\,arcsec from the SMA position. 
$z_{phot}$ and $z_{LW}$ are in good agreement that the source has a redshift 
$z \simeq 2.5$. 
Again, this source would have been securely identified using 
all four types of statistical association on the basis of the original AzTEC position.

\noindent
\textbf{AzTEC13.} This object was not associated with any optical or IRAC counterpart using either 
the SMA or AzTEC position. A weak radio detection yields $z_{LW} \simeq 4.7$, but no stellar mass 
can be given in Table 5. 

\noindent
\textbf{AzTEC14.} Like AzTEC13 this object was not associated with any optical or IRAC counterpart using 
either the SMA or AzTEC position. The weak radio flux density measurement yields 
$z_{LW} \simeq 3.4$, but no stellar mass 
can be given in Table 5. 

\noindent
\textbf{AzTEC15.} A robust single identification 1.05\,arcsec away from the SMA position. 
This source could not have been identified on the basis of the AzTEC position because 
the SMA centroid is shifted by more than 10\,arcsec.
$z_{phot}$ and $z_{LW}$ suggest $z \simeq 3$.

\noindent
\textbf{COSLA-5.} This object has two possible optical counterparts less than 
1.5\,arcsec from the PdBI position. The first one is 1.3\,arcsec away with $z_{phot}=0.85$, and is the identification 
adopted by Smolcic et al. (2012). However, because our mm/radio redshift estimate yields $z_{LW} \simeq 3.44$, we conclude that this 
cannot be the correct counterpart (see Fig.\,4). The second possible optical counterpart is 1.1\,arcsec away, but is only visible in the
$z'$- and $K_s$-bands, and so no reliable optical/infrared photometric redshift could be derived. We thus cautiously 
adopt $z_{LW} = 2.5$, and do not give a stellar mass estimate in Table 5.

\noindent
\textbf{COSLA-6N.} This object was not associated with any optical or IRAC counterpart on the basis of either the 
LABOCA or PdBI position. The weak radio flux measurement suggests $z_{LW} \simeq 3.7$, but no stellar mass estimate 
can be given in Table 5. 

\noindent
\textbf{COSLA-6S.} This object has an optical counterpart 0.5\,arcsec from the PdBI position, 
for which Smolcic et al. (2012) derived $z_{phot}=0.48$. However, once again because our mm/radio redshift 
estimate yields $z_{LW} \simeq 4$, and completely excludes $z < 1$, we conclude that this cannot be the correct 
identification (although clearly it could be a lensing galaxy; see Fig.\,4). 
We thus adopt $z_{LW} \simeq 4$ as the best estimate of the redshift of the sub-mm source, 
but cannot provide a stellar mass estimate in Table 5. 

\noindent
\textbf{COSLA-8.} This object has no secure optical nor IRAC counterpart. It was associated by Smolcic et al. (2012) with an 
optical object 1\,arcsec from the PdBI peak which was found to have $z_{phot}=1.83^{+0.4}_{-1.31}$ based on two $\sim 3 \sigma$ 
data points. Given the unreliability of this measurement, we choose here to adopt our mm/radio redshift estimate,
but in fact this is perfectly consistent with the redshift given by Smolcic et al. (2012).

\noindent
\textbf{COSLA-16N.} A robust single identification 0.70\,arcsec from the PdBI position. 
$z_{phot}$ and $z_{LW}$ are in good agreement that the source has a redshift 
$z \simeq 2.25$. 
This source would have been securely identified using 
all four types of statistical association on the basis of the original LABOCA position.

\noindent
\textbf{COSLA-17N.} A robust single identification 0.17\,arcsec from the PdBI position, but this 
would not have been secured on the basis of the LABOCA position.

\noindent
\textbf{COSLA-17S.} This object was not associated with any optical or IRAC counterpart. A weak radio 
flux measurement leads to $z_{LW} \simeq 4$, but we cannot provide a stellar mass estimate in Table 5. 
We note that Smolcic et al. (2012) adopted a redshift $z \simeq 0.7$, but such a redshift is implausible
for this source (see Fig.\,4).

\noindent
\textbf{COSLA-18.} A robust single identification 0.16\,arcsec from the PdBI position. 
$z_{phot}$ and $z_{LW}$ are in good agreement that the source has a redshift 
$z \simeq 2$.
This source would have been securely identified using 
all four types of statistical association on the basis of the original LABOCA position.

\noindent
\textbf{COSLA-19.} This object was not associated with any optical or IRAC counterpart. 
A weak radio 
flux measurement leads to $z_{LW} \simeq 3.5$, but we cannot provide a stellar mass estimate in Table 5. 

\noindent
\textbf{COSLA-23N.} A robust single identification 0.11\,arcsec from the PdBI position. 
$z_{phot}$ and $z_{LW}$ are in good agreement that the source has a redshift 
$z \simeq 4$.
 This object would have been correctly identified 
using all but the 24\,$\mu$m identification technique on the basis of the original LABOCA position.

\noindent
\textbf{COSLA-23S.} This object was not associated with any optical or IRAC counterpart. 
Smolcic et al. (2012) found an optical counterpart $\simeq 0.9$\,arcsec from the PdBI peak
with a redshift of $z_{phot}=2.58^{+1.52}_{-2.48}$ based on one $\sim 3 \sigma$ data point. 
We derive a mm/radio redshift estimate of $z_{LW} = 4.80$, and take it to be a more reliable redshift estimate, but 
cannot provide a stellar mass estimate in Table 5.

\noindent
\textbf{COSLA-35.} A robust single identification 0.17\,arcsec from the PdBI position. 
$z_{phot}$ and $z_{LW}$ are in excellent agreement that the source has a redshift 
$z \simeq 3$.
This object would have been correctly identified 
using all but the 24\,$\mu$m identification technique on the basis of the original LABOCA position.

\noindent
\textbf{COSLA-38.} The PdBI coordinates for this object are $\simeq 15$\,arcsec distant from the original LABOCA 
centroid, placing this object at the edge of the PdBI beam. In addition, the quoted PdBI flux density is higher
than the original LABOCA flux density, raising the possibility that, for whatever reason, it is not 
the same source. For this reason we decided to exclude it from the main analysis, and so it does not appear in Table 5.

\noindent
\textbf{COSLA-47.} A robust single identification 0.18\,arcsec from the PdBI position. 
$z_{phot}$ and $z_{LW}$ are in reasonable agreement that the source has a redshift 
$z \simeq 3$. This object would have been tentatively identified on the basis of $i-K$ colour
given the original LABOCA position.

\noindent
\textbf{COSLA-54.} A robust single identification 0.50\,arcsec from the PdBI position.
$z_{phot}$ and $z_{LW}$ are in excellent agreement that the source has a redshift 
$z \simeq 3$.
This object could not have been identified on the basis of the LABOCA position.

\noindent
\textbf{COSLA-128.} This object was not associated with any optical or IRAC counterpart given 
the PdBI position. 
We adopt $z_{LW} = 4.90$, but cannot provide a stellar mass estimate in Table 5. 
We note that Smolcic et al. (2012) adopted a redshift $z \simeq 0.1$, but such a redshift is implausible
for this source (see Fig.\,4).

%\clearpage

\section{Multi-wavelength identifications}

In this appendix we first illustrate, in Fig.\,B1, the galaxy identifications secured utilising the accurate 
positions provided for the (sub-)mm sources by the SMA and PdBI interferometric observations, overlaying the
SMA and PdBI positions on CFHT optical, UltraVISTA near-infrared and IRAC 8\,$\mu$m image stamps.

We then provide Table\,B1, which summarises the results of our attempt to establish galaxy identifications
based on multi-frequency associations with the original single-dish (AzTEC and LABOCA) positions.

Finally in Tables\,B2 and B3 we provide the optical-infrared photometry for the 18 secure 
galaxy identifications (based on the interferometric positions) which 
was used to estimate the photometric redshifts and stellar masses given in Table\,5.

\begin{figure*}
\begin{center}
\includegraphics[scale=0.75, trim = 30mm 30mm 30mm 0mm]{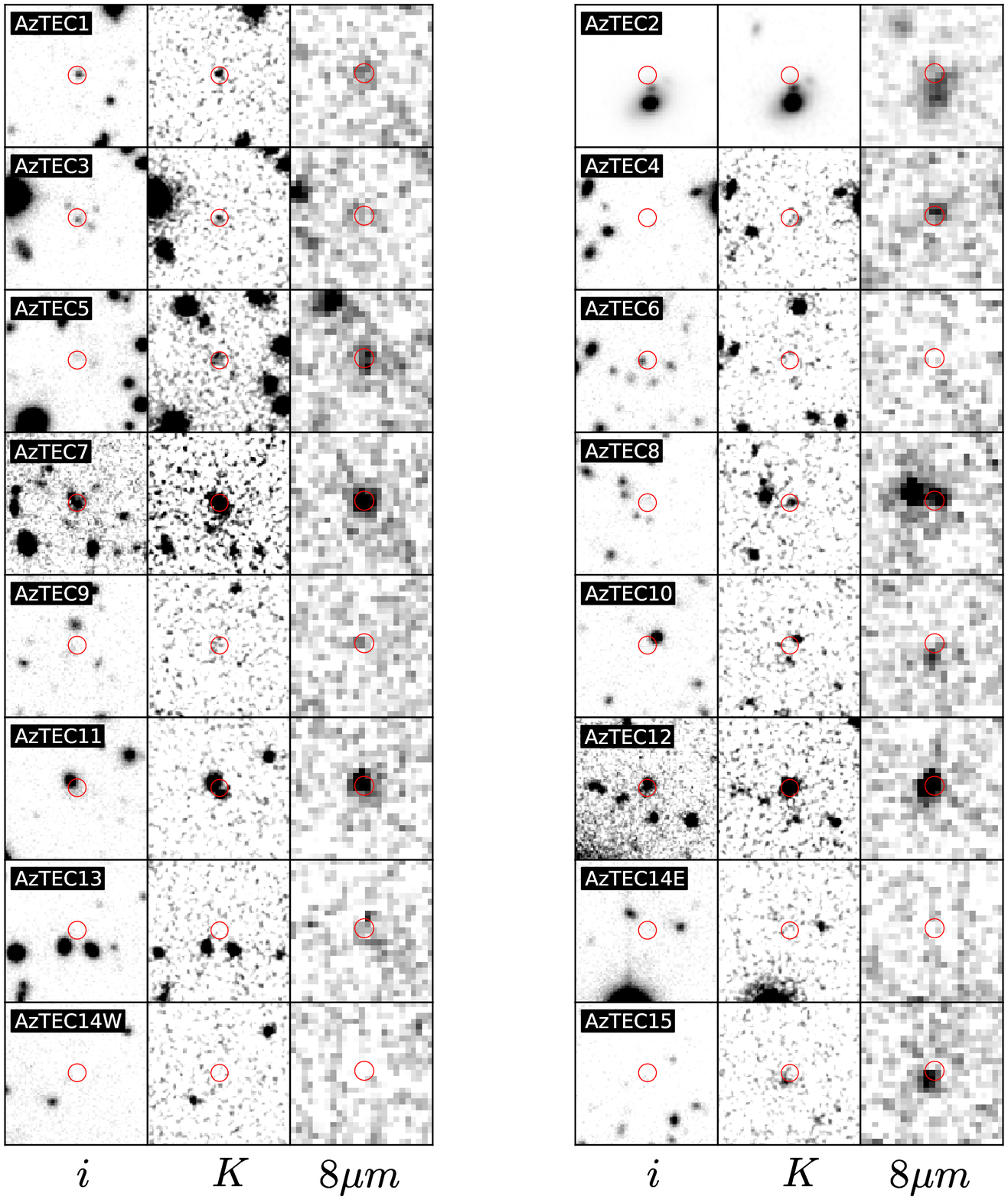}
\end{center}
\caption{CFHTLS $i$-band (in the case of AzTEC7 \& AzTEC12, Subaru $i'$-band), UltraVISTA $K_s$-band and IRAC $8\,\mu$m band stamps ($15 \times 15 $\,arcsec) 
for AzTEC objects. Red circles are 2\,arcsec in diameter and are centred on the SMA positions.}
\label{fig:stamps1}
\end{figure*}

\begin{figure*}
\begin{center}
\includegraphics[scale=0.75, trim = 30mm 30mm 30mm 0mm]{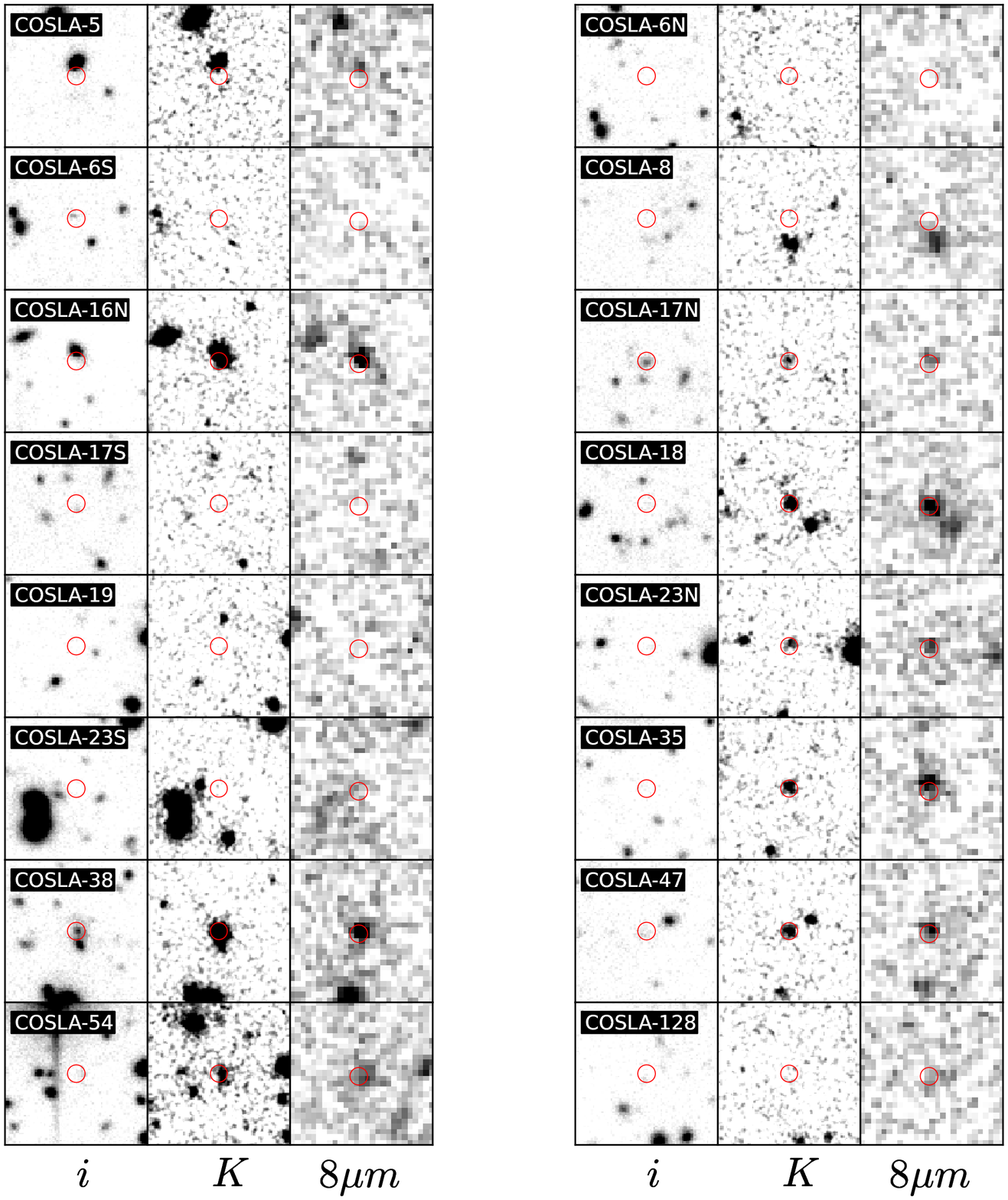}
\end{center}
\caption{CFHTLS $i$-band, UltraVISTA $K_s$-band and IRAC $8\,\mu$m band stamps ($15 \times 15$\,arcsec) for COSLA objects. Red circles are 2\,arcsec in diameter and are centred on the PdBI
positions.}
\label{fig:stamps2}
\end{figure*}

\begin{table*}
%\begin{minipage}{240mm} 
\caption{The results of our attempt to establish galaxy identifications for the (sub-)mm sources based on statistical
associations between the original single-dish (sub-)mm positions
and potential counterparts in the multi-wavelength imaging. RA and DEC refer
to the position of the $K$-band counterpart (except in the case of AzTEC2 where the position refersto the radio counterpart), 
and `Offset' is the distance in arcsec from this position 
and the original single-dish (sub-)mm source position. We sought counterparts based on positional offset and 
$i-K$ colour, 8\,$\mu$m flux density, 24\,$\mu$m flux density, and radio 1.4\,GHz flux density as described in Section 5.2.
For each method the probability that the counterpart could have been found by chance is given by the signficance level 
$p$ (see Dunlop et al. 1989; Ivison et al. 2007).
Objects highlighted in bold indicate the 16 sources for which the identification chosen here is 
confirmed as correct by the improved positional accuracy provided by the SMA and PdBI interferometric 
observations.
COSLA-23 
(as identified in the LABOCA map) was matched to an object close to the position of COSLA-23N 
(as identified by PdBI). No significant association was found with COSLA-23S.}
\setlength{\tabcolsep}{0.9 mm}
\include{p_statistics}
\label{tab:p}
%\end{minipage}
\end{table*}

\begin{landscape}
\begin{table}
%\begin{minipage}{240mm} 
\caption{Optical CFHTLS, near-infrared UltraVISTA and IRAC AB magnitudes with errors calculated using 2\,arcsec-diameter aperture measurements corrected to `total' 
using the relevant on-image PSF. Errors and flux limits are 
           the $1\sigma$ and $2\sigma$ values respectively.}
\setlength{\tabcolsep}{0.7 mm} 
\include{photometry1}
\label{tab:photo1}
%\end{minipage}
\end{table}

\begin{table}
%\begin{minipage}{240mm} 
\caption{Optical Subaru, near-infrared UltraVISTA and IRAC AB magnitudes with errors calculated using 2\,arcsec-diameter aperture measurements corrected to `total' using 
the relevant on-image PSF. Errors are the $1\sigma$ values.}
\setlength{\tabcolsep}{0.7 mm} 
\include{photometry2}
\label{tab:photo2}
%\end{minipage}
\end{table}
\end{landscape}

\section{Comparison of single-dish and interferometic flux densities and identifications}

In this appendix we illustrate the extent to which flux densities and redshifts resulting from the galaxy 
identification process depend on whether one works with the original single dish (sub-)mm fluxes and positions, or instead 
adopts the corresponding information derived from the interferometric (PdBI and SMA) observations.

\begin{figure*}
\begin{center}
\includegraphics[scale=0.50,angle=270]{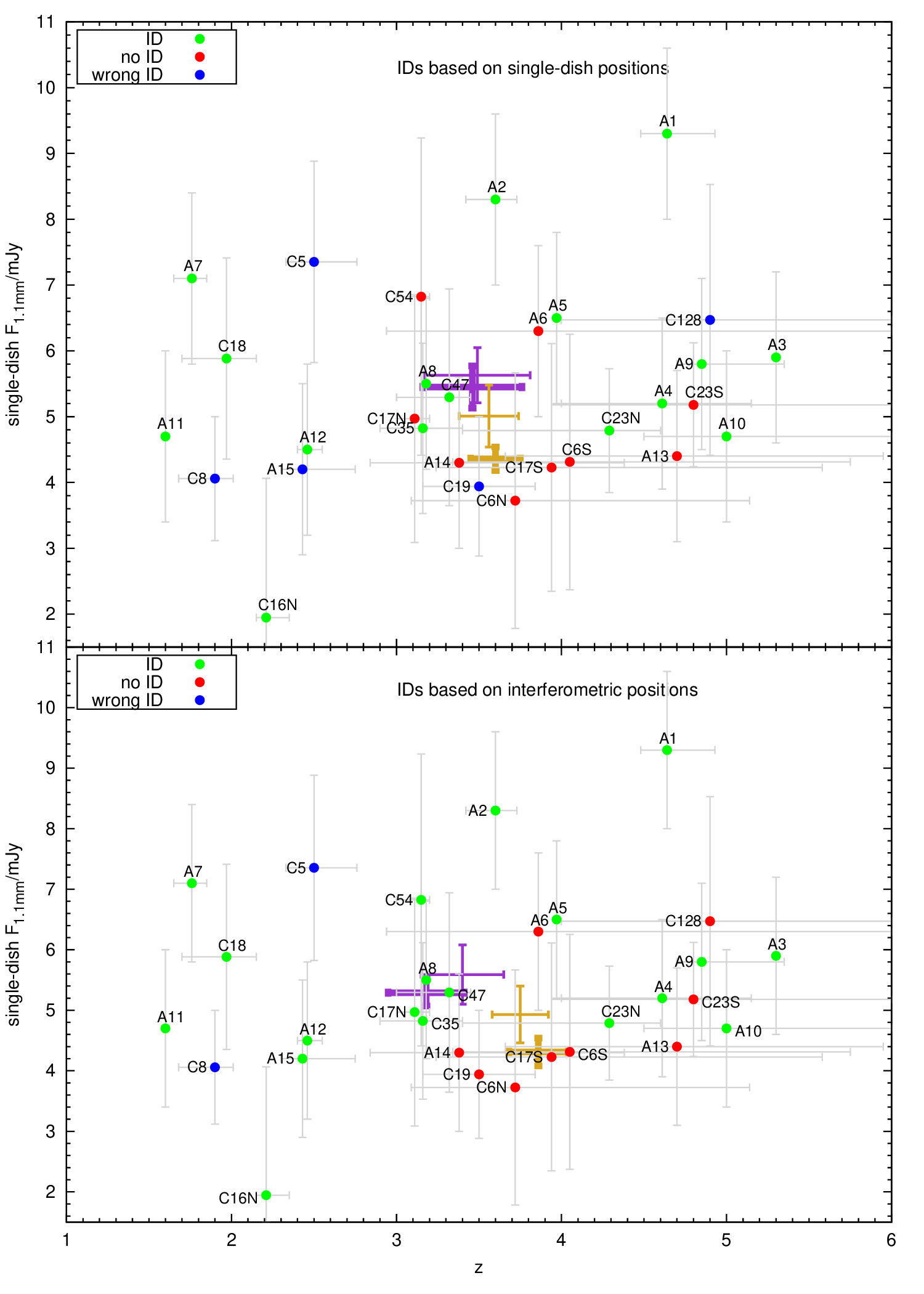}
\end{center}
\caption{Single-dish 1.1\,mm flux densities plotted against redshift. Flux densities are taken directly from the 
1.1\,mm AzTEC observations or scaled from the LABOCA 870\,$\mu$m measurements
using $F_{870\, \mu m}/F_{1.1\,{\rm mm}}=1.7$ (Michalowski et al. 2010). Green dots show objects which were correctly identified using the 
single-dish positions (upper panel) or interferometric positions (lower panel). Red dots indicate 
the unidentified sources, while blue dots indicate sources which formally have statistically acceptable identifications
which we are confident are not in fact the correct galaxy counteparts (usually due to a severe mismatch between, 
$z_{phot}$ and $z_{LW}$ as produced by, for example, galaxy-galaxy lensing).
The violet points with error bars show median (thicker errorbars) and mean (thinner errorbars) values for all 
the identified sources. The brown points with error bars indicate the corresponding average values for the 
unidentified sources.}
\label{fig:twographs1}
\end{figure*}

\begin{figure*}
\begin{center}
\includegraphics[scale=0.50,angle=270]{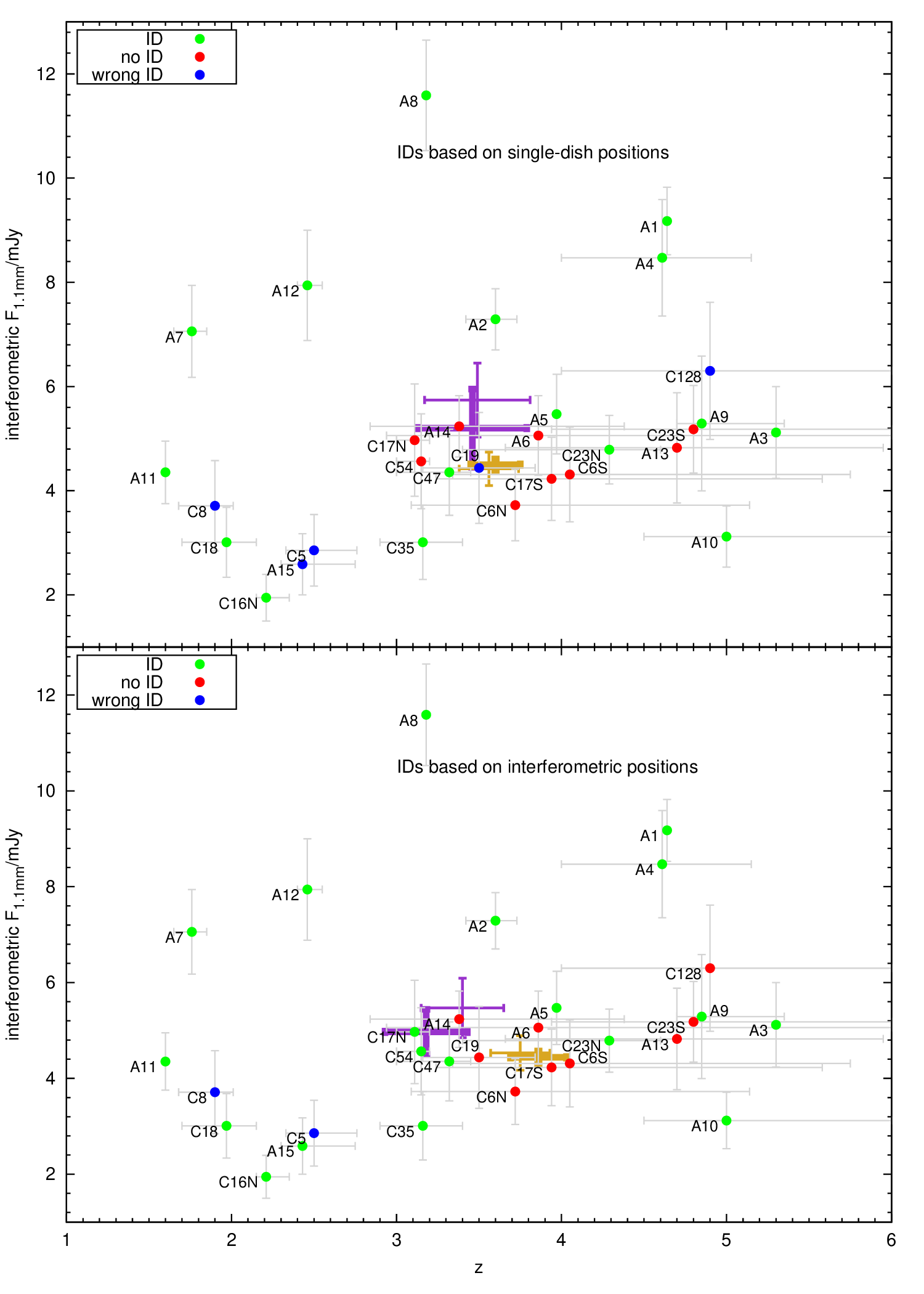}
\end{center}
\caption{Interferometric 1.1\,mm flux densities plotted against redshift. Flux densities are scaled from the SMA 890\,$\mu$m 
measurements 
using $F_{890\,\mu m}/F_{1.1\,{\rm mm}}=1.7$, and scaled from the PdBI 1.3\,mm measurements using 
$F_{1.3mm}/F_{1.1\,{\rm mm}}=0.7$.
Green dots show objects which were correctly identified using the 
single-dish positions (upper panel) or interferometric positions (lower panel). Red dots indicate 
the unidentified sources, while blue dots indicate sources which formally have statistically acceptable identifications
which we are confident are not in fact the correct galaxy counteparts (usually due to a severe mismatch between, 
$z_{phot}$ and $z_{LW}$ as produced by, for example, galaxy-galaxy lensing).
The violet points with error bars show median (thicker error bars) and mean (thinner error bars) values for all 
the identified sources. The brown points with error bars indicate the corresponding average values for the 
unidentified sources.}
\label{fig:twographs2}
\end{figure*}

\clearpage

\section{Spectral Energy Distribution fits and photometric error estimates}
In this final appendix we provide, for each robust galaxy identification with 
optical-infrared data of sufficient quality, the photometric data overlaid with the 
best-fitting SED models, along with the associated plot of $\chi^2$ versus redshift $z$, 
after marginalising over all other fitted parameters. The redshift corresponding to 
the minimum $\chi^2$ is the value of $z_{phot}$ tabulated in Table\,5, while the uncertainty 
in redshift is derived from the redshifts corresponding to $\Delta \chi^2 = 1$ above the 
minimum $\chi^2$ value.

\begin{figure*}
\begin{center}
\begin{tabular}{cc}
\includegraphics[scale=0.38]{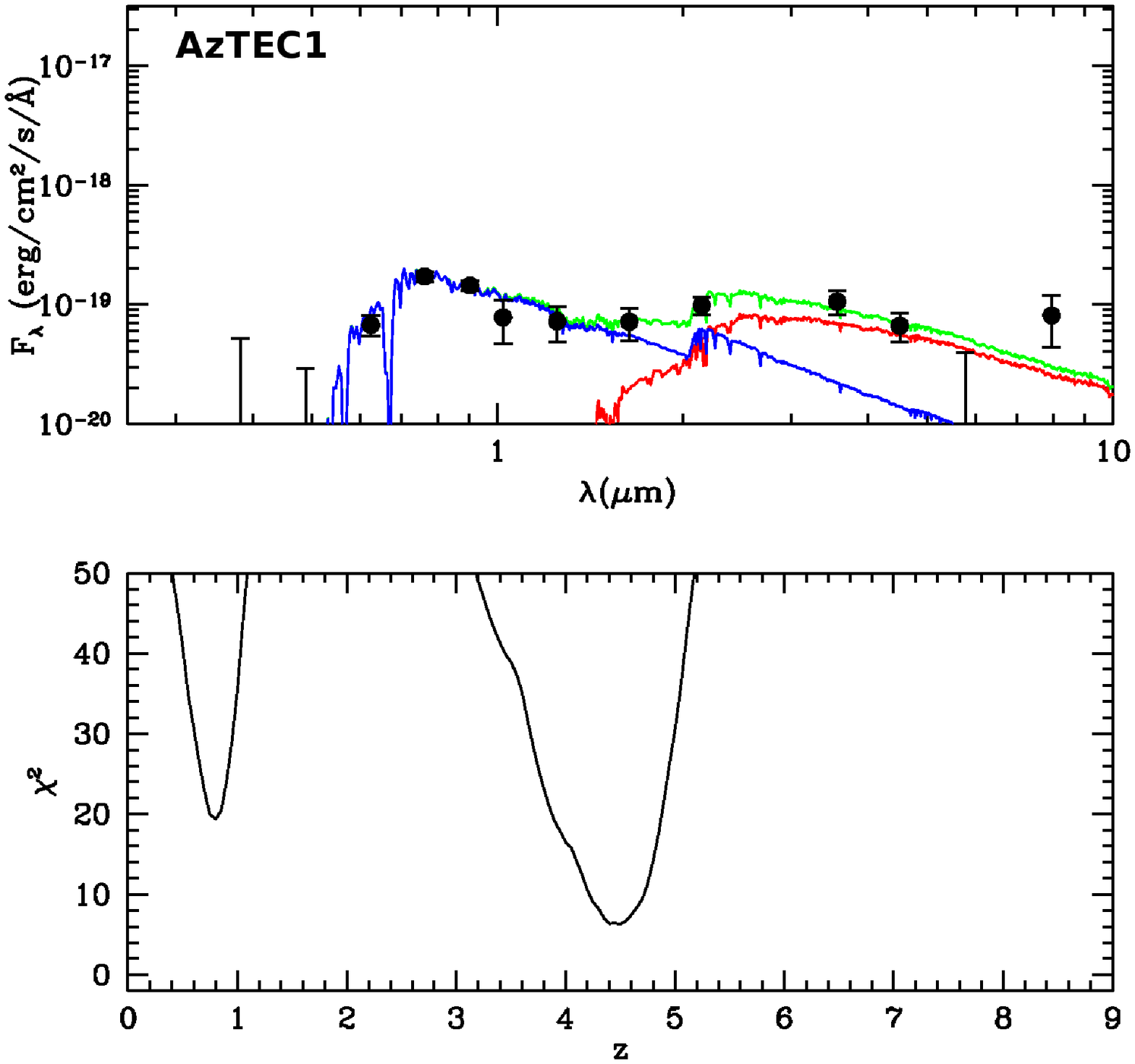}
\includegraphics[scale=0.38]{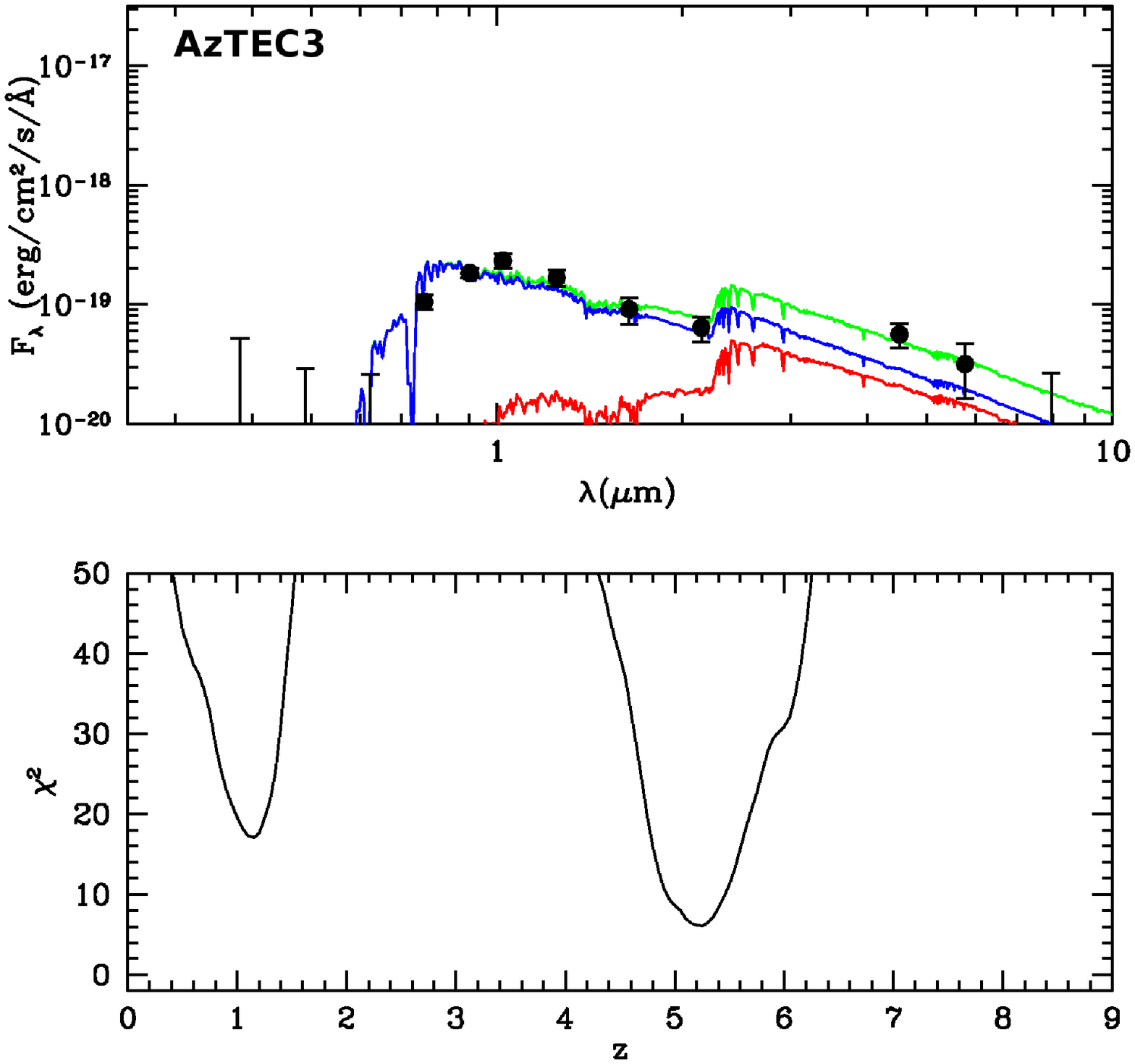}\\
\includegraphics[scale=0.38]{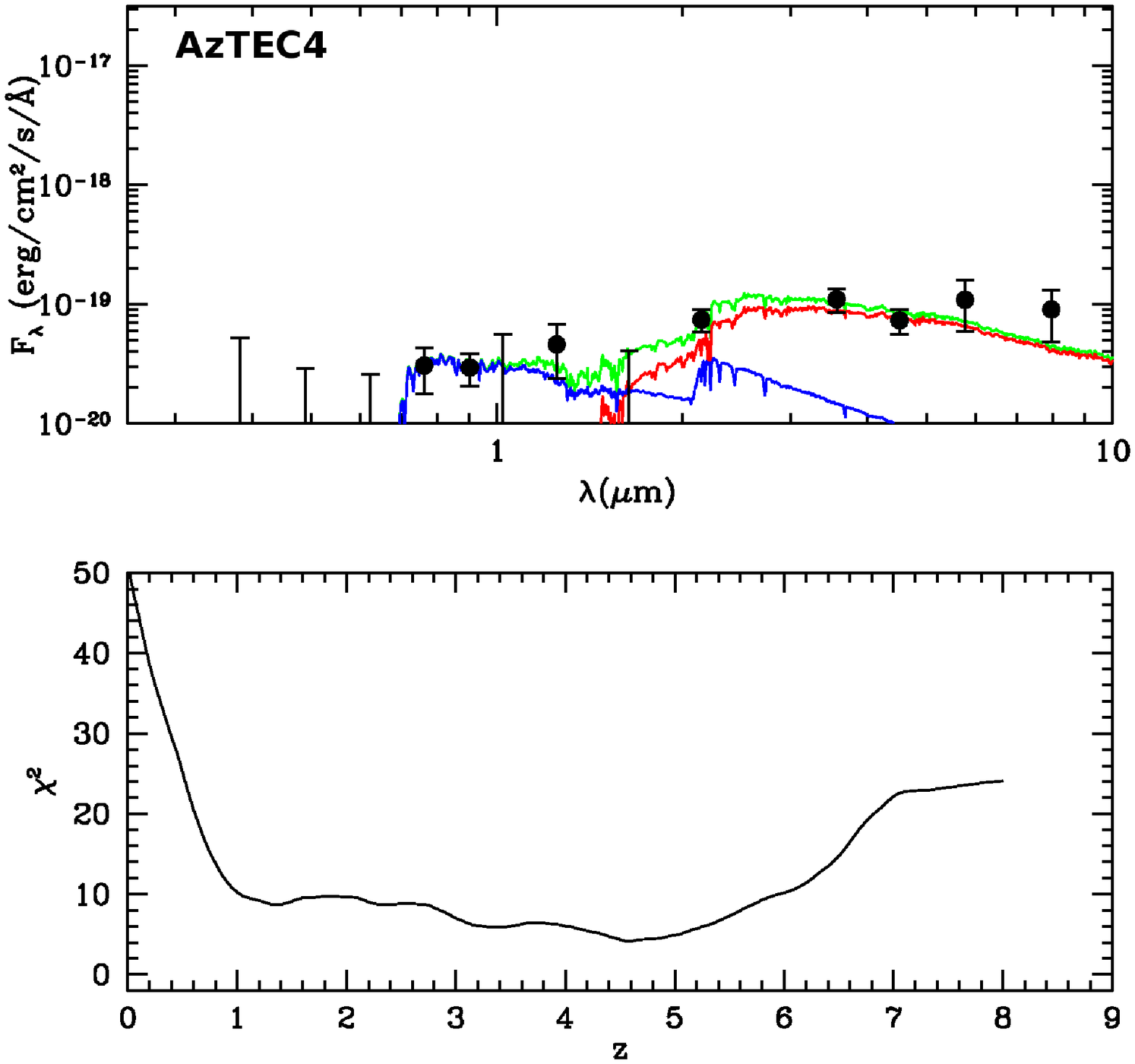}
\includegraphics[scale=0.38]{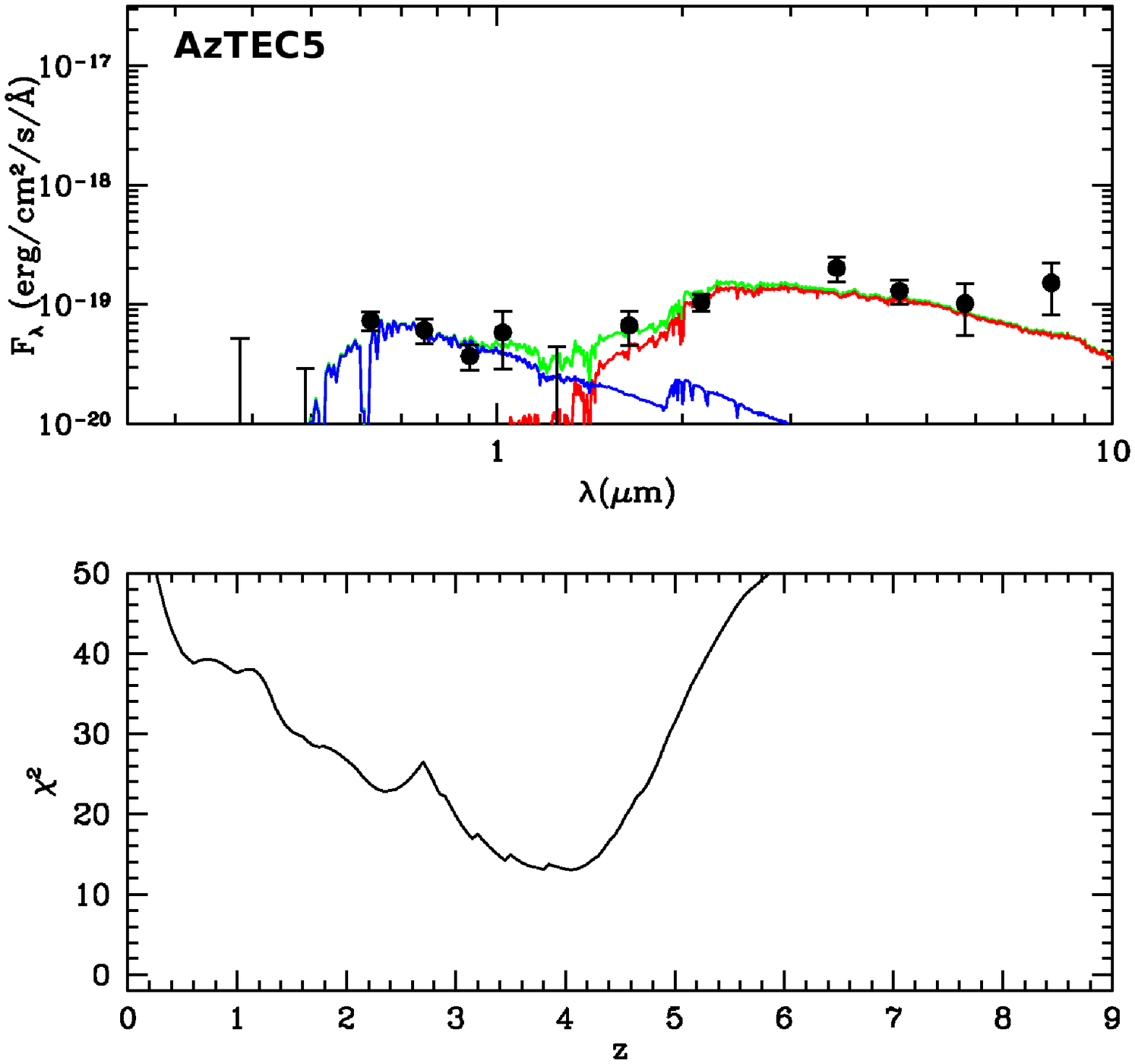}\\
\includegraphics[scale=0.38]{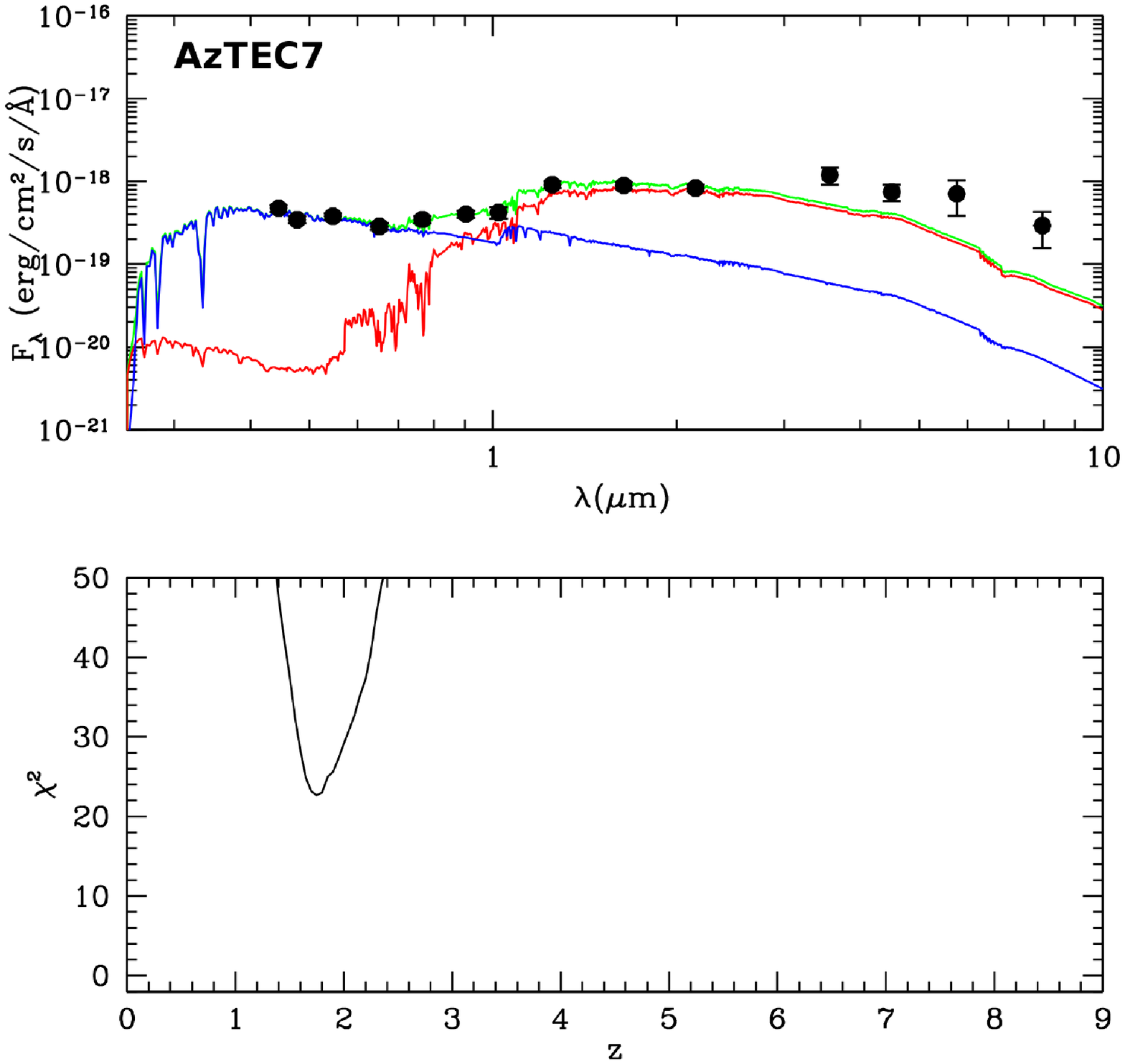}
\includegraphics[scale=0.38]{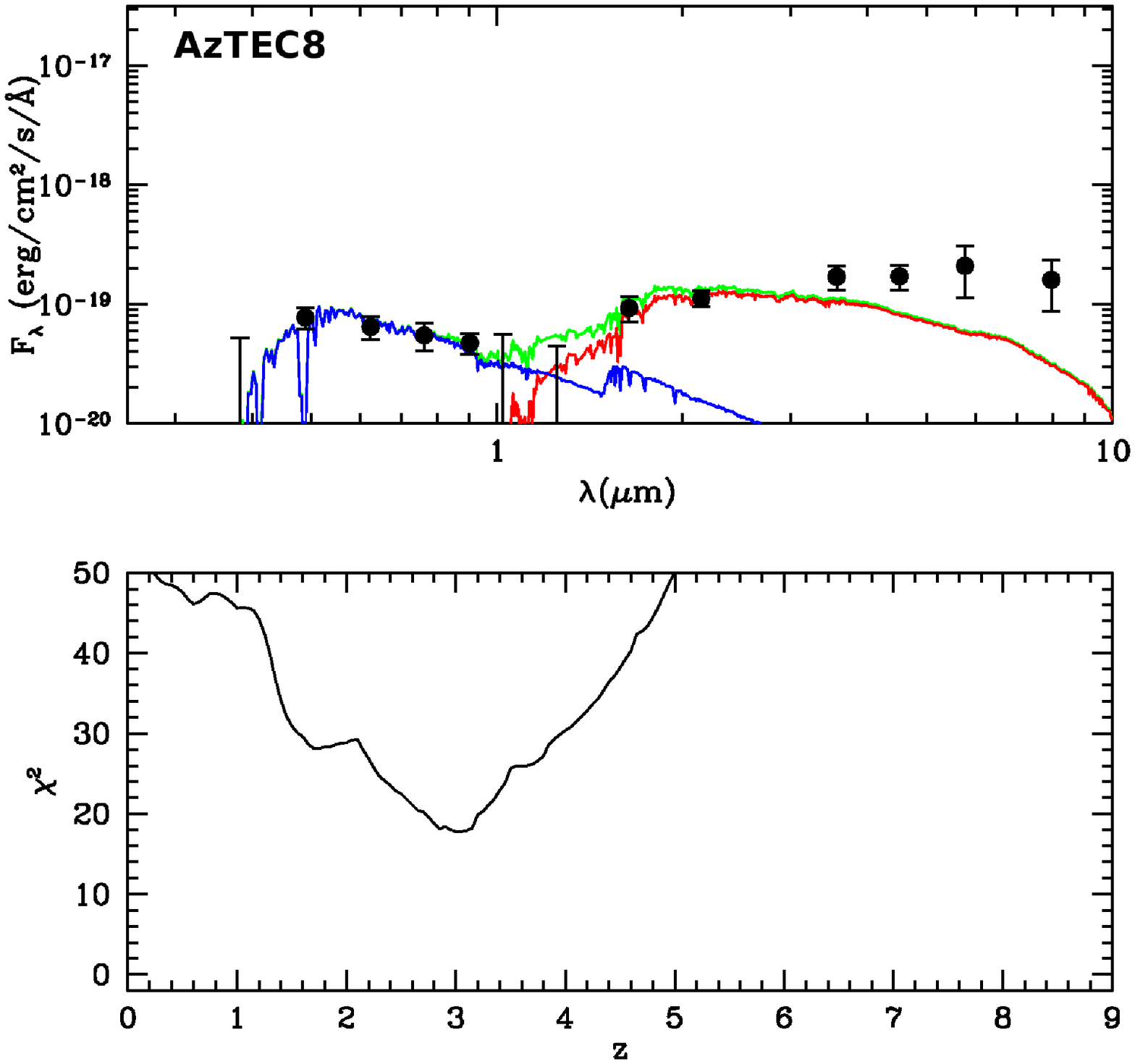}
\end{tabular}
\end{center}
\caption{Best-fitting spectral energy distributions (SEDs) and plots of $\chi^2$ versus redshift for sources where optical IDs were found (Tables B2
and B3). In all cases the star formation history was modelled with two instantaneous bursts, where the blue, red and green
lines indicate the young, old and composite stellar populations respectively.}
\label{fig:SED1}
\end{figure*}

\setcounter{figure}{0}

\begin{figure*}
\begin{center}
\begin{tabular}{cc}
\includegraphics[scale=0.38]{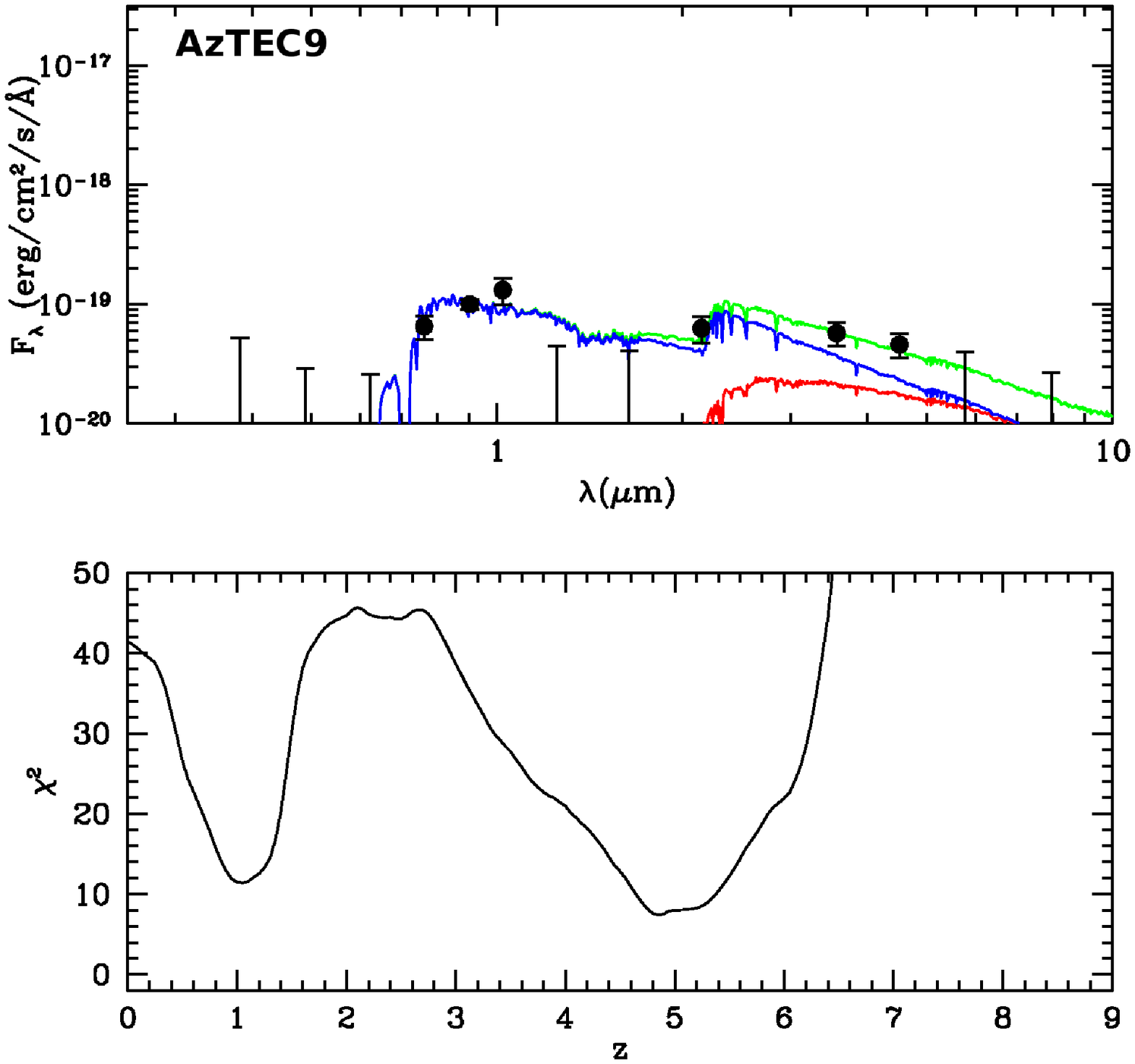}
\includegraphics[scale=0.38]{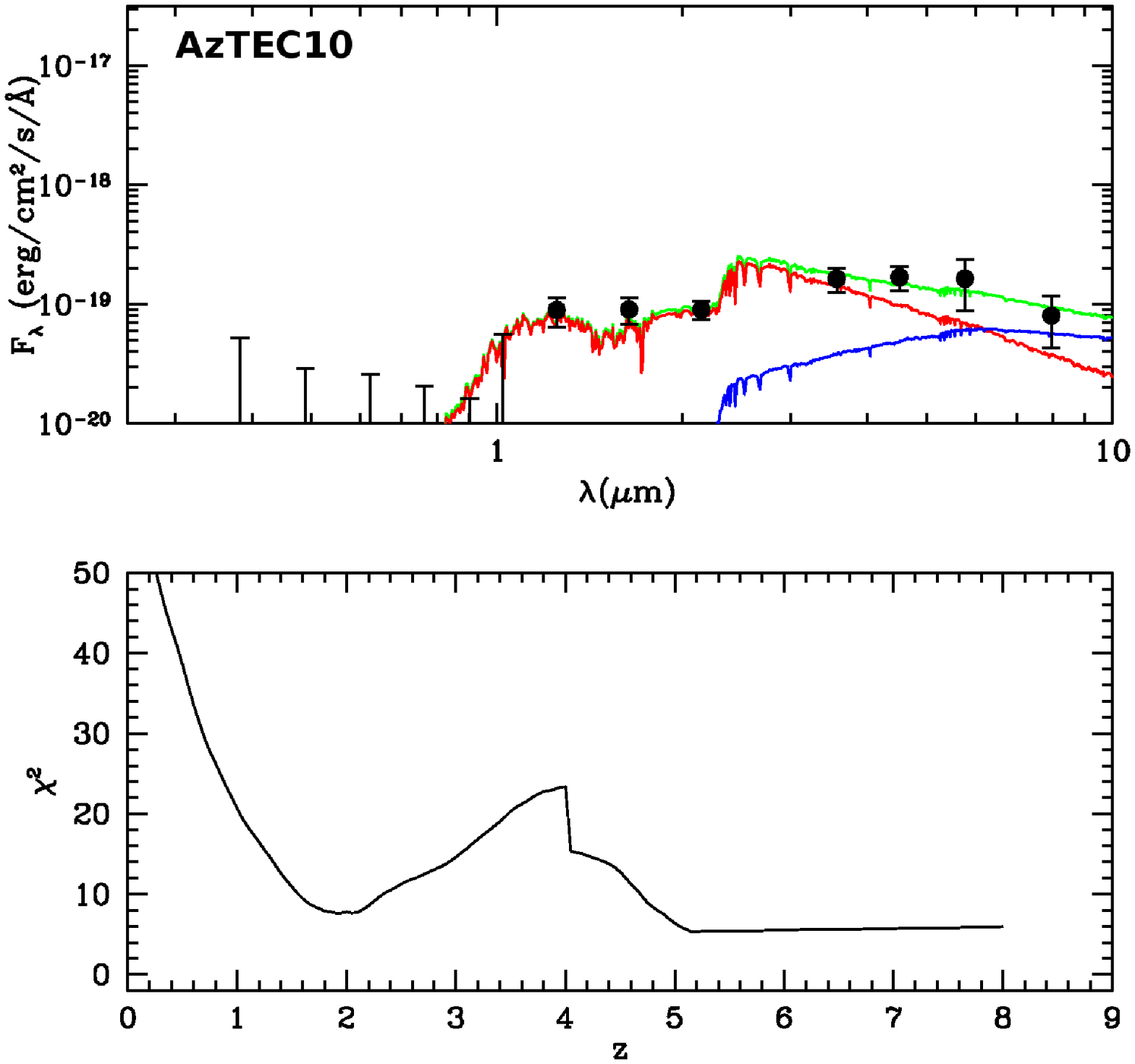}\\
\includegraphics[scale=0.38]{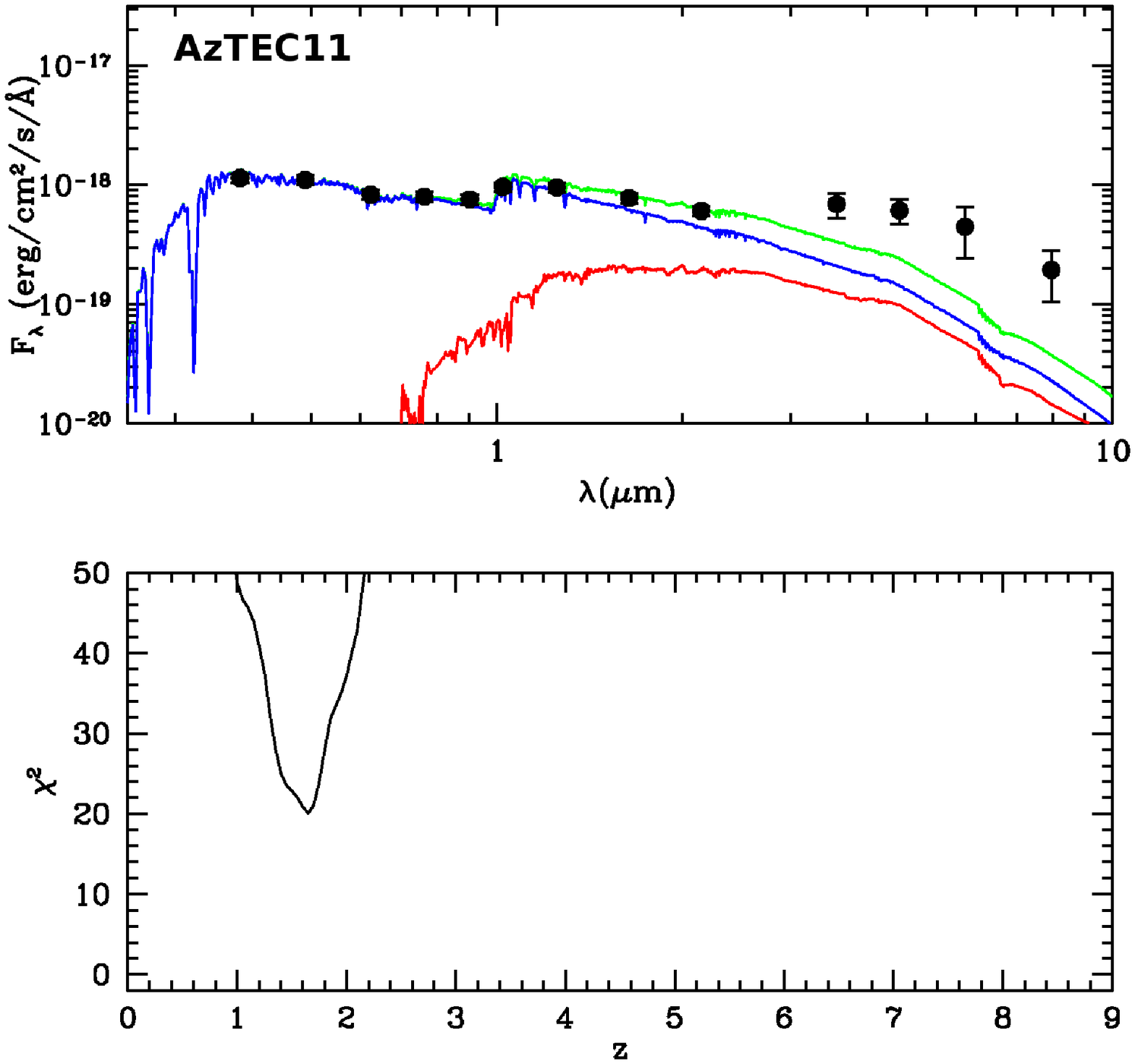}
\includegraphics[scale=0.38]{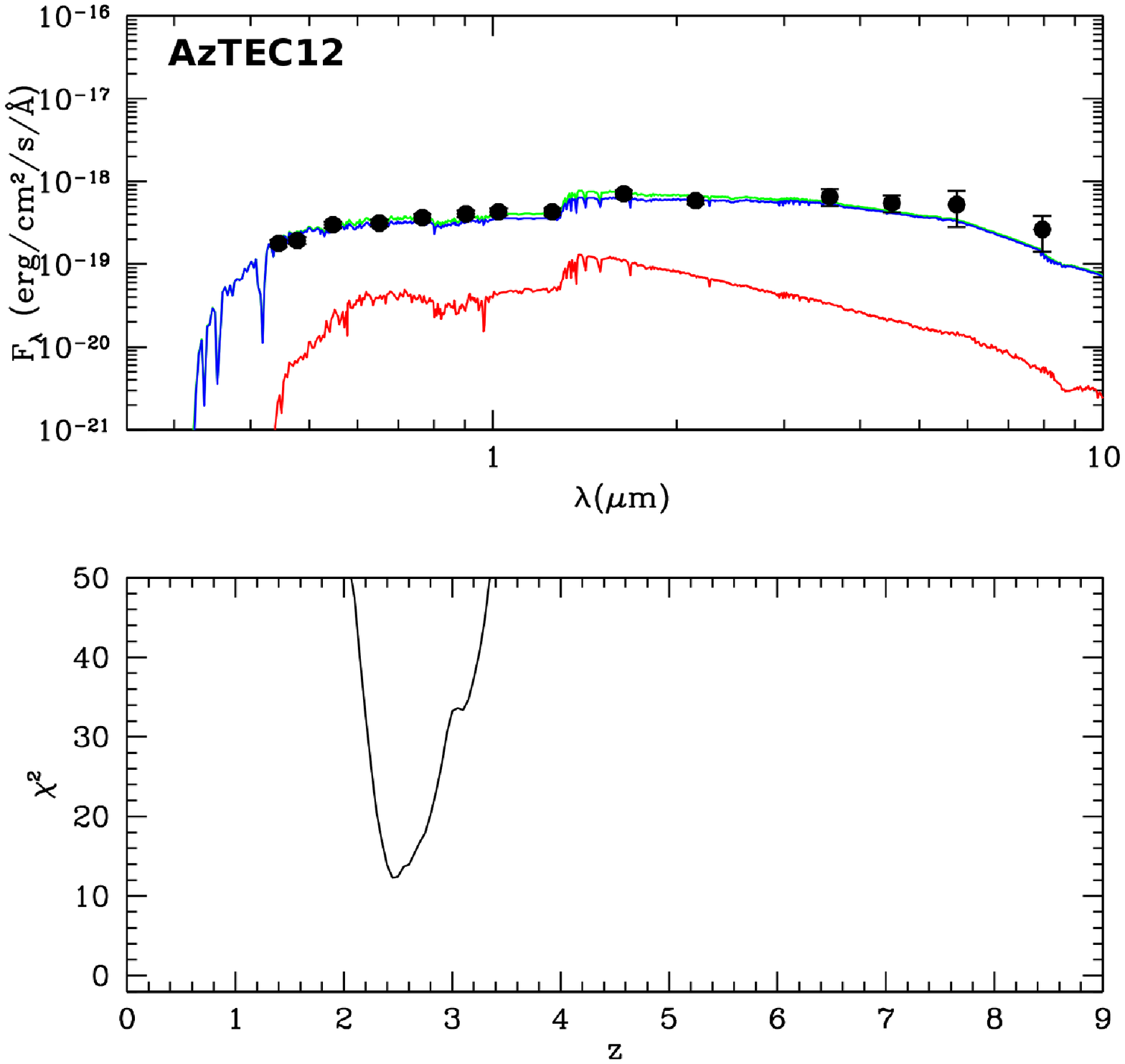}\\
\includegraphics[scale=0.38]{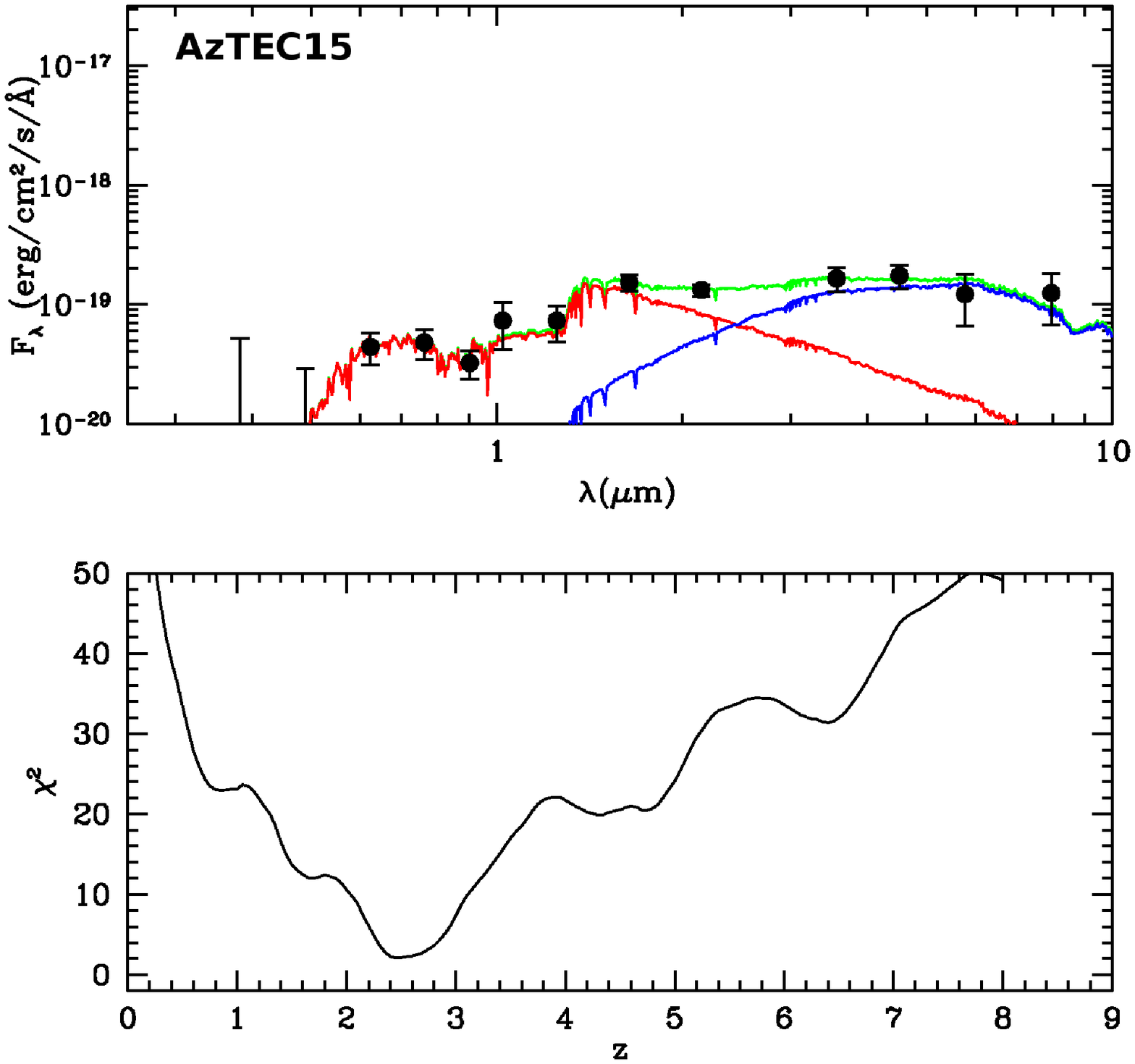}
\includegraphics[scale=0.38]{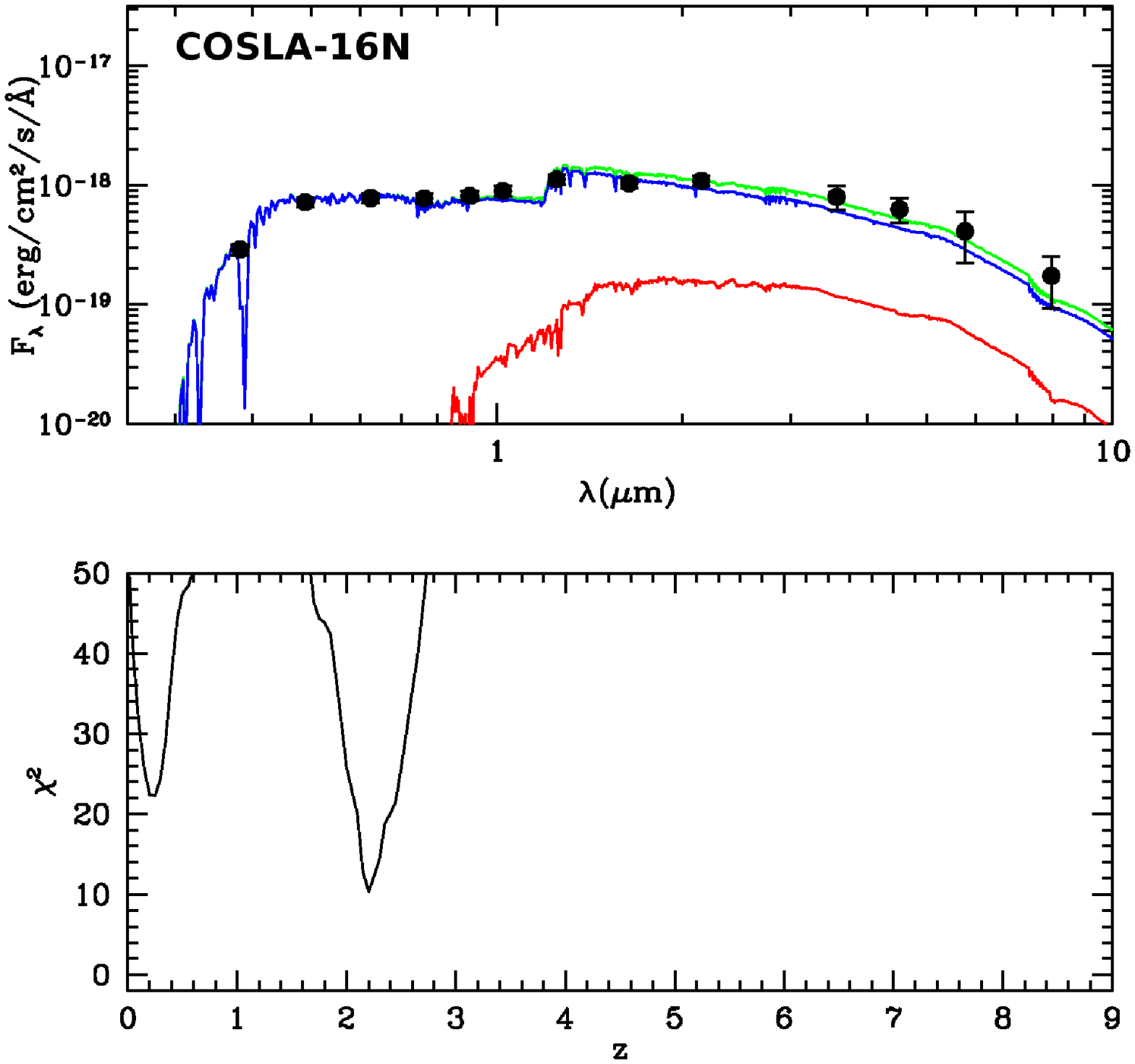}
\end{tabular}
\end{center}
\caption{continued.}
\label{fig:SED2}
\end{figure*}

\setcounter{figure}{0}

\begin{figure*}
\begin{center}
\begin{tabular}{cc}
\includegraphics[scale=0.38]{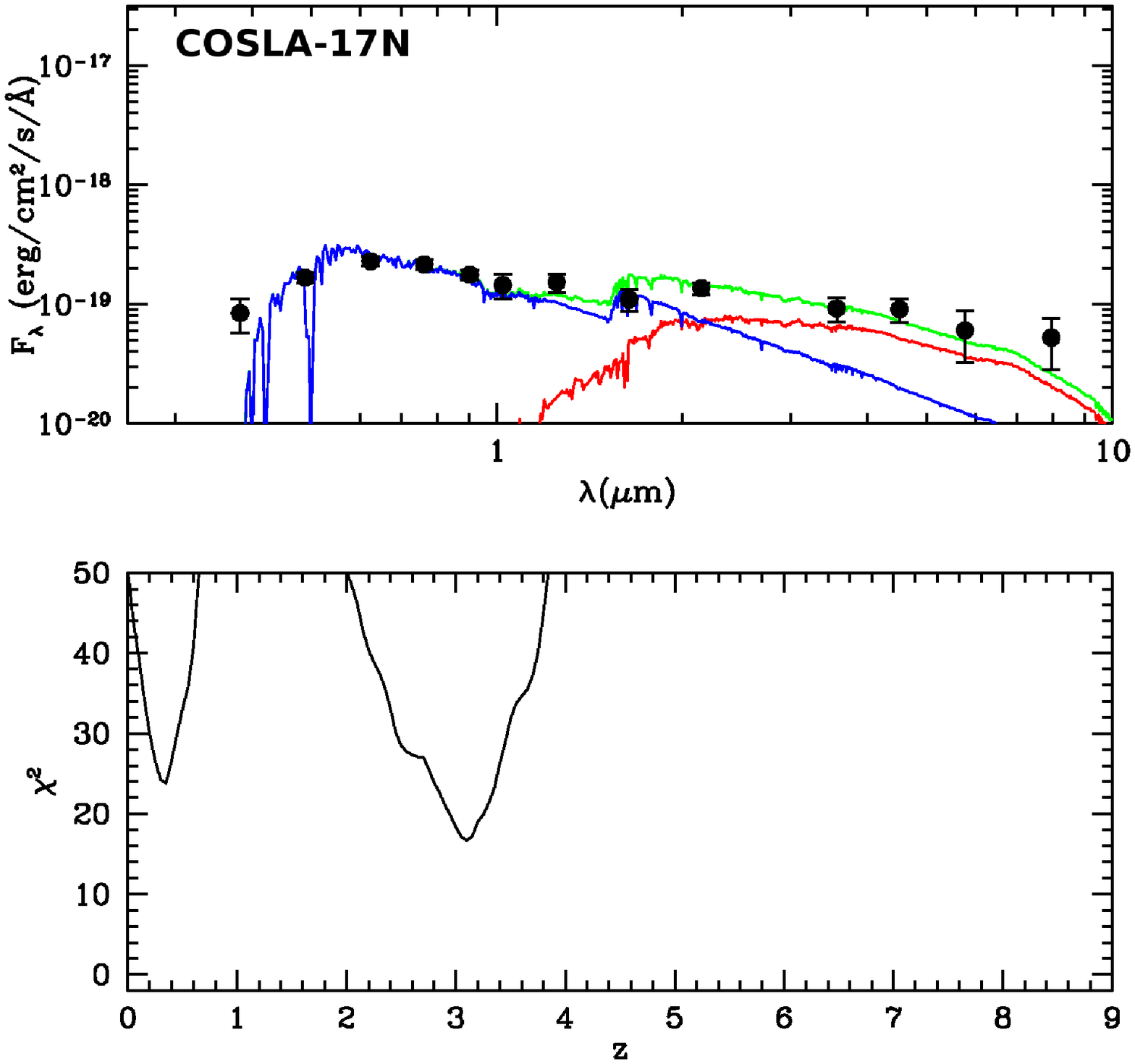}
\includegraphics[scale=0.38]{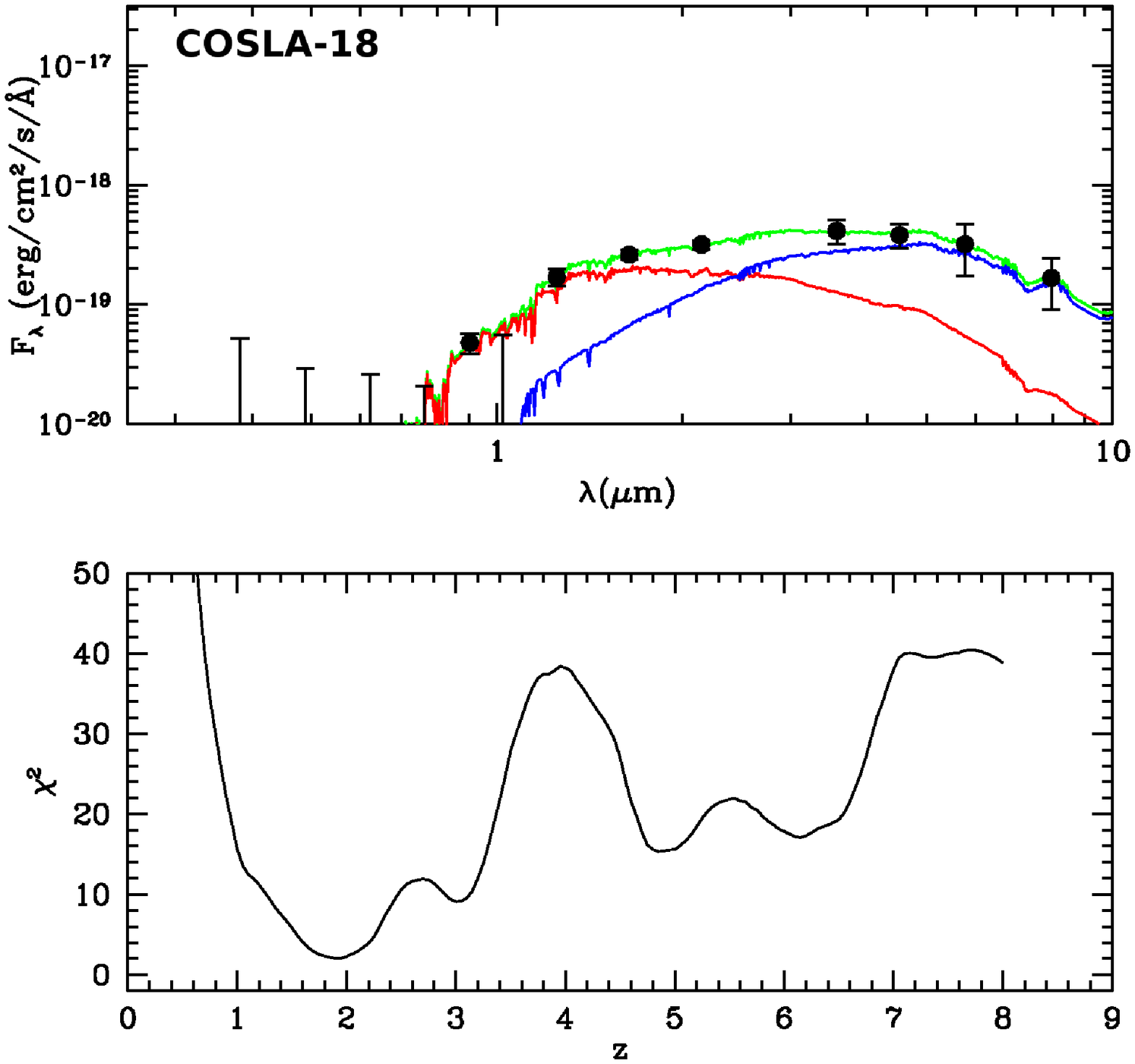}\\
\includegraphics[scale=0.38]{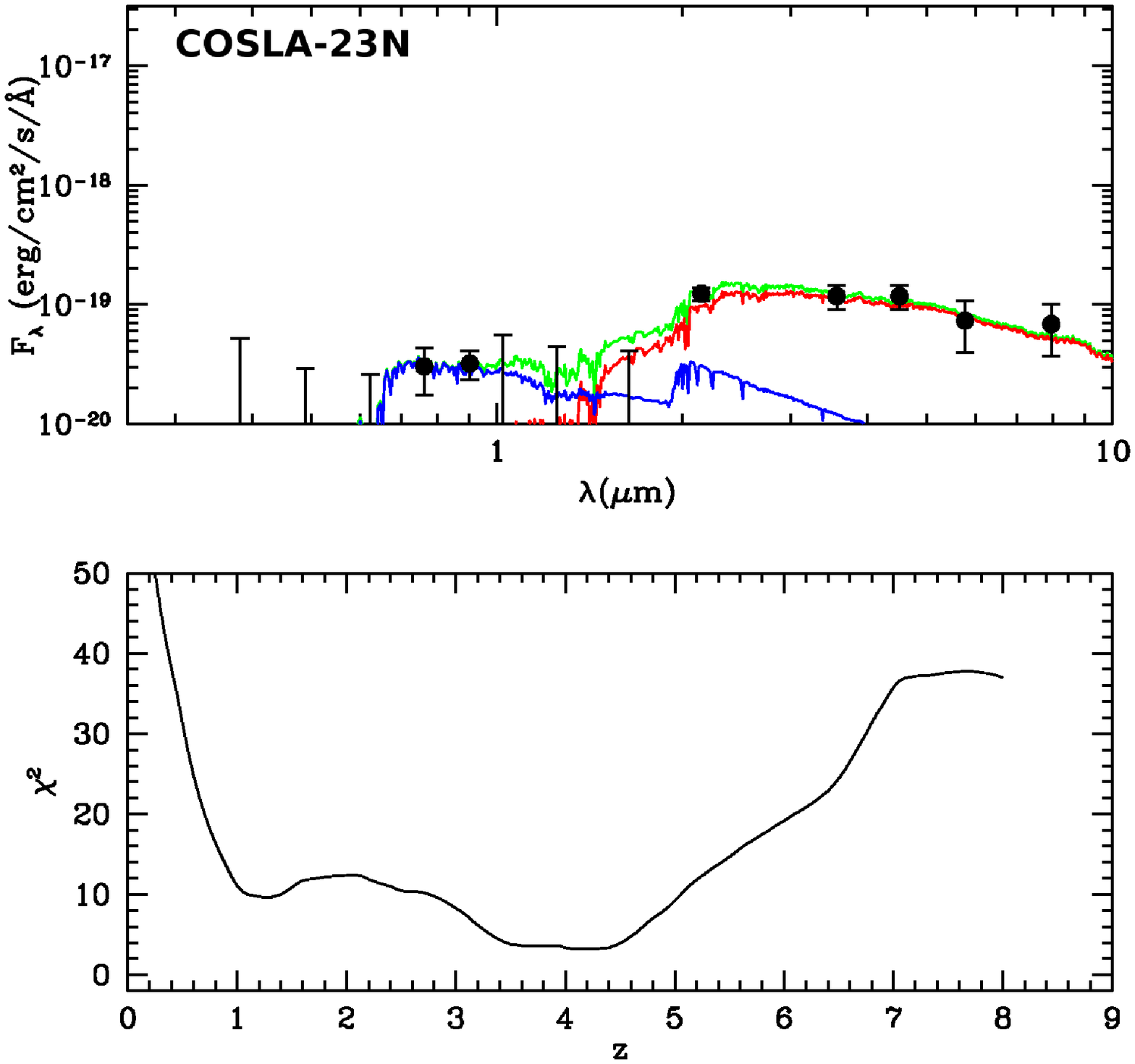}
\includegraphics[scale=0.38]{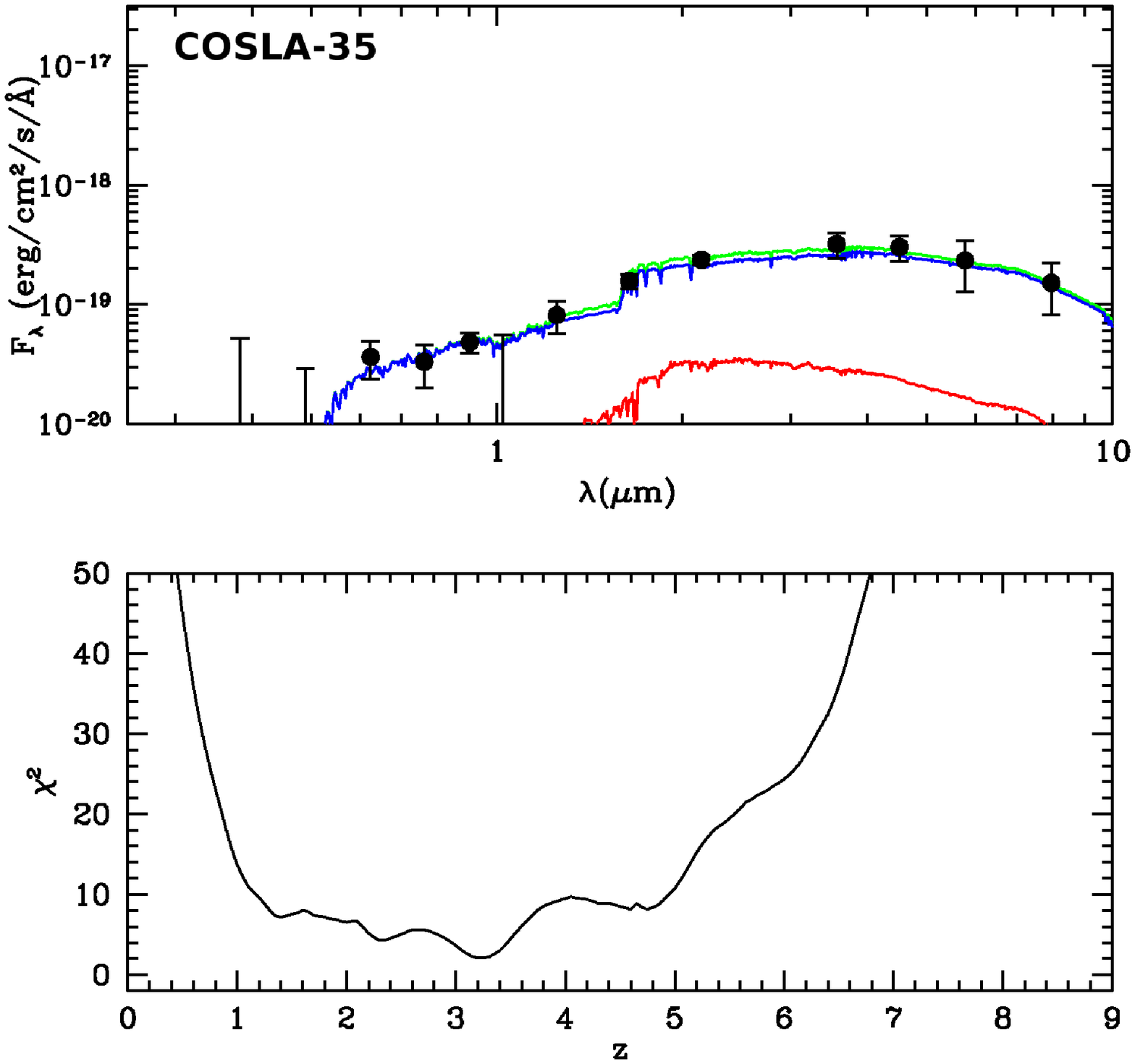}\\
\includegraphics[scale=0.38]{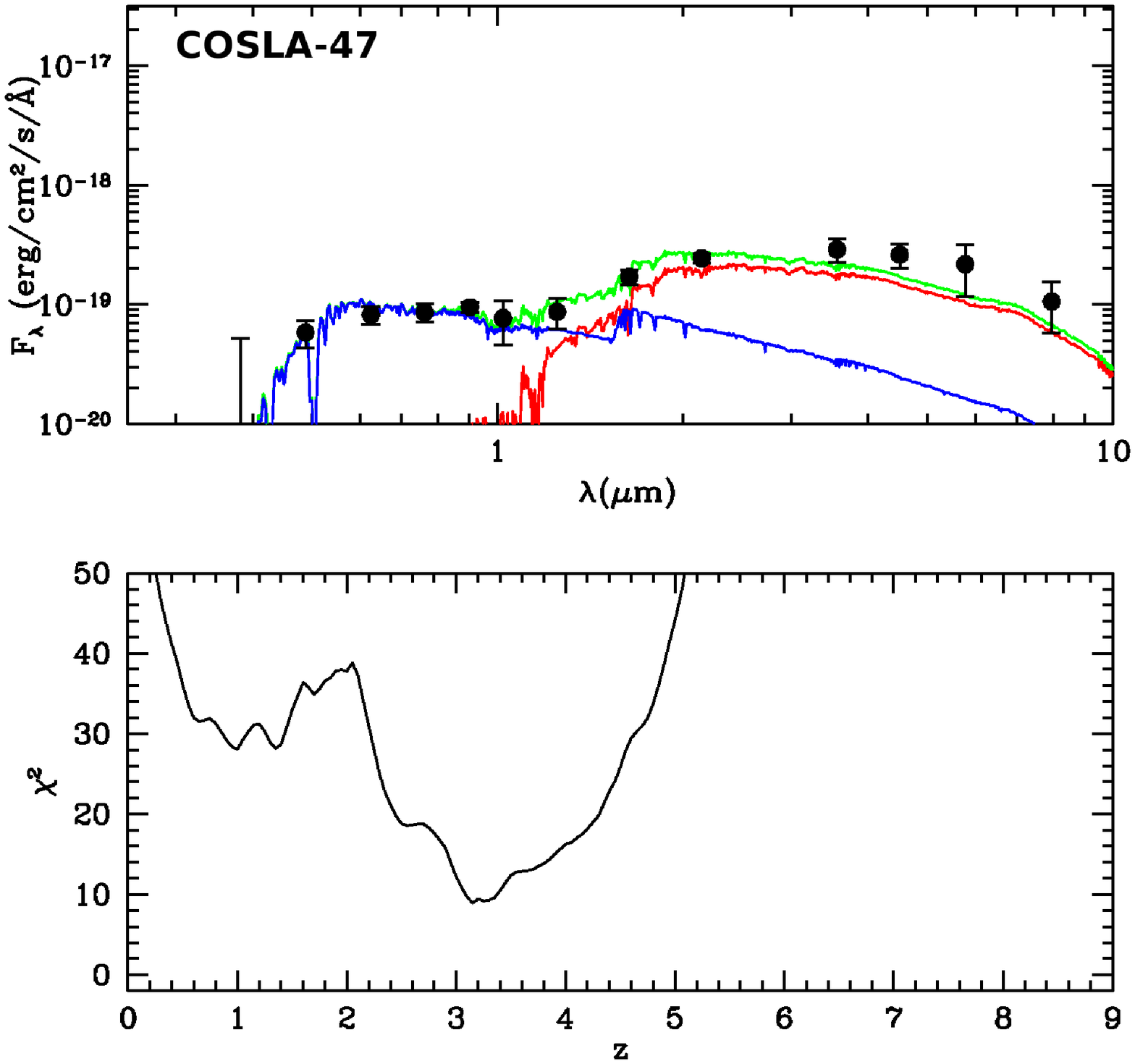}
\includegraphics[scale=0.38]{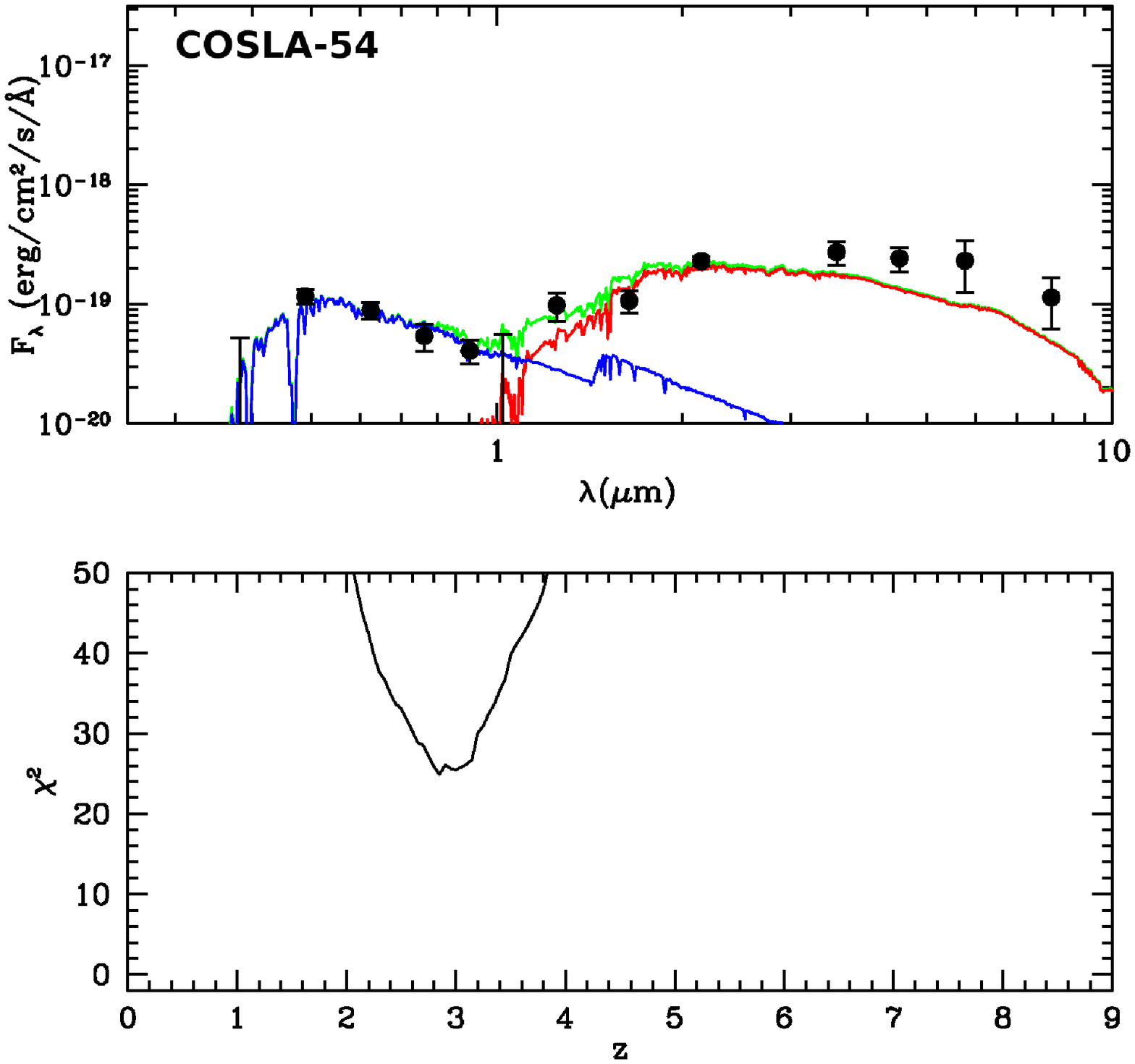}
\end{tabular}
\end{center}
\caption{continued.}
\label{fig:SED3}
\end{figure*}

\label{lastpage}

\end{document}

%% file: p_statistics.tex
\begin{normalsize}
\begin{center}
\begin{tabular}{lcccccccccccc}
\hline
ID & RA    & Dec   & Offset & $K$  & $i-K$ & $p_{i-K}$ & $S_{8\mu m}$  & $p_{8\mu m}$ & $S_{24\mu m}$ & $p_{24\mu m}$ & $S_{VLA}$ & $p_{VLA}$ \\
   & /deg & /deg & /arcsec  & /AB & /AB  &           & /$\mu$Jy    &              & /mJy       &              & /mJy  &  \\
\hline
\bf{AzTEC1} & 149.92859 & 2.49393 & 3.5 & 23.44 & 1.60 & $>0.1$ & $14.0\pm2.4$ & 0.036 & - & - & - & - \\    
\bf{AzTEC2} & 150.03343 & 2.43671 & 0.1 & $>24.57$ & - & - & - & - & $0.181\pm0.027$ & 0.002 & $0.076\pm0.014$ & 0.001 \\
\bf{AzTEC3} & 150.08629 & 2.58898 & 2.1 & 23.94 & 1.16 & $>0.1$ & $10.5\pm2.3$ & 0.059 & - & - & - & - \\   
\bf{AzTEC4} & 149.88196 & 2.51215 & 4.3 & 23.76 & 3.16 & 0.031 & $17.5\pm2.0$ & 0.083 & - & - & - & - \\
\bf{AzTEC5} & 150.08240 & 2.53456 & 1.7 & 23.38 & 2.79 & 0.041 & $23.4\pm2.2$ & 0.028 & $0.189\pm0.013$ & 0.017 & $0.126\pm0.015$ & 0.002 \\
\bf{AzTEC7} & 150.07529 & 2.80841 & 2.7 & 21.13 & 3.18 & 0.003 & $57.3\pm2.6$ & 0.025 & $0.441\pm0.012$ & 0.006 & $0.132\pm0.022$ & 0.003 \\
\bf{AzTEC8} & 149.99721 & 2.57804 & 4.8 & 23.30 & 2.98 & 0.072 & $34.6\pm2.5$ & 0.065 & - & - & - & - \\
\bf{AzTEC9} & 149.98870 & 2.45840 & 1.7 & 24.15 & 1.52 & $>0.1$ & - & - & - & - & $0.068\pm0.013$ & 0.002 \\
\bf{AzTEC10} & 149.87819 & 2.67563 & 1.9 & 23.54 & 4.24 & 0.031 & $17.3\pm2.3$ & 0.031 & $0.086\pm0.016$ & 0.021 & - & - \\
\bf{AzTEC11} & 150.03726 & 2.66956 & 3.2 & 21.48 & 1.90 & 0.036 & $42.0\pm2.5$ & 0.043 & $0.488\pm0.011$ & 0.008 & $0.302\pm0.045$ & 0.002 \\
\bf{AzTEC12} & 150.14708 & 2.73144 & 1.4 & 21.51 & 2.74 & 0.004 & $56.9\pm2.4$ & 0.010 & $0.261\pm0.011$ & 0.007 & $0.098\pm0.016$ & 0.002 \\
AzTEC15 & 150.05586 & 2.57334 & 5.1 & 19.90 & 2.16 & 0.014 & $26.2\pm2.2$ & $>0.1$ & - & - & - & - \\
COSLA-5 & 150.24872 & 2.28574 & 3.3 & 19.92 & 2.63 & 0.003 & $26.1\pm2.2$ & 0.060 & - & - & - & - \\
COSLA-8 & 150.10641 & 2.25154 & 4.0 & 22.04 & 3.98 & 0.023 & $26.4\pm2.2$ & 0.080 & $0.560\pm0.017$ & 0.012 & $0.112\pm0.010$ & 0.006 \\
\bf{COSLA-16} & 150.21494 & 2.55951 & 3.1 & 20.83 & 2.57 & 0.009 & $36.4\pm2.5$ & 0.049 & $0.339\pm0.025$ & 0.016 & $0.122\pm0.013$ & 0.004 \\
\bf{COSLA-18} & 150.17992 & 2.08863 & 2.9 & 22.18 & 5.14 & 0.018 & $35.3\pm2.0$ & 0.044 & $0.320\pm0.069$ & 0.022 & $0.078\pm0.014$ & 0.005 \\
COSLA-19 & 150.03380 & 2.19506 & 6.2 & 20.92 & 2.27 & 0.059 & - & - & - & - & - & - \\
\bf{COSLA-23} & 150.04231 & 2.22635 & 1.8 & 23.21 & 3.72 & 0.025 & $14.4\pm2.4$ & 0.048 & $0.135\pm0.035$ & 0.066 & $0.059\pm0.011$ & 0.003 \\
\bf{COSLA-35} & 150.09857 & 2.36537 & 3.9 & 22.49 & 4.35 & 0.037 & $31.8\pm2.5$ & 0.075 & $0.168\pm0.017$ & 0.049 & $0.043\pm0.011$ & 0.010 \\
\bf{COSLA-47} & 150.13901 & 2.43378 & 6.6 & 22.46 & 3.33 & 0.070 & $22.3\pm2.4$ & $>0.1$ & - & - & - & - \\
COSLA-128 & 150.40825 & 2.39440 & 8.0 & 17.63 & 1.19 & $>0.1$ & $17.2\pm2.3$ & 0.020 & $0.864\pm0.032$ & 0.015 & $0.172\pm0.048$ & 0.010 \\ 
\hline
\end{tabular}
\end{center}
\end{normalsize}

%% file: photometry1.tex
\begin{scriptsize}
\begin{tabular}{lcccccccccccccccc}
\hline
ID & RA & DEC & $u$ & $g$ & $r$ & $i$ & $z$ & $Y$ & $J$ & $H$ & $Ks$ & $3.6\mu m$ & $4.5\mu m$ & $5.8\mu m$ & $8.0\mu m$\\      
\hline
AzTEC1 & 149.92859 & 2.49394 & $>27.69$ & $>27.79$ & $26.55\pm0.22$ & $25.10\pm0.10$ & $24.92\pm0.07$ & $25.32\pm0.43$ & $24.96\pm0.36$ & $24.38\pm0.33$ & $23.45\pm0.18$ & $22.27\pm0.21$ & $22.27\pm0.29$ & $>22.29$ & $20.82\pm0.41$\\
AzTEC3 & 150.08620 & 2.58900 & $>27.69$ & $>27.79$ & $>27.41$ & $25.63\pm0.16$ & $24.66\pm0.06$ & $24.13\pm0.16$ & $24.05\pm0.17$ & $24.12\pm0.27$ & $23.92\pm0.26$ & $23.95\pm0.18$ & $22.45\pm0.09$ & $>22.29$ & $>22.01$\\
AzTEC4 & 149.88196 & 2.51216 & $>27.69$ & $>27.79$ & $>27.41$ & $26.97\pm0.46$ & $26.64\pm0.32$ & $>25.34$ & $25.45\pm0.53$ & $>24.75$ & $23.76\pm0.23$ & $22.23\pm0.20$ & $22.16\pm0.26$ & $21.20\pm0.47$ & $20.70\pm0.38$\\
AzTEC5 & 150.08240 & 2.53456 & $>27.69$ & $>27.79$ & $26.46\pm0.20$ & $26.22\pm0.25$ & $26.40\pm0.26$ & $25.64\pm0.55$ & $>25.21$ & $24.46\pm0.35$ & $23.38\pm0.17$ & $21.57\pm0.12$ & $21.54\pm0.15$ & $21.27\pm0.50$ & $20.14\pm0.24$\\
AzTEC8 & 149.99721 & 2.57804 & $>27.69$ & $26.93\pm0.22$ & $26.60\pm0.23$ & $26.34\pm0.28$ & $26.13\pm0.21$ & $>25.34$ & $>25.21$ & $24.10\pm0.26$ & $23.31\pm0.16$ & $21.76\pm0.02$ & $21.24\pm0.02$ & $20.49\pm0.04$ & $20.08\pm0.08$\\
AzTEC9 & 149.98870 & 2.45840 & $>27.69$ & $>27.79$ & $>27.41$ & $26.15\pm0.24$ & $25.32\pm0.10$ & $24.75\pm0.28$ & $>25.21$ & $>24.75$ & $23.94\pm0.27$ & $22.94\pm0.08$ & $22.67\pm0.11$ & $>22.29$ & $>22.01$\\
AzTEC10 & 149.87819 & 2.67563 & $>27.69$ & $>27.79$ & $>27.41$ & $>27.21$ & $>27.06$ & $>25.34$ & $24.73\pm0.30$ & $24.13\pm0.27$ & $23.55\pm0.19$ & $21.80\pm0.01$ & $21.26\pm0.02$ & $20.76\pm0.05$ & $20.83\pm0.13$\\
AzTEC11 & 150.03726 & 2.66957 & $24.54\pm0.03$ & $24.05\pm0.02$ & $23.83\pm0.02$ & $23.43\pm0.02$ & $23.12\pm0.01$ & $22.59\pm0.04$ & $22.16\pm0.03$ & $21.80\pm0.03$ & $21.48\pm0.03$ & $20.24\pm0.02$ & $19.86\pm0.02$ & $19.67\pm0.03$ & $19.87\pm0.04$\\
AzTEC15 & 150.05388 & 2.57634 & $>27.69$ & $>27.79$ & $27.01\pm0.32$ & $26.48\pm0.31$ & $26.54\pm0.29$ & $25.39\pm0.46$ & $24.95\pm0.36$ & $23.56\pm0.17$ & $23.13\pm0.13$ & $21.79\pm0.02$ & $21.22\pm0.01$ & $21.08\pm0.07$ & $20.35\pm0.10$\\
COSLA-16N & 150.21490 & 2.55930 & $26.03\pm0.11$ & $24.51\pm0.03$ & $23.89\pm0.02$ & $23.46\pm0.02$ & $23.04\pm0.01$ & $22.67\pm0.05$ & $21.99\pm0.03$ & $21.48\pm0.03$ & $20.84\pm0.02$ & $20.08\pm0.01$ & $19.83\pm0.01$ & $19.76\pm0.05$ & $19.99\pm0.08$\\
COSLA-17N & 150.40340 & 2.18600 & $27.37\pm0.34$ & $26.10\pm0.11$ & $25.22\pm0.07$ & $24.85\pm0.08$ & $24.70\pm0.06$ & $24.65\pm0.25$ & $24.15\pm0.19$ & $23.92\pm0.23$ & $23.10\pm0.13$ & $22.42\pm0.05$ & $21.93\pm0.06$ & $21.84\pm0.31$ & $21.29\pm0.25$\\
COSLA-18 & 150.17990 & 2.08860 & $>27.69$ & $>27.79$ & $>27.41$ & $>27.21$ & $26.12\pm0.21$ & $>25.34$ & $24.03\pm0.17$ & $22.98\pm0.10$ & $22.18\pm0.06$ & $20.79\pm0.01$ & $20.37\pm0.01$ & $20.03\pm0.07$ & $20.03\pm0.08$\\
COSLA-23N & 150.04230 & 2.22640 & $>27.69$ & $>27.79$ & $>27.41$ & $26.98\pm0.46$ & $26.55\pm0.29$ & $>25.34$ & $>25.21$ & $>24.75$ & $23.21\pm0.14$ & $22.16\pm0.04$ & $21.65\pm0.04$ & $21.63\pm0.26$ & $21.00\pm0.19$\\
COSLA-35 & 150.09850 & 2.36530 & $>27.69$ & $>27.79$ & $27.22\pm0.38$ & $26.89\pm0.43$ & $26.11\pm0.20$ & $>25.34$ & $24.83\pm0.33$ & $23.53\pm0.16$ & $22.50\pm0.08$ & $21.07\pm0.01$ & $20.62\pm0.02$ & $20.37\pm0.09$ & $20.14\pm0.09$\\
COSLA-47 & 150.13890 & 2.43380 & $>27.69$ & $27.24\pm0.28$ & $26.34\pm0.19$ & $25.85\pm0.19$ & $25.37\pm0.11$ & $>25.34$ & $24.76\pm0.31$ & $23.44\pm0.15$ & $22.47\pm0.08$ & $21.18\pm0.02$ & $20.78\pm0.02$ & $20.45\pm0.09$ & $20.53\pm0.13$\\
COSLA-54 & 149.65830 & 2.23570 & $>27.69$ & $26.49\pm0.15$ & $26.25\pm0.17$ & $26.35\pm0.28$ & $26.29\pm0.24$ & $>25.34$ & $24.63\pm0.28$ & $23.95\pm0.23$ & $22.53\pm0.08$ & $21.24\pm0.02$ & $20.86\pm0.02$ & $20.38\pm0.09$ & $20.45\pm0.12$\\
\hline
\end{tabular}
\end{scriptsize}

%% file: photometry2.tex
\begin{scriptsize}
\begin{tabular}{lcccccccccccccccc}
\hline
ID & RA & DEC & $B_j$ & $g+$ & $V_j$ & $r+$ & $i+$ & $z+$ & $Y$ & $J$ & $H$ & $Ks$ & $3.6\mu m$ & $4.5\mu m$ & $5.8\mu m$ & $8.0\mu m$\\      
\hline
AzTEC7 & 150.07529 & 2.80842 & $25.16$$\pm$$0.03$ & $25.35$$\pm$$0.07$ & $24.96$$\pm$$0.06$ & $24.88$$\pm$$0.06$ & $24.32$$\pm$$0.04$ & $23.80$$\pm$$0.05$ & $23.48$$\pm$$0.18$ & $22.20$$\pm$$0.05$ & $21.64$$\pm$$0.05$ & $21.13$$\pm$$0.06$ & $19.64$$\pm$$0.02$ & $19.64$$\pm$$0.03$ & $19.16$$\pm$$0.08$ & $19.42$$\pm$$0.13$\\
AzTEC12 & 150.14708 & 2.73144 & $26.22$$\pm$$0.10$ & $25.98$$\pm$$0.10$ & $25.21$$\pm$$0.06$ & $24.78$$\pm$$0.03$ & $24.26$$\pm$$0.06$ & $23.79$$\pm$$0.03$ & $23.46$$\pm$$0.10$ & $23.02$$\pm$$0.10$ & $21.89$$\pm$$0.04$ & $21.51$$\pm$$0.04$ & $20.29$$\pm$$0.01$ & $19.97$$\pm$$0.01$ & $19.49$$\pm$$0.02$ & $19.54$$\pm$$0.04$\\
\hline
\end{tabular}
\end{scriptsize}